\documentclass[prl,twocolumn,english,superscriptaddress]{revtex4-1}

\usepackage[T1]{fontenc}
\usepackage{newtxtext}
\usepackage{newtxmath}

\usepackage{amsmath}
\usepackage{mathtools}
\usepackage{amsfonts}
\usepackage{bm}
\usepackage{xfrac}

\usepackage{color}
\usepackage{xcolor}
\usepackage[colorlinks,
            citecolor=blue,
            linkcolor=blue,
            urlcolor=blue]{hyperref}

\usepackage{graphicx}
\usepackage[all]{hypcap} 
            
\makeatletter
\newsavebox{\@brx}
\newcommand{\llangle}[1][]{\savebox{\@brx}{\(\m@th{#1\langle}\)}%
  \mathopen{\copy\@brx\kern-0.5\wd\@brx\usebox{\@brx}}}
\newcommand{\rrangle}[1][]{\savebox{\@brx}{\(\m@th{#1\rangle}\)}%
  \mathclose{\copy\@brx\kern-0.5\wd\@brx\usebox{\@brx}}}
\makeatother

\begin{document}


%
%

\title{Witnessing quantum criticality and entanglement in the triangular antiferromagnet KYbSe$_2$}

\author{A. O. Scheie} 
\email{scheieao@ornl.gov}
\affiliation{Neutron Scattering Division, Oak Ridge National Laboratory, Oak Ridge, TN 37831, USA}

\author{E. A. Ghioldi}
\affiliation{Department of Physics and Astronomy, University of Tennessee, Knoxville, TN, USA}
\affiliation{Instituto de Física Rosario (CONICET) and Universidad Nacional de Rosario, Boulevard 27 de Febrero 210 bis, (2000) Rosario, Argentina}
	
\author{J. Xing}
\address{Materials Science and Technology Division, Oak Ridge National Laboratory, Oak Ridge, Tennessee 37831, USA}

\author{J. A. M. Paddison}
\address{Materials Science and Technology Division, Oak Ridge National Laboratory, Oak Ridge, Tennessee 37831, USA}

\author{N. E. Sherman}
\address{Department of Physics, University of California, Berkeley, California 94720, USA}
\address{Materials Sciences Division, Lawrence Berkeley National Laboratory, Berkeley, California 94720, USA}

\author{M. Dupont}  
\address{Department of Physics, University of California, Berkeley, California 94720, USA}
\address{Materials Sciences Division, Lawrence Berkeley National Laboratory, Berkeley, California 94720, USA}

\author{L. D. Sanjeewa}
\address{University of Missouri Research Reactor (MURR), Columbia, MO, 65211, USA}
\address{Department of Chemistry, University of Missouri, Columbia, MO, 65211, USA}

\author{Sangyun Lee}
\affiliation{Los Alamos National Laboratory, Los Alamos, New Mexico 87545, USA}
\affiliation{Quantum Science Center, Oak Ridge National Laboratory, TN 37831, USA}

\author{A.J. Woods}
\affiliation{Los Alamos National Laboratory, Los Alamos, New Mexico 87545, USA}
\affiliation{Quantum Science Center, Oak Ridge National Laboratory, TN 37831, USA}

\author{D. Abernathy}
\affiliation{Neutron Scattering Division, Oak Ridge National Laboratory, Oak Ridge, TN 37831, USA}
    
\author{D. M. Pajerowski}  
\affiliation{Neutron Scattering Division, Oak Ridge National Laboratory, Oak Ridge, TN 37831, USA}

\author{T. J. Williams}  
\affiliation{Neutron Scattering Division, Oak Ridge National Laboratory, Oak Ridge, TN 37831, USA}

\author{Shang-Shun Zhang} 
\affiliation{School of Physics and Astronomy and William I. Fine Theoretical Physics
Institute, University of Minnesota, Minneapolis, MN 55455, USA}

\author{L. O. Manuel}
\affiliation{Instituto de Física Rosario (CONICET) and Universidad Nacional de Rosario, Boulevard 27 de Febrero 210 bis, (2000) Rosario, Argentina}

\author{A. E. Trumper}
\affiliation{Instituto de Física Rosario (CONICET) and Universidad Nacional de Rosario, Boulevard 27 de Febrero 210 bis, (2000) Rosario, Argentina}

\author{C. D. Pemmaraju}
\affiliation{Stanford Institute for Materials and Energy Sciences, SLAC National Accelerator Laboratory, Stanford, CA 94025, USA}

\author{A. S. Sefat} 
\affiliation{Materials Science and Technology Division, Oak Ridge National Laboratory, Oak Ridge, Tennessee 37831, USA}

\author{D. S. Parker} 
\affiliation{Materials Science and Technology Division, Oak Ridge National Laboratory, Oak Ridge, Tennessee 37831, USA}

\author{T. P. Devereaux}
\affiliation{Stanford Institute for Materials and Energy Sciences, SLAC National Accelerator Laboratory, Stanford, CA 94025, USA}
\affiliation{Department of Materials Science and Engineering, Stanford University, Stanford, CA 94305, USA}

\author{R. Movshovich}
\affiliation{Los Alamos National Laboratory, Los Alamos, New Mexico 87545, USA}
\affiliation{Quantum Science Center, Oak Ridge National Laboratory, TN 37831, USA}

\author{J. E. Moore}
\address{Department of Physics, University of California, Berkeley, California 94720, USA}
\address{Materials Sciences Division, Lawrence Berkeley National Laboratory, Berkeley, California 94720, USA}
\affiliation{Quantum Science Center, Oak Ridge National Laboratory, TN 37831, USA}

\author{C. D. Batista} 
\email{cbatist2@utk.edu}
\affiliation{Department of Physics and Astronomy, University of Tennessee, Knoxville, TN, USA}
\affiliation{Shull Wollan Center - A Joint Institute for Neutron Sciences, Oak Ridge National Laboratory, TN 37831. USA}

\author{D. A. Tennant}
\affiliation{Neutron Scattering Division, Oak Ridge National Laboratory, Oak Ridge, TN 37831, USA}
\affiliation{Quantum Science Center, Oak Ridge National Laboratory, TN 37831, USA}
\affiliation{Shull Wollan Center - A Joint Institute for Neutron Sciences, Oak Ridge National Laboratory, TN 37831. USA}
	


\begin{abstract}
	The Heisenberg triangular lattice quantum spin liquid and the phase transitions to nearby magnetic orders have received much theoretical attention, but clear experimental manifestations of these states are rare. This work investigates a new spin-half Yb$^{3+}$ delafossite material, KYbSe$_2$, whose inelastic neutron scattering spectra reveal a diffuse continuum with a sharp lower bound. Applying entanglement witnesses to the data reveals  significant multipartite entanglement spread between its neighbors, and analysis of its magnetic exchange couplings shows close proximity to the triangular lattice Heisenberg quantum spin liquid. Key features of the data are reproduced by Schwinger-boson theory and tensor network calculations with a significant second-neighbor coupling $J_2$. The strength of the dynamical structure factor at the $K$ point shows a scaling collapse in $\hbar\omega/k_\mathrm{B}T$ down to 0.3 K, indicating a second-order quantum phase transition. Comparing this to previous theoretical work suggests that the proximate phase at larger $J_2$ is a gapped $\mathbb{Z}_2$ spin liquid, resolving a long-debated issue. We thus show that KYbSe$_2$ is close to a spin liquid phase, which in turn sheds light on the theoretical phase diagram itself.
\end{abstract}

\maketitle


\section{Introduction}  

A quantum spin liquid (QSL) is an elusive state of matter where magnetic degrees of freedom on a lattice are in a highly entangled, fluctuating ground state with exotic quasiparticle excitations~\cite{Knolle2019_review,broholm2019quantum,Savary_2016review,Zhou2017}. The quasiparticles are of singular interest for, {\it e.g.}, quantum information applications~\cite{broholm2019quantum,tokura2017emergent} but have been, together with the extended entanglement, frustratingly difficult to identify experimentally.

The search for a QSL is a very active field of research with many candidate QSL materials: from organic materials~\cite{Yamashita2008,Itou2008} to 2D Kagome minerals~\cite{han2012fractionalized} to rare earth pyrochlores \cite{Gaudet_2019,Gao2019}. However,
despite tremendous effort, no materials have unambiguously been shown to realize a genuine QSL. This is partly because many studies focus on ``negative evidence'' such as lack of magnetic order, lack of coherent excitations, {\it etc.}, which are not unique to QSL states. Instead, to conclusively identify an experimental QSL, ``positive evidence'' is needed: experimental evidence of either (i) a highly entangled ground state, or (ii) exotic quasiparticles---both key properties of a QSL.

Beginning with Anderson's resonating valence bond state~\cite{Anderson1973}, the two-dimensional (2D) triangular geometry has long been studied as a platform for QSLs. Although the simplest spin-$1/2$ model with nearest-neighbor antiferromagnetic Heisenberg interactions orders magnetically in a $120^{\circ}$ phase, the magnetic frustration makes the order weak~\cite{PhysRevLett.99.127004}. The magnetic order can be further destabilized by additional interactions such as a next-nearest-neighbor exchange coupling. In that case, it has been found that a realistic strength as small as $\approx 10\%$ of the main interaction is enough to destroy magnetic order and bring the system into a QSL phase~\cite{PhysRevB.92.041105,PhysRevB.92.140403,PhysRevB.93.144411,PhysRevB.94.121111,PhysRevB.95.035141,PhysRevB.96.075116,PhysRevLett.123.207203}
(which is continuously connected to a QSL phase driven by nearest neighbor anisotropic exchange~\cite{Zhu_2018}).
Determining the nature of the QSL phase is a theoretical challenge, with proposals ranging from gapped $\mathbb{Z}_2$ and gapless $U(1)$ Dirac to chiral~\cite{PhysRevB.92.041105,PhysRevB.92.140403,PhysRevB.93.144411,PhysRevB.94.121111,PhysRevB.95.035141,PhysRevB.96.075116,PhysRevLett.123.207203}, with no clear consensus within the community. In order to discern among possible QSL states, experiments are called for.



\begin{figure*}
	\centering\includegraphics[width=0.82\textwidth]{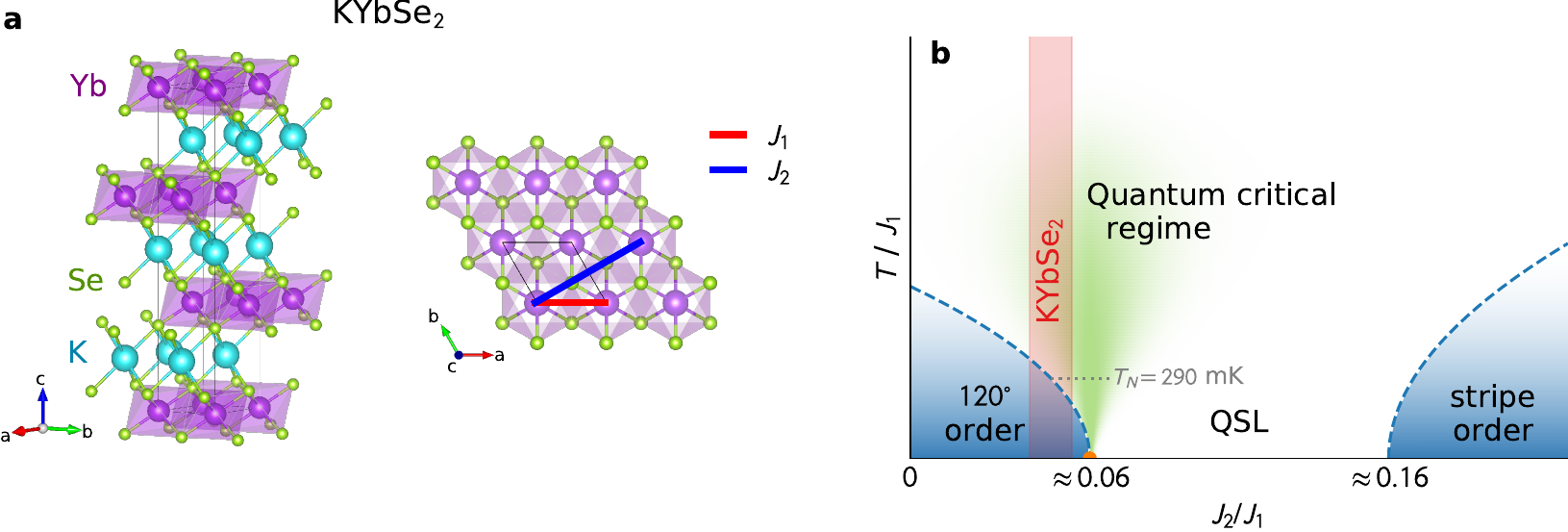}
	\caption{\textbf{\textsf{Crystal structure and phase diagram of KYbSe$_2$.}} Panel $\textbf{\textsf{a}}$ shows the crystal structure with a side view of the stacked triangular layers and a top view showing the Yb$^{3+}$ triangular lattice mediated by Se$^{2-}$ ions. Panel $\textbf{\textsf{b}}$ shows a schematic phase diagram of the triangular lattice Heisenberg antiferromagnet as a function of second neighbor exchange strength $J_2$. This includes a zero temperature $120^{\circ}$ ordered phase for $J_2/J_1\lesssim 0.06$, a zero temperature stripe ordered phase for $J_2/J_1\gtrsim 0.16$, and an intermediate QSL phase~\cite{PhysRevB.92.041105,PhysRevB.92.140403,PhysRevB.93.144411,PhysRevB.94.121111,PhysRevB.95.035141,PhysRevB.96.075116,PhysRevLett.123.207203}. Near the quantum critical points we expect quantum critical regime extending at finite temperature.}
	\label{flo:KYS_crystal}
\end{figure*}

In the last decade, Yb$^{3+}$ based materials  have become popular as QSL candidates because of the  Yb$^{3+}$ effective $S=1/2$ state. 
Most recently, a class of delafossite materials have been proposed as relatively disorder-free QSL candidates, including NaYbO$_2$~\cite{Ding_2019_NYO,Bordelon2019,Ranjinth2019}, NaYbS$_2$~\cite{Baenitz_2018,sarkar2019quantum}, NaYbSe$_2$~\cite{Ranjith2019_2,Dai_2021} and CsYbSe$_2$~\cite{xie2021field}. Each of these materials shows diffuse excitations and no long-range magnetic order down to 0.4~K or lower, but because neither are unique to QSL states (both are also caused by spin glass~\cite{Zhang_2019}, random singlet phases~\cite{Zhu_2017_YMGO}, or 2D magnetic order only in the zero temperature limit), they remain QSL candidates only.

Here we investigate a new member of the Yb$^{3+}$ delafossite family: KYbSe$_2$ which forms a layered triangular lattice of magnetic Yb$^{3+}$ ions, see Fig.~\ref{flo:KYS_crystal}(a). This material shows no long-range order above $400$~mK~\cite{xing2021_KYS}, and finite-field ordered phases similar to NaYbO$_2$~\cite{Bordelon2019} and NaYbS$_2$~\cite{sarkar2019quantum}. Thus it appears promising as a quantum spin liquid candidate. We successfully apply entanglement witnesses one-tangle, two-tangle, and quantum Fisher information (QFI) to KYbSe$_2$~\cite{scheie2021witnessing}, and detect the presence of quantum entanglement at low temperatures. Using a combination of density-functional theory, Onsager reaction field theory, Schwinger bosons, and tensor network approaches to model KYbSe$_2$, we find that its physics is well-captured by a microscopic spin-$1/2$ Hamiltonian with nearest and next-nearest neighbor Heisenberg interactions on the triangular lattice in proximity to the QSL phase [see Fig.~\ref{flo:KYS_crystal}(b)]. Finally, the neutron spectrum displays signatures of quantum criticality and fractionalized spinon quasiparticles. Together, these results show KYbSe$_2$ to be proximate to a spin liquid with positive evidence for the two key features: quantum entanglement and exotic quasiparticles.

\section{Experiments}

\subsection{Cold Neutron Chopper Spectrometer (CNCS)}

\begin{figure*}
	\centering\includegraphics[width=0.98\textwidth]{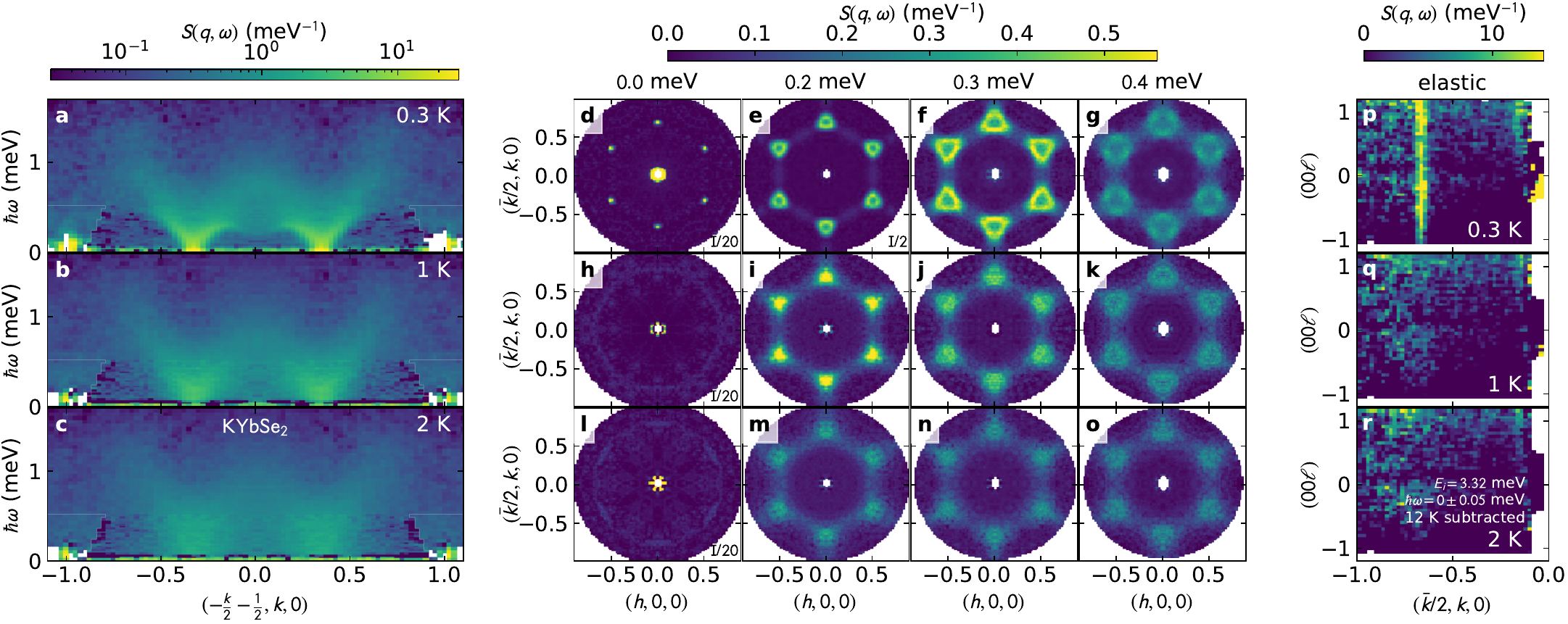}
	\caption{\textbf{\textsf{Neutron spectrum of KYbSe$_2$ at $0.3$~K (top row), $1$~K (middle row) and $2$~K (bottom row).}} The left panels show energy-dependent scattering along $(-k/2-1/2,k,0)$ which includes where the dispersion touches zero energy. These plots comprise data with $E_i=1.55$~meV below $\hbar\omega=0.5$~meV, and $E_i=3.32$~meV above $\hbar\omega=0.5$~meV. Note the roton-like mode at $0.3$~K and the diffuse high energy spectrum. The center panels show constant energy slices measured with $E_i=1.55$~meV. Panel \textbf{\textsf{d}} shows elastic intensity associated with $(1/3,1/3)$ static magnetism which disappears at higher temperatures. The right panels plot this elastic intensity as a function of $\ell$, which reveals almost no dependence on $\ell$, and thus $2$D correlations.}
	\label{flo:KYS_CNCS}
\end{figure*}

We measured the low-energy KYbSe$_2$ single crystal neutron spectrum on the CNCS spectrometer~\cite{CNCS} at Oak Ridge National Laboratory's Spallation Neutron Source~\cite{mason2006spallation} between $0.3$~K and $2$~K using a ${}^3$He refrigerator (for details, see the methods section). The data are shown in Fig.~\ref{flo:KYS_CNCS}.

In the elastic channel, quasi-Bragg intensities appear between $1$~K and $0.3$~K which look like $(1/3,1/3)$ Bragg peaks signaling 120$^{\circ}$ correlations. They 
have no dependence upon $\ell$ [Fig.~\ref{flo:KYS_CNCS}(p)] which evidences truly $2$D static correlations and weak inter-plane exchange. (As an aside, this weak inter-plane exchange is expected given the fragility of the crystal inter-plane bonds: KYbSe$_2$ planes readily flake off when the crystals are not handled gently.)
Fitting the in-plane scattering to extract the correlation length using the (101) peak to define the resolution width, we find the magnetic peaks are much broader than the nuclear Bragg peaks with a fitted correlation length of $47\pm 10$~\AA~ at $0.3$~K ($\approx 10$ unit cells in the plane).
In the supplemental information, we show these quasi-Bragg intensities become well-defined Bragg peaks below a $T_N=290$~mK, evidencing a transition to 120$^{\circ}$ long range magnetic order.

In the inelastic channel, two features stand out in the low-temperature KYbSe$_2$ spectrum: a diffuse continuum of excitations, and a pronounced $0.2$~meV energy minimum at $M=(1/2,0,0)$. Both of these features are seen in the triangular lattice compound Ba$_3$CoSb$_2$O$_9$~\cite{Macdougal_2020,Zhou_2012,Ito2017,Ma_2016}. The ``roton-like'' minimum at $M$ is a generic feature of the $2$D quantum triangular lattice Heisenberg antiferromagnet and is a nonlinear effect (i.e., not captured by linear spin wave theory)~\cite{Zheng_2006,Starykh_2006,Chernyshev_2009}. Fits to the KYbSe$_2$ roton mode [see Supplementary Information] show a mode maximum of $0.288(12)$~meV, and a roton minimum $0.200(13)$~meV at $M$. This indicates that strong quantum effects are at work in KYbSe$_2$.

The continuum, meanwhile, extends up to $1.6$~meV, over five times the roton mode bandwidth. This is far too high in energy to be a two-magnon continuum, which is limited to twice the single-magnon bandwidth. Integrating the scattering intensity over the entire Brillouin zone shows that $\sim 60\%$ of the magnetic scattering intensity is found above $0.4$~meV, compared to only $\sim 29\%$ between $0.05$~meV and $0.4$~meV, showing that the continuum scattering carries twice the spectral weight of the ``single-magnon'' intensity. 
Perhaps most interestingly, the continuum in KYbSe$_2$ comes all the way down to the sharp low-energy modes [Fig.~\ref{flo:KYS_CNCS}(a)]. 
The KYbSe$_2$ diffuse continuum with a sharp lower bound is reminiscent of the Van Hove singularity observed in $1$D spin chains---which are known to have highly entangled ground states with fractionalized spinon excitations~\cite{scheie2021witnessing,laurell2020dynamics,Lake2013}. This well-defined lower bound to the continuum distinguishes KYbSe$_2$ from other QSL candidates, such as NaCaNi$_2$F$_7$~\cite{plumb2019continuum}, YbMgGaO$_4$~\cite{Shen2016,Paddison2017}, and herbertsmithite~\cite{han2012fractionalized} which are diffuse everywhere. 
This also distinguishes KYbSe$_2$ from NaYbSe$_2$, which does not have a lower bound to its continuum  \cite{Dai_2021}. Whether this signals a genuine QSL in NaYbSe$_2$ or the effect of its 3\% site disorder is unclear.

\subsection{Wide Angular-Range Chopper Spectrometer (ARCS)}

In order to understand how ``quantum'' the KYbSe$_2$ spins are,  we measured the crystal electric field (CEF) excitations using the ARCS spectrometer~\cite{ARCS} at Oak Ridge National Lab's Spallation Neutron Source. We fitted a single-ion CEF Hamiltonian to the excitations using \textit{PyCrystalField}~\cite{PyCrystalField} software; data and fits 
are shown in Fig.~\ref{flo:KYS_CEF} [Details on the CEF fitting procedure are given in the Supplemental Information].

\begin{figure}
	\centering\includegraphics[width=0.48\textwidth]{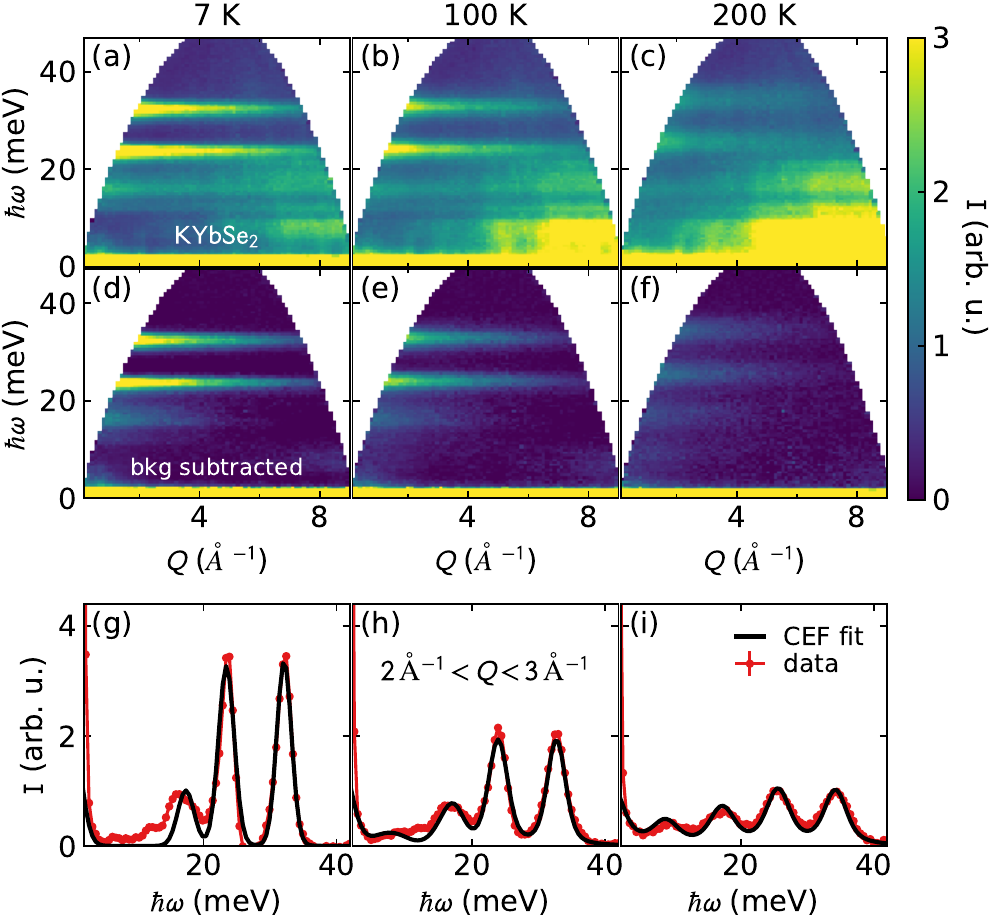}
	\caption{\textbf{\textsf{Crystal field spectrum of KYbSe$_2$.}} The top row shows the measured intensity at $E_i=50$~meV at $7$~K, $100$~K, and $200$~K. The middle row shows the same data with the self-consistent $300$~K background subtracted. The bottom row shows a cut through the data between $2$ and $3$ \AA${}^{-1}$ compared to the fitted CEF Hamiltonian}
	\label{flo:KYS_CEF}
\end{figure}

The best fit crystal field Hamiltonian shows a ground state doublet 
\begin{equation}
	| \psi_{\pm} \rangle = 0.78(3) \big| \mp \frac{5}{2} \big\rangle \mp 0.44(4) \big| \pm \frac{1}{2} \big\rangle - 0.44(3) \big| \pm \frac{7}{2} \big\rangle
\end{equation}
with a first excited state at $17.1(3)$~meV. This ground state doublet gives a weak easy plane $g$ tensor $g_{xx}=g_{yy}=3.0(2)$, and $g_{zz}=1.8(6)$. As the large $g_{xx}$ and $g_{yy}$ indicate, the ground state doublet allows for significant quantum tunnelling from effective spin operator $J_{\pm}$. 
Thus, the Yb$^{3+}$ spins in KYbSe$_2$ can be treated like a spin-$1/2$ system.


\section{Entanglement Witnesses}

Diffuse neutron excitations suggest---but do not prove---proximity to QSL behavior, which makes their mere observation ambiguous.
Fortunately, entanglement witnesses provide a way out of this quandary: by quantifying entanglement in KYbSe$_2$ we can 
rule out trivial phases like random singlet or valence bond solid states.

We apply three entanglement witnesses to the KYbSe$_2$ data (same as in Refs.~\cite{scheie2021witnessing,laurell2020dynamics}):
one-tangle $\tau_1$, which quantifies entanglement of a spin with the entire system~\cite{Wootters_1998, PhysRevA.61.052306};
the two-tangle $\tau_2$, which quantifies the total bipartite entanglement derived from quantum concurrence~\cite{Roscilde_2004,Amico_2006}; and QFI which gives a lower bound on multi-partite entanglement~\cite{Hauke2016}. 
For details of these calculations, see the methods section.

One-tangle is calculated from the static spin  at zero temperature and ranges between zero (unentangled state) and one (maximally entangled state). 
In the supplemental information, we extract the $T \rightarrow 0$ static moment from fits to the Yb$^{3+}$ zero temperature heat capacity nuclear Schottky anomaly, giving a local static ordered moment of $0.58(1) \> {\rm \mu_B}$ per ion. 
Comparing this to the maximal ground state static moment  from the crystal field fit $\mu = 1.48(8)$~meV, this is only 39(2)~\% of the maximal static moment. Projecting it onto an effective S=1/2, the one-tangle $\tau_1 = 0.85(2)$. This evidences substantial spin entanglement in KYbSe$_2$.

\begin{figure}
	\centering\includegraphics[width=0.47\textwidth]{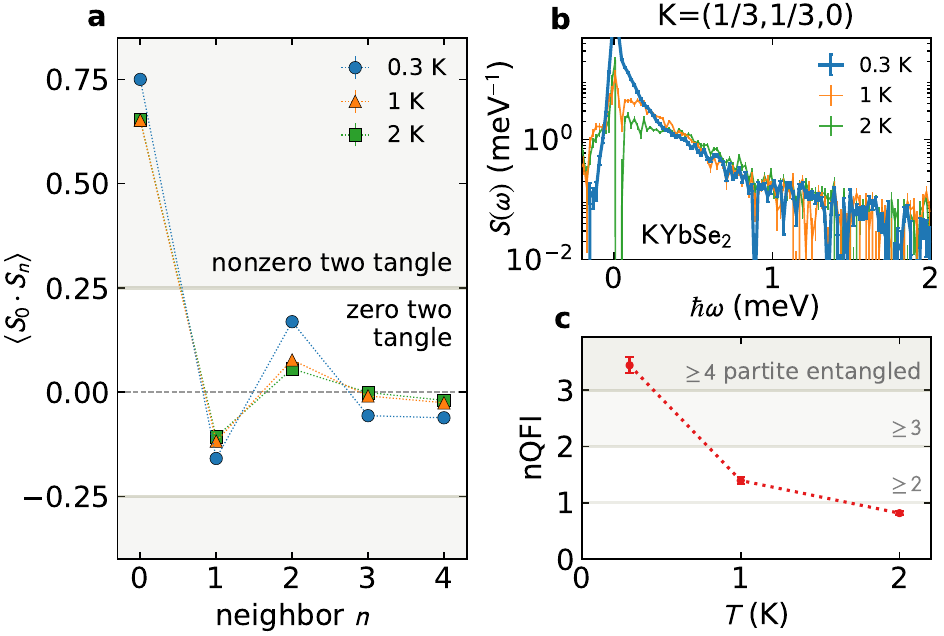}
	\caption{\textbf{\textsf{KYbSe$_2$ entanglement witnesses.}} Panel \textbf{\textsf{a}} shows the real-space spin spin correlations for the first four neighbors used to calculate two-tangle: a measure of bipartite entanglement~\cite{Roscilde_2004}. None of the neighbor spin-spin correlators exceed the classical threshold, which means two-tangle is zero at all temperatures. Panel \textbf{\textsf{b}} shows the intensity as a function of energy at $K=(1/3,1/3,0)$ (integrated over $\pm 0.025$~RLU), with the normalized quantum Fisher information (nQFI) shown below in panel \textbf{\textsf{c}}. 
		At $0.3$~K, 
		nQFI exceeds the threshold for $\geq 4$-partite entanglement per spin, indicating a highly entangled state.
		Thus no classical arrangement of spins could produce the observed KYbSe$_2$ spectrum.
		\label{flo:Entanglement}}
\end{figure}

Two tangle is calculated from the Fourier transform to real space of the frequency integrated $S(\boldsymbol{q},\omega)$ and is shown in Fig.~\ref{flo:Entanglement}. We find that none of the neighboring spin correlators exceed the classical $\langle\boldsymbol{S}_i\cdot\boldsymbol{S}_j\rangle$ threshold, and thus two-tangle is zero for all temperatures in KYbSe$_2$. This makes sense given quantum monogamy~\cite{PhysRevLett.96.220503} and six equivalent nearest neighbors for every site to distribute its entanglement. The significance of this will become apparent shortly.

The third entanglement witness, QFI, is calculated from an energy integral at a specific point in $Q$~\cite{Hauke2016}. For KYbSe$_2$ we evaluate QFI at $(1/3,1/3)$, the wavevector associated with the strongest correlations. The scattering and nQFI are shown in Fig. \ref{flo:Entanglement}.  
At $1$~K and $2$~K, $\mathrm{nQFI}=1.39(6)$ and $0.82(4)$ respectively, indicating nonzero witnessed entanglement below 1~K. At $0.3$~K, $\mathrm{nQFI}=3.4(2)$, which shows multipartite entanglement with an entanglement depth of four spins in a highly correlated ground state. 
Importantly, this nonzero entanglement appears for all six $Q$ vectors along the nearest neighbor bond directions.

Clearly, these entanglement witnesses reveal appreciable spin entanglement in KYbSe$_2$, but the combination of two-tangle and QFI is particularly revealing. The zero two-tangle shows that the entanglement is spread out over nearest neighbors rather than pairing with a particular neighbor in singlets. This is what one expects for a highly-entangled ground state (c.f. vanishing two-tangle for the Kitaev spin liquid~\cite{PhysRevLett.98.247201}). Meanwhile the QFI shows at least bipartite entanglement within the $(1/3,1/3)$ correlations. Both of these rule out classical glassiness or random singlet formation. Instead, they point to many sites entangled together at the lowest temperatures---as one would expect for a QSL.

\section{Microscopic modeling}

To better understand the features observed in KYbSe$_2$, and find a microscopic model for the compound, we use a combination of theoretical techniques such as density-functional theory (which showed a magnetic insulating state, discussed in the Methods section), the Onsager reaction field, Schwinger bosons, and tensor networks.


\subsection{Onsager reaction field:\\estimating the exchange ratios}


First, we employ the Onsager reaction field (ORF)~\cite{Paddison_2020} to fit the energy-integrated paramagnetic scattering shown in Fig.~\ref{flo:SchwingerBosons}. This approach neglects quantum fluctuations, but in the paramagnetic regime it is accurate up to a temperature-dependent energy scale normalization~\cite{Huberman_2008} which in our case is unknown. Despite this limitation, ORF does give relative anisotropy and ratios between exchanges. Using the $g$-tensor derived from crystal electric field fits and allowing for first and second neighbor exchange, we find the off-diagonal anisotropic exchange is small and the nearest neighbor exchange is isotropic to within uncertainty [see Methods] making KYbSe$_2$ a very good approximation to a triangular lattice Heisenberg antiferromagnet described by the microscopic $J_1-J_2$ Hamiltonian,
\begin{equation}
	\hat{\mathcal{H}} = J_1\sum_{\langle i,j\rangle}\hat{\boldsymbol{S}}_i\cdot\hat{\boldsymbol{S}}_j+J_2\sum_{\langle\langle i,j\rangle\rangle}\hat{\boldsymbol{S}}_i\cdot\hat{\boldsymbol{S}}_j.
	\label{eq:hamiltonian}
\end{equation}
What is more, the fitted $J_2/J_1=0.047(7)$. This is extremely close to the predicted phase boundary between $120^{\circ}$ magnetic order and a QSL phase on the triangular lattice Heisenberg antiferromagnet: $J_2/J_1\approx 0.06$~\cite{PhysRevB.92.041105,PhysRevB.92.140403,PhysRevB.93.144411,PhysRevB.94.121111,PhysRevB.95.035141,PhysRevB.96.075116,PhysRevLett.123.207203}. Thus, ORF fits show KYbSe$_2$ has nearly isotropic Heisenberg exchange and is very close to a quantum spin liquid phase.


\subsection{Schwinger bosons: comparing the neutron spectrum}

To understand the inelastic neutron spectrum, we turn to a Schwinger Boson (SB)  theory beyond the mean field level~\cite{Arovas1988,Auerbach1994,Ghioldi_2018}. This is a parton formulation where
the Heisenberg model is expressed in terms of interacting spin-$1/2$ bosons or spinons, whose  condensation leads to long-range magnetic ordering~\cite{Arovas1988,Auerbach1994}. For details, see the Methods section.

\begin{figure}
	\centering\includegraphics[width=0.48\textwidth]{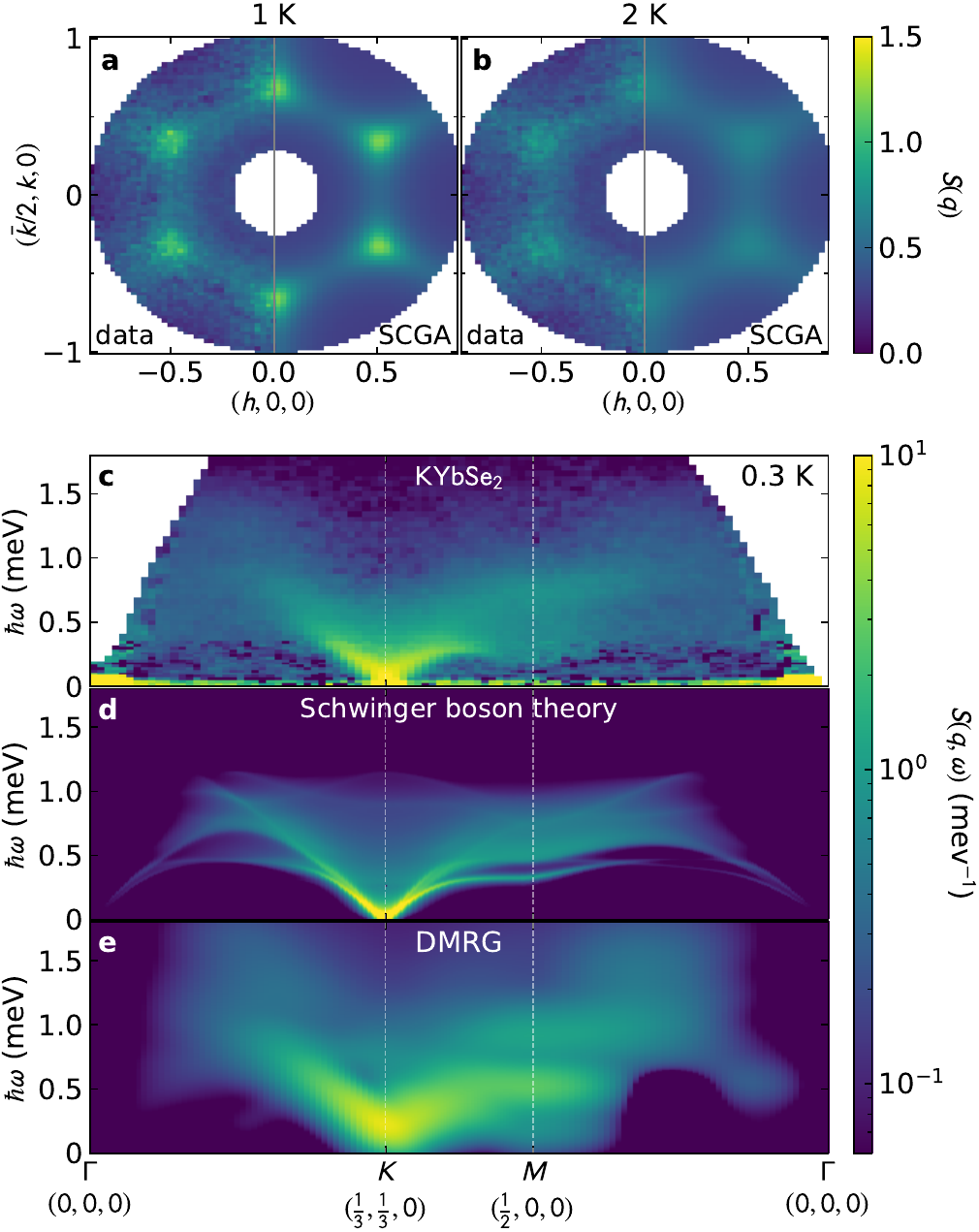}
	\caption{\textbf{\textsf{Comparison between experimental KYbSe$_2$ scattering and theoretical simulations.}} Panels \textbf{\textsf{a}} and \textbf{\textsf{b}} show Onsager reaction field (ORF) fits to energy-integrated paramagnetic KYbSe$_2$ scattering at $1$~K and $2$~K. In each panel, the data is on the left and the fit is on the right. Panels \textbf{\textsf{c}}-\textbf{\textsf{e}} show neutron scattering along high-symmetry directions. \textbf{\textsf{c}} shows the experimental data for KYbSe$_2$ and \textbf{\textsf{d}} shows the zero-temperature simulated spectrum from Schwinger boson calculations with $J_1=0.56(3)$~meV and $J_2/J_1=0.05$. Panel \textbf{\textsf{e}} shows tensor network simulations
		of a triangular lattice with the same $J_1$ and $J_2$ on a cylinder with a circumference of 6 sites and length 36 sites. 
		On a qualitative level, the theory captures the continuum excitations observed in experiment.}
	\label{flo:SchwingerBosons}
\end{figure}

The dynamical spin structure factor $S(\boldsymbol{q},\omega)$ at $T=0$ 
using SB ~\cite{Ghioldi_2018} for  $J_2/J_1 = 0.05$ is shown in Fig.~\ref{flo:SchwingerBosons}(d). On a qualitative level, this result captures the features seen in the experimental data: the strong dispersive cone emanating from $K$, the continuum scattering at higher energies, the diffuse high-energy feature at $M$, and the pronounced low-energy ``roton-like'' mode at $M$. We note that the downturn of the roton-like mode is much less pronounced in the SB result because of the lack of $1/N$ corrections to the internal vertices and the single-spinon propagator~\cite{Ghioldi_2018}.
However, the most remarkable aspect of this comparison is that the SB approach captures the intensity modulation of the continuum scattering at higher energies, which is determined by the two-spinon continuum of the SB theory. This correspondence points to the continuum scattering in KYbSe$_2$ originating from its proximity to a deconfined spin liquid state with fractionalized spinon excitations. 

The measured continuum scattering extends up to higher energies than SB predicts: $\approx 1.6$~meV, approximately three times the fitted $J_1=0.56(3)$~meV (see Supplemental Information). We attribute this discrepancy to the lack of $4$-spinon contributions arising from Feynman diagrams which have not been included in the SB calculation~\cite{Ghioldi_2018}. Note that the KYbSe$_2$ continuum extent does match the predicted continuum extent near the $J_2/J_1\approx 0.06$ transition point as calculated by Gutzwiller projected variational Monte Carlo~\cite{Ferrari_2019}.

\subsection{Tensor networks: full spectrum model}

The third technique we use to model the diffuse inelastic neutron scattering is based on tensor networks [see the Methods section]. A related approach was recently used to interpret and describe the scattering of CsYbSe$_2$~\cite{xie2021field}, and provides a full quantum picture of the neutron spectrum. The downside to this technique is finite size effects, which cause broadened modes and gaps in the low energy spectrum. Nevertheless, qualitative comparisons can be made.

The simulated data along high symmetry directions of the Brillouin zone for $J_2/J_1=0.05$ is shown in Fig.~\ref{flo:SchwingerBosons}(e). The overall features of the experimental data are reproduced in the simulations: the asymmetric dispersive modes emanating from $K$, the diffuse continuum extending to high energies, and even the broad $1$~meV feature at $M$. This shows that the triangular lattice Heisenberg $J_1$-$J_2$ model is indeed an appropriate model for KYbSe$_2$. Further microscopic simulations show that most of the high energy scattering remains unchanged as $J_2$ is increased and the system enters the QSL phase, showing that the high-energy scattering can be interpreted as bound spinons of a proximate spin liquid.

\section{Critical Scaling}

\begin{figure}
	\centering\includegraphics[width=0.48\textwidth]{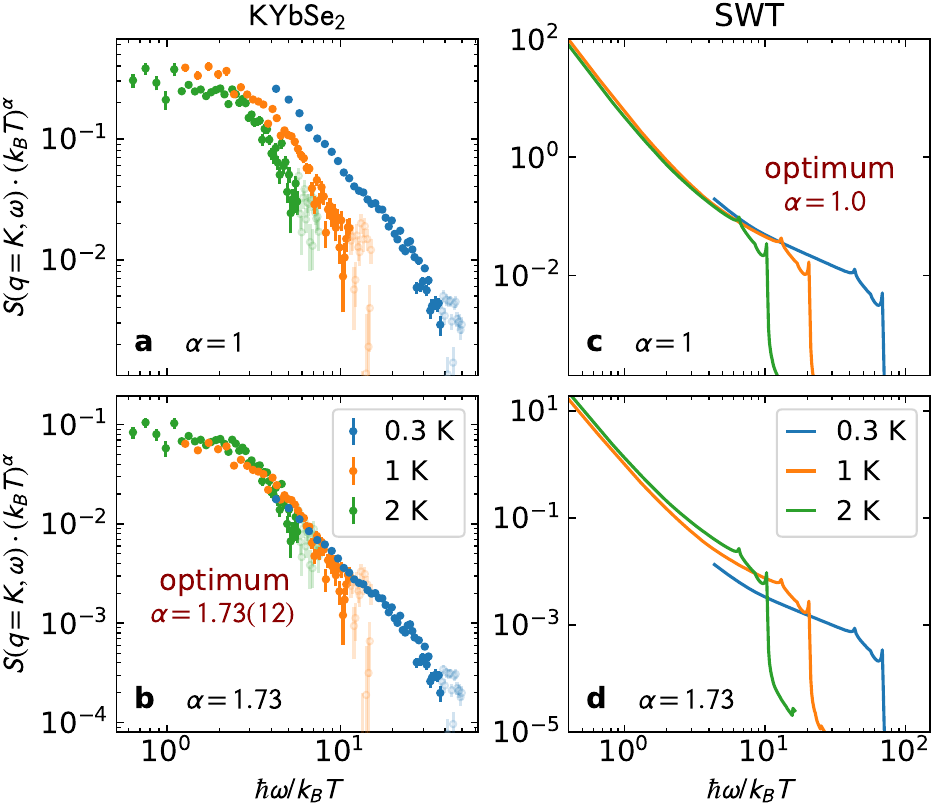}
	\caption{\textbf{\textsf{Critical scaling in KYbSe$_2$, showing data at the $K$ point at three different temperatures scaled by $\hbar\omega/k_\mathrm{B}T$ and $S(\boldsymbol{q}=K,\omega)(k_\mathrm{B}T)^{\alpha}$.}}
		Measured KYbSe$_2$ spectra are on the left column, and calculated spin wave theory (SWT) are on the right column. When $\alpha=1.73(12)$, the KYbSe$_2$ data from the three temperatures follow the same curve, suggesting quantum critical scaling. SWT spectra, meanwhile, overlap when $\alpha = 1.0$. This suggests fundamentally different behavior in KYbSe$_2$ that cannot be captured by non-interacting magnons. 
		(Fitted energy range was $\hbar \omega \leq 1.3$~meV; data above this are plotted in lighter colors.) 
	}
	\label{flo:CriticalScaling}
\end{figure}

So far, the entanglement witnesses and theoretical comparisons indicate that KYbSe$_2$ is close to the $J_1/J_2$ QSL quantum critical point. If this is true, we should see quantum critical scaling in the finite temperature neutron spectrum~\cite{Lake2005,schroder2000onset,Chakravarty_1989,Sachdev_1992}.
Plotting scattered intensity times $(k_B T)^{\alpha}$ versus $\hbar\omega/k_\mathrm{B}T$, shown in Fig. \ref{flo:CriticalScaling}, we see a critical exponent $\alpha=1.73(12)$ over more than a decade in $\omega/T$. Theoretically, the semiclassical spin wave scattering from an ordered Heisenberg triangular lattice predicts an exponent $\alpha=1$. The observed scattering is unquestionably inconsistent with this [Fig. \ref{flo:CriticalScaling}(a)]. Thus this scaling shows that 
the KYbSe$_2$ inelastic spectrum is dominated by non-magnon quasiparticles, confirming the interpretation above of fractionalized spinons.

Elastic Bragg scattering and heat capacity  show a transition to long range magnetic order below $T_N=290$~mK (see supplemental information), showing that KYbSe$_2$ is on the $120^{\circ}$ side of the phase boundary. Nevertheless, 
the critical scaling is strong evidence that KYbSe$_2$ is within the quantum critical regime at finite $T$. 

This scaling holds over a single decade in $\hbar \omega / k_B T$, which may not be enough to definitively establish power law behavior. Nevertheless, if it holds over a larger range, it 
has important implications regarding the nature of the QSL state. Indeed, the gapped $\mathbb{Z}_2$ QSL state proposed by Sachdev~\cite{Sachdev92} is the only liquid which  can be continuously connected with $120^{\circ}$ N\'eel ordered state, as it does not break any symmetries and has lowest energy modes at the $K$-points~\cite{Wang06}. 
(The low energy excitations of the other possibility, a $\pi$-flux state, are gapped at the $K$-points and gapless at the $M$-points, inconsistent with the observations.) The resulting quantum critical point is expected to have a dynamically generated $O(4)$ symmetry~\cite{Azaria90,Chubukov94}.


\section{Conclusion}

These results show that KYbSe$_2$ is within the quantum critical fan of a quantum spin liquid state. CEF fits show an isotropic $J=1/2$ doublet with strong quantum effects, and ORF simulations show a $J_2/J_1$ ratio  within the $120^{\circ}$ ordered phase but very close to the QSL quantum critical point $J_2/J_1\approx 0.06$. Entanglement witnesses reveal an entangled ground state with distributed entanglement, just as  was shown in the 1D case to indicate proximity to quantum criticality \cite{Hauke2016}. Finally, there are strong signs of quantum criticality in the neutron spectrum: (i) the majority spectral weight in the continuum, (ii) the sharp lower continuum bound reminiscent of the $1$D spinon spectrum, (iii) strong correspondence to SB and tensor network simulations near the transition to a spin liquid, and (iv) critical scaling incompatible with semiclassical excitations all indicate that the KYbSe$_2$ excitations are fractionalized spinons of a QSL phase. Thus, despite the existence of magnetic order at the lowest temperature, we propose KYbSe$_2$ as a model for triangular lattice QSL physics at finite energies and temperatures (exactly like many 1D spin chains---c.f. KCuF$_3$ \cite{Lake2005}).

These results have implications beyond just this material. As noted earlier, triangular lattice CsYbSe$_2$ and NaYbSe$_2$  also show features of a QSL phase: with CsYbSe$_2$ possibly more toward the $J_2=0$ limit~\cite{xie2021field}, and NaYbSe$_2$ $J_2/J_1$ possibly within the QSL phase (Yb site disorder notwithstanding)~\cite{Dai_2021}.  This suggests that the periodic table can be used to ``tune'' $J_2/J_1$ such that the delafossite lattice can be brought into and out of a QSL phase depending on the A-site element. This gives a remarkably controlled way to study QSL materials. Another possible way to ``tune'' $J_2/J_1$ could be through hydrostatic pressure---there are even reports of superconductivity in NaYbSe$_2$ under pressure~\cite{Jia_2020,zhang2020pressure}, which suggests pressure does more than just shift magnetic exchange constants.

The family of Yb$^{3+}$ delafossites are a remarkable platform for 2D triangular lattice Heisenberg systems. By controlling $J_2/J_1$, we are able to systematically approach a QSL from the 120$^{\circ}$ ordered phase, which gives a clear pathway towards an experimentally verifiable QSL state.  Scaling behavior in $\hbar \omega / k_B T$ with a nontrivial exponent, i.e., a value inconsistent with gapless spin wave excitations, is observed in the spin correlations down to the lowest temperature measured (0.3 K), with a correlation length of at least ten unit cells.   

While a weakly first-order transition with a long correlation length is possible, the natural interpretation of the results in this work is that the phase transition from 120$^{\circ}$ to a QSL is second order, which combines with previous theoretical work to constrain strongly the nature of the QSL.  One of the frontiers in quantum condensed matter physics is to understand the possible phase transitions between topological and broken-symmetry phases, and the combined experimental and theoretical analysis of KYbSe$_2$ helps clarify one piece of this frontier.
%

\section{Methods}

\subsection{CNCS experiment}\label{app:CNCS}

\begin{figure}
	\centering\includegraphics[width=0.3\textwidth]{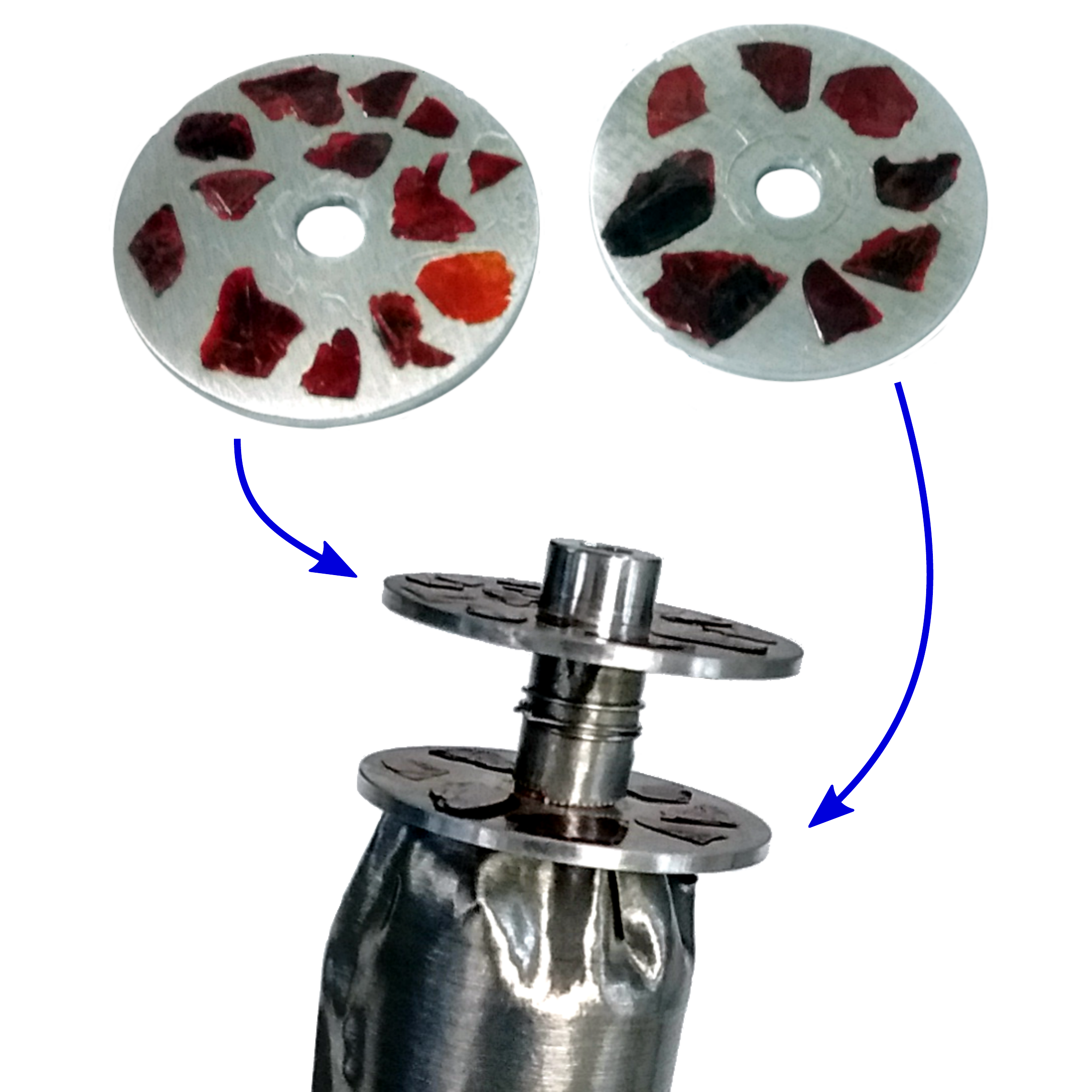}
	\caption{KYbSe$_2$ sample used to measure the low-energy spin excitations on CNCS. 20 crystals were coaligned and glued to two aluminum plates (top) which were then screwed to a copper rod (bottom). 
	{  The different crystals are different shades of red because of their different thicknesses.}
	}
	\label{flo:SampleMount}
\end{figure}

We measured the low-energy spin excitations with the CNCS spectrometer at Oak Ridge National Laboratory's Spallation Neutron Source.
The sample for this experiment consisted of 20 coaligned plate-like crystals glued to aluminum discs (see Fig. \ref{flo:SampleMount}), for a total mass of 200 mg KYbSe$_2$ in the $(hk0)$ scattering plane. The sample was mounted in a ${}^3$He refrigerator and measured with double-disc chopper frequency 300.0 Hz (high-flux mode, 9 degree opening on the double disk).
All CNCS data were corrected for the isotropic Yb$^{3+}$ form factor~\cite{BrownFF}.

\begin{figure}
	\centering\includegraphics[width=0.48\textwidth]{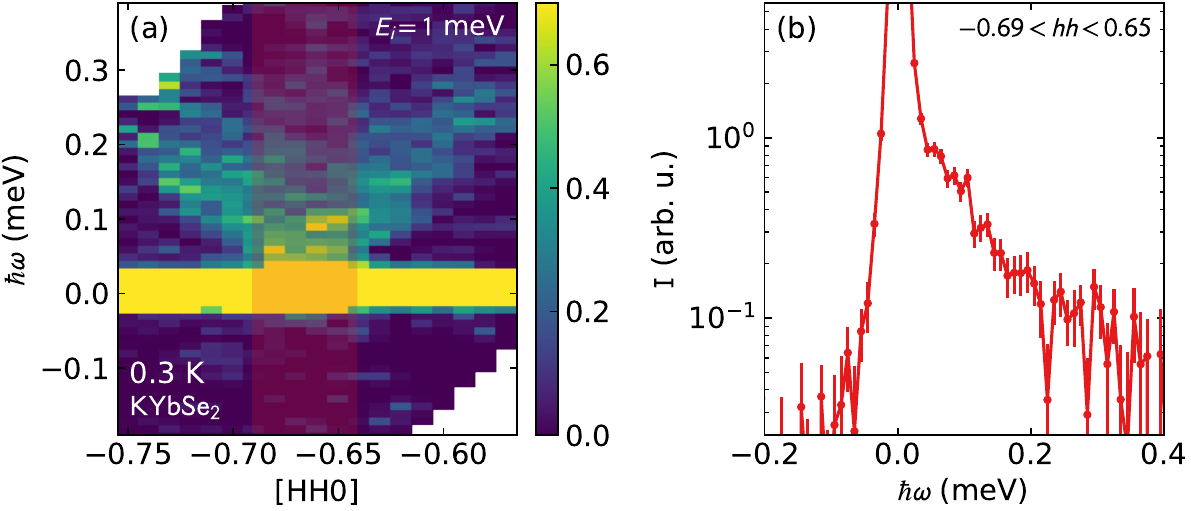}
	\caption{High resolution KYbSe$_2$ scattering at $(1/3,1/3,0)$. Panel (a) shows a slice through the data showing the gapless dispersion. Panel (b) shows a 1D cut indicated by the faint vertical red bar in panel (a). Both plots show the dispersion to be gapless at 0.3~K to within 0.04 meV.}
	\label{flo:KHighRes}
\end{figure}

The spectrum was measured over 180$^{\circ}$ rotation at $E_i = 3.32$~meV and $E_i = 1.55$~meV at base temperature and at 12~K. At 1~K and 2~K, we measured only over 60$^{\circ}$ and used $-3m$ crystal symmetry to fold the scattering over and cover the full range of reciprocal space. In comparing intensity of nuclear Bragg peaks, we did find some degree of obverse-reverse twinning of the crystal array, such that some crystal planes were rotated 60$^{\circ}$ from those below. This did not affect the in-plane scattering due to the lack of scattering dependence upon $\ell$.
The sample thermometer at base temperature read 270 mK, but because this thermometer was not exactly on the sample we round up the effective base temperature to 300 mK.
To probe a possible gap at $K$, we also measured a rotation scan over 15$^{\circ}$ at $E_i=1.0$~meV, for a resolution FWHM of 20 $\rm \mu$eV at $\hbar \omega  = 0$. These data are shown in Fig. \ref{flo:KHighRes}, and reveal a gapless excitation spectrum at 0.3~K to within 40 $\rm \mu$eV.

\subsubsection{Background subtraction}

\begin{figure*}
	\centering\includegraphics[width=\textwidth]{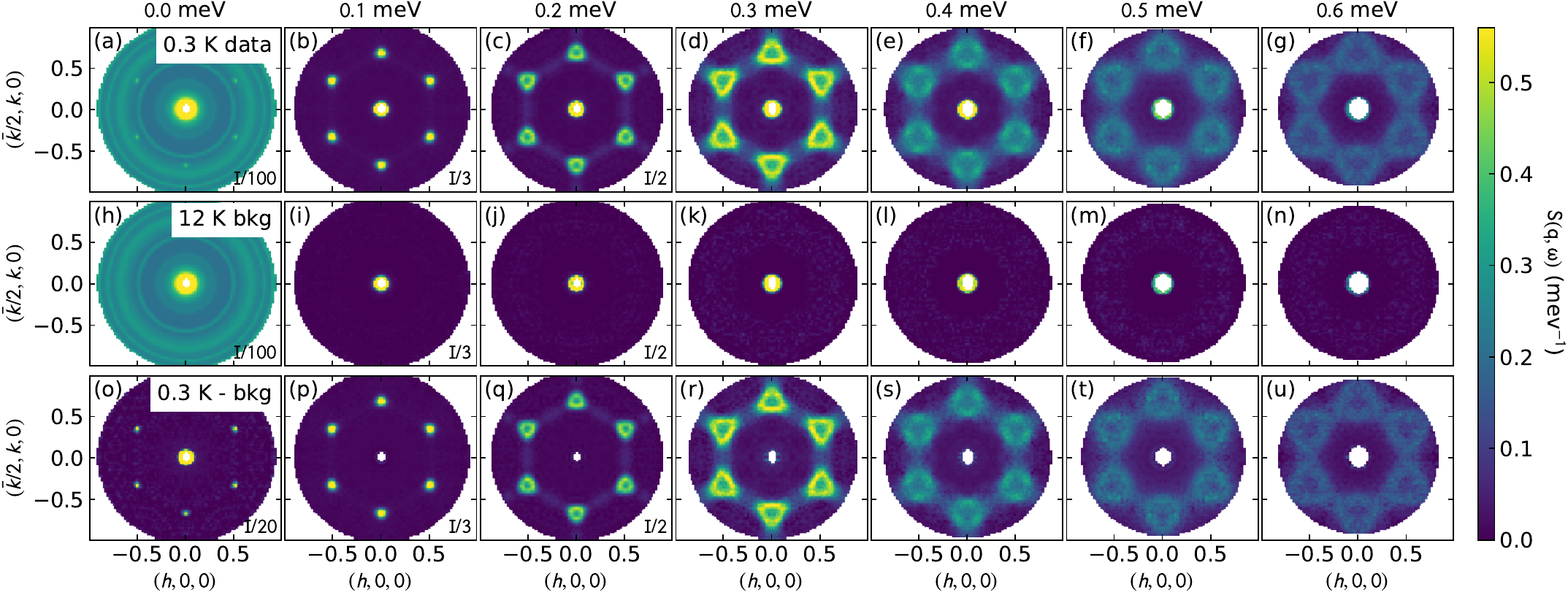}
	\caption{KYbSe$_2$ background subtraction for CNCS data. The top row shows the raw data at 0.3~K. The middle row shows the phenomenological background generated from the 12~K scattering data. The bottom row shows the data with the background subtracted, eliminating artifacts near $Q=0$ and $\hbar \omega = 0$.}
	\label{flo:KYS_CNCS_Background}
\end{figure*}

For the CNCS experiment, a phenomenological background was created and subtracted using the 12~K scattering data. At 12~K, the spin excitations become totally diffuse paramagnetic excitations. To model and eliminate these, we took the median intensity at each constant energy slice to be the approximate value of paramagnetic intensity, and subtracted this value from each pixel at that energy transfer. Then, we set any negative intensities to zero, and subtracted this background from the data. This median-value subtraction was not done for elastic scattering because paramagnetic intensity has negligible elastic contributions. Thus, for elastic data the 12~K was directly subtracted from lower temperatures. We find that this procedure effectively eliminates artifacts in the data while leaving magnetic intensity unchanged, as shown in Fig. \ref{flo:KYS_CNCS_Background}. Finally, because entanglement witnesses require a total sum rule satisfying $S(S+1) = 0.75$ for an effective $J=1/2$ system, we normalized the background-subtracted 300~mK KYbSe$_2$ scattering such that the total scattering is $\langle S ^2 \rangle = 0.75$.

\subsubsection{Critical scaling fits}

To fit the critical exponent in Fig. \ref{flo:CriticalScaling}, we used data at $\hbar \omega / k_B T$ above the ``knee'' where the power law behavior starts. Using this data range, we minimized the $\chi^2$ of the scaled data fitted to a power law in $(\hbar \omega / k_B T)^{\alpha}$, varying $\alpha$ and rescaling the data in each iteration. This resulted in a fitted $\alpha = 1.73(12)$.

\subsection{ARCS experiment}

The sample for the ARCS measurement was 3~g of plate-like crystals ground into a powder. We measured the inelastic neutron scattering at incident energies $E_i=35$~meV, 50~meV, and 130~meV and at temperatures 7~K, 100~K, 200~K, and 300~K (for $E_i=50$~meV only). For details of the crystal field fits, see see the Supplemental Information.

\subsection{Onsager reaction field fits}

The magnetic diffuse scattering from the CNCS experiment was analyzed using the Onsager reaction field approach of Ref.~\onlinecite{Paddison_2020}.
Fits to the single-crystal diffuse-scattering data sets were performed using the Migrad algorithm in the \textsc{Minuit} program \citep{James_1975}. The fitted data sets comprised the $(hk0)$, $(h0l)$ and $(hhl)$ scattering planes measured at 1\,K and 2\,K. All data were energy-integrated over $E>0.05$\,meV. We minimize the sum of squared residuals, defined as
\begin{equation}
\chi^{2}=\sum_{\mathrm{d}}\sum_{i\in\mathrm{d}}\left(\frac{I_{i}^{\mathrm{data}}-sI_{i}^{\mathrm{calc}}-B}{\sigma_{i}}\right)^{2},
\end{equation}
where $d$ denotes a data set, $I_{i}^{\mathrm{data}}$ is the intensity of data point $i$, $I_{i}^{\mathrm{calc}}$ is the corresponding calculated intensity [see Supplemental Information], $\sigma_{i}$ is the corresponding uncertainty, and $s$ and $B$ denote, respectively, fitted intensity scale and offset factors determined at each iteration using linear-least-squares relations.

The $J_{\alpha\beta}(\mathbf{Q})$
are elements of an interaction matrix given by
\begin{equation}
\mathsf{J}(\mathbf{Q})=-\left(\begin{array}{ccc}
aJ_{X}+bJ_{A} & cJ_{A} & 0\\
cJ_{A} & aJ_{X}-bJ_{A} & 0\\
0 & 0 & aJ_{Z}
\end{array}\right),\label{eq:interaction_matrix}
\end{equation}
in which
\begin{align}
a & =2[\cos2\pi(h+k)+\cos 2\pi h+\cos 2\pi k], \\
b & =2\cos 2\pi(h+k)-\cos 2\pi h-\cos 2\pi k,\\
c & =\sqrt{3}(\cos 2\pi k-\cos 2\pi h), 
\end{align}
where $h$ and $k$ are noninteger Miller indices.
We find a best fit Hamiltonian
\begin{align}
    J_X = 2.33(10) \> {\rm K} \quad & \quad J_Z = 2.28(10) \> {\rm K} \nonumber \\
    J_A = -0.018(8) \> {\rm K} \quad & \quad J_2 = 0.11(2) \> {\rm K} 
\label{ec}
\end{align}
where $J_X$ and $J_Z$ are the $ab$-plane and $c$-axis nearest neighbor exchange respectively, $J_A$ is off-diagonal exchange~\cite{Paddison_2020}, $J_2$ is second neighbor Heisenberg exchange, and spins have been treated as classical vectors of unit length. These values show off-diagonal exchange $J_A$ being much smaller than the Heisenberg terms $J_X$ and $J_Y$, showing that KYbSe$_2$ can be effectively modeled by the $J_1-J_2$ Heisenberg model of Eq.~\eqref{eq:hamiltonian} in the main text.

To check the robustness of the results, we performed three checks. First, to check for the possibility of local $\chi^{2}$ minima, we performed 20 separate fits initialized with different parameter values in the range $[-1:1]$\,K. No local minima were found to give acceptable agreement with the experimental data, and the parameters reported in the text correspond to the minimum $\chi^{2}$ we obtained. Second, we considered the effect of including an additional symmetry-allowed off-diagonal exchange interaction, $J_B$ \cite{Paddison_2020}. This parameter refined to a zero value within uncertainty, and  has negligible effect on the results. Third, we considered the effect of the obverse-reverse twinning of the crystal array, and found that including this effect in the calculation had negligible effect on the fit quality or parameter values.

\subsection{Schwinger boson calculations}

Here we describe the main steps of the Schwinger boson calculations.
The triangular antiferromagnetic Heisenberg model is given in Eq.~\eqref{eq:hamiltonian}. 
The spin operators can be expressed in terms of SB operators, $\hat{\boldsymbol S}_i = \frac{1}{2}  {\boldsymbol b}_{i}^{\dag} \  {\boldsymbol \sigma} \ {\boldsymbol b}_{i}$, 
where ${\boldsymbol b}_{i}^{\dag}=(b_{i\uparrow}^{\dag}, \ b_{i\downarrow}^{\dag} )$, and ${\boldsymbol \sigma} \equiv (\sigma^x, \ \sigma^y, \ \sigma^z)$ is the vector of Pauli matrices. The spin-$1/2$ representation of the spin operator is enforced by  the constraint $ b_{i \uparrow}^{\dag}b_{i \uparrow}^{}+ b_{i \downarrow}^{\dag}b_{i \downarrow}^{}=1$.

The Heisenberg interaction can be expressed in terms of the bond operators $ A_{ij}^{} = \frac{1}{2} (b_{i\uparrow}b_{j\downarrow} - b_{i\downarrow}b_{j\uparrow})$ and $ B_{ij}^{} = \frac{1}{2} (b_{j\uparrow}^{\dag} b_{i\uparrow}^{} + b_{j\downarrow}^{\dag} b_{i\downarrow}^{})$:
\begin{equation}
 \hat{\boldsymbol S}_i \cdot  \hat{\boldsymbol S}_j = S^2 \left(1-2\alpha \right) - 2(1-\alpha) A_{ij}^{\dag} A_{ij}^{} + 2\alpha :\! B_{ij}^{\dag} B_{ij}^{}\!:,
\end{equation}
where the real parameter $\alpha$ fixes the decoupling scheme  of a path integral formulation over coherent states (for KYbSe$_2$,  we set $\alpha=0.45$), where the bond fields
$A_{ij}$ and $B_{ij}$ are introduced via a Hubbard-Stratonovich transformation. At the saddle-point level (uniform and static bond fields), the theory describes non-interacting spin-$1/2$ spinons, whose condensation  leads to long-range magnetic ordering~\cite{Arovas1988,Auerbach1994}.  The inclusion of fluctuations of the bond fields  that mediate the spinon-spinon interaction  drastically modifies the excitation spectrum in the sense that the true collective modes (magnons) of the antiferromagnetically ordered phase  emerge  as two-spinon bound states and the two-spinon continuum is strongly renormalized~\cite{Ghioldi_2018,Zhang19}. 

\subsection{Tensor network calculation}

\begin{figure}
	\centering\includegraphics[width=\columnwidth]{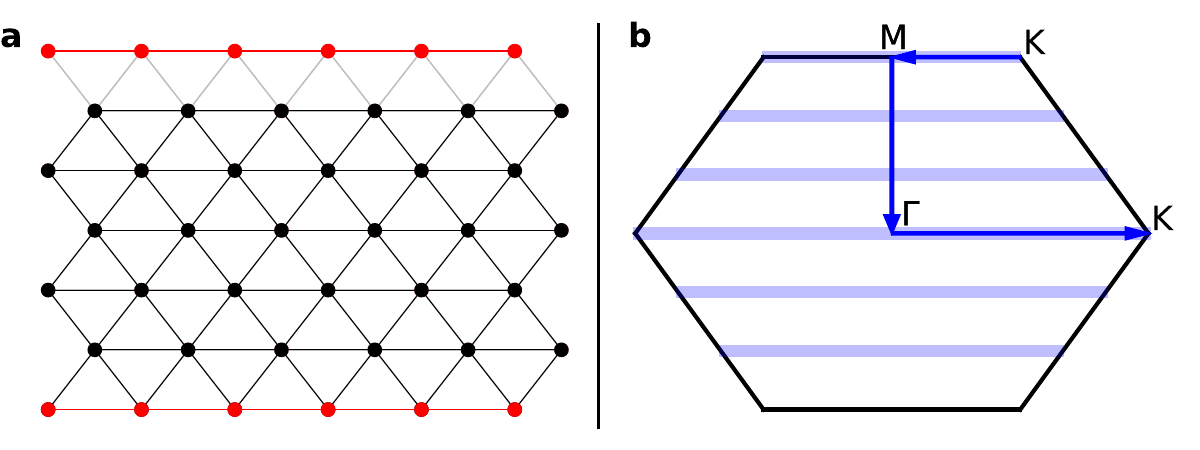}
	\caption{\textbf{\textsf{Illustration of the geometry used and corresponding Brillouin zone for the tensor network simulations.}} Panel \textsf{\textbf{a}} is a $6\times 6$ lattice that we make a cylinder by identifying the top and bottom rows shown in red. Panel \textsf{\textbf{b}} is the Brillouin zone for this geometry, with the blue shaded region showing the allowed momenta, and the arrows show the path we take to generate Fig.~\ref{flo:SchwingerBosons}(e).}
	\label{flo:lattice_bz}
\end{figure}

For the tensor network calculation we studied the spin-$1/2$ triangular antiferromagnetic Heisenberg model defined in Eq.~\eqref{eq:hamiltonian}. In these simulations we wrapped the triangular lattice into a cylinder with a circumference of $C=6$ and length $L=36$ sites. There is some freedom in how one identifies sites in the triangular lattice to form a cylinder geometry, and this choice leads to different allowed momentum values $\boldsymbol{q}$ in the Brillouin zone. We use the XC6 boundary conditions explained in \cite{Szasz20}. In Fig.~\ref{flo:lattice_bz}(a), we show a sample of this lattice with $C=6$ and $L=6$. In Fig.~\ref{flo:lattice_bz}(b) we show the allowed $\boldsymbol{q}$ values for the XC6 boundary conditions, as well as the path we take to generate Fig.~\ref{flo:SchwingerBosons}(e). We note that this choice of boundary conditions yields the maximum number of allowed $\boldsymbol{q}$ points in the path through the high symmetry $\boldsymbol{q}$ points of interest.

To calculate the dynamical structure factor, $S(\boldsymbol{q},\omega)$, we first calculate the time-dependent spin-spin correlation function given by
\begin{equation}
    G(\boldsymbol{x},t)\coloneqq\bigl\langle\hat{\boldsymbol{S}}_{\boldsymbol{x}}(t)\cdot\hat{\boldsymbol {S}}_{\boldsymbol{c}}(0) \bigr\rangle,
\end{equation}
where $\langle \cdot \rangle$ is the expectation value in the ground state, and $\hat{\boldsymbol {S}}_{\boldsymbol{c}}$ is the spin operator at the center site in the lattice. Due to the rotational symmetry of this model, we only look at the $z$-component of the spin using the identity
\begin{equation}
    \bigl\langle\hat{\boldsymbol{S}}_{\boldsymbol{x}}(t)\cdot\hat{\boldsymbol {S}}_{\boldsymbol{c}}(0)\bigr\rangle=3\bigl\langle \hat{S}^z_{\boldsymbol{x}}(t)\hat{S}^z_{\boldsymbol{c}}(0)\bigr\rangle,
\end{equation}
and we drop the pre-factor of 3 in the calculations. In defining $\hat{S}^z_{\boldsymbol{c}}$, we subtract $\bigl\langle\hat{S}^z_{\boldsymbol{c}}\bigr\rangle$ to remove potential disconnected contributions present from finite precision. The dynamical structure factor is then related to this quantity through the Fourier transform,
\begin{equation}
    S\bigl(\boldsymbol{q},\omega\bigr) = \frac{1}{2\pi \sqrt{N}}\int_{-\infty}^{\infty}\mathrm{d}t\sum\nolimits_{\boldsymbol{x}} \, \mathrm{e}^{-i\bigl(\boldsymbol{q}\cdot\boldsymbol{x}-\omega t\bigr)}G(\boldsymbol{x},t),
\end{equation}
where the sum is over all $N=LC$ lattice sites, and $\boldsymbol{x}$ measures the distance from the center site. The quantity $G(\boldsymbol{x},t)$ has the following properties,
\begin{align}
    G(-\boldsymbol{x},t) &= G(\boldsymbol{x},t),\\
    \mathsf{Re} G(\boldsymbol{x},-t) &= \mathsf{Re} G(\boldsymbol{x},t),\\
    \mathsf{Im} G(\boldsymbol{x},-t) &= -\mathsf{Im} G(\boldsymbol{x},t).
\end{align}
Meaning we only need positive times, and can write
 \begin{align}
    S\bigl(\boldsymbol{q},\omega\bigr) &= \frac{1}{\pi \sqrt{N}}\int_0^{\infty}\mathrm{d}t\sum\nolimits_{\boldsymbol{x}} \cos\bigl(\boldsymbol{q}\cdot\boldsymbol{x}\bigr)  \nonumber\\ &\times\Bigl(\cos(\omega t)\mathsf{Re} G(\boldsymbol{x},t) - \sin(\omega t)\mathsf{Im} G(\boldsymbol{x},t)\Bigr).
\end{align}
One major advantage of performing the Fourier transform this way is this ensures $S(\boldsymbol{q},\omega)$ is real, even when the time integral is truncated to a finite upper limit. Due to the finite system size, the resulting spectral function will be a sum of delta-functions, and not the desired analytic function in the thermodynamic limit. To remedy this, we broaden these delta functions with a Gaussian distribution with a width $\eta$. This is achieved by scaling $G(\boldsymbol{x},t)$ by a Gaussian,
\begin{equation}
    G(\boldsymbol{x},t)\longrightarrow\mathrm{e}^{-\eta t^2}G(\boldsymbol{x},t),
\end{equation}
before integrating.
 
To perform this calculation, we use the Density Matrix Renormalization Group (DMRG) algorithm \cite{schollwock2011} to find the ground state, and then use the Time Dependent Variational Principle (TDVP) \cite{Vanderstraeten2019} for the time evolution. In this work, we used a maximum bond-dimension $\chi=512$, a time step $\delta t=0.1$, a maximum time $t_\mathrm{max}=60$, and a Gaussian width of $\eta=0.02 J_1$. The introduction of a finite $t_\mathrm{max}$ corresponds to a frequency resolution $\omega \sim 1/t_\mathrm{max}$, for which lower frequencies are not reliable, and The finite system size introduces a gap in the spectrum that scales as $\Delta \sim 1/C$, even if the system is gapless in the thermodynamic limit. For this study, we utilize the ITensor library \cite{itensor}.

\subsection{Density Functional Theory simulation}

\begin{figure}
	\centering\includegraphics[width=\columnwidth]{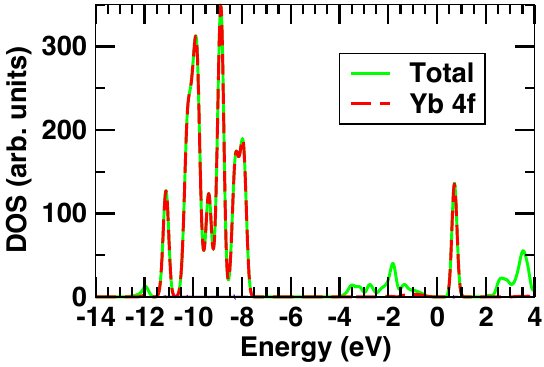}
	\caption{{{DFT electronic density of states in KYbSe$_2$ calculated using the SCAN functional with an additional Hubbard\textit{-U} correction of \textit{U=}8 eV on the Yb 5f states. The Fermi energy is set to 0 eV.}}}
	\label{fig:dft}
\end{figure}

Localized \textit{f-}electron magnetism in Mott-Hubbard systems has traditionally been a challenge for \textit{ab initio} Density Functional Theory~\cite{Hohenberg1964,Kohn1965} (DFT) owing to systematic self-interaction and static-correlation errors~\cite{Cohen2008} in semi-local approximate exchange correlation (XC) functionals. When \textit{f-}shell magnetism is not important, DFT simulations of rare-earth compounds often relegate the \textit{f-}electrons to a core-shell~\cite{Duan2018} within a pseudopotential approximation. Such a description yields satisfactory structural property predictions especially in the Lathanides where the \textit{f-}shell-ligand hybridization is weak~\cite{Duan2018}.  In KYbSe$_2$, magnetism is of primary relevance and so \textit{f-}electrons need to be treated explicitly as valence electrons. While semi-local GGA XC functionals within the traditional Kohn-Sham scheme often describe open \textit{f-}shell insulators as metals, modern XC functionals when deployed within a generalized Kohn-Sham (GKS)~\cite{Seidl1996,Perdew2017} framework are able to mitigate \textit{f-}shell self-interaction~\cite{Perdew1981} errors  (SIE) and yield a qualitatively correct accounting of the transport gap. In KYbSe$_2$, we find that a meta-GGA\textit{+U} approach where the SCAN~\cite{Sun2015a} meta-GGA functional is employed in conjunction with an on-site Hubbard\textit{-U} correction~\cite{Anisimov1997} of \textit{U}=8 eV is able to describe the system as insulating with one unpaired \textit{f-}electron per Yb site (see Fig ~\ref{fig:dft}). Non-local screened hybrid functionals in the HSE06~\cite{Heyd2003} family are similarly able to describe the band gap in this system. Once an insulating ground state is obtained, the size of the gap can be tuned by varying the fraction of non-local Fock exchange in the XC functional.

 Furthermore we find that the band gap in KYbSe$_2$ is largely insensitive to the specific magnetic ordering between different Yb sites. This is in line with the expectation that 4\textit{f-}shell hybridization is weak and the GKS band gap between a Se ligand dominated valence band and a narrow Yb 4\textit{f} conduction band is almost purely determined by on-site Coulomb  repulsion between Yb \textit{f-}electrons. However, a quantitative accounting of the energies of different low-energy magnetic orderings in KYbSe$_2$ is complicated by the previously documented multiple-minima problem~\cite{Deilynazar2015,Payne2019} encountered in orbital-dependent XC functional approaches to modeling \textit{d-} and \textit{f-}electron systems. We find that in trying to stabilize specific in-plane magnetic orderings in KYbSe$_2$, the GKS self-consistency cycle can get trapped in any one of a plethora of local minima associated with an overall similar magnetization density. These stationary points are further found to be separated by energies comparable to the inter-site J couplings ($\sim$1K) that one wishes to extract, making unambiguous identification of the lowest energy minimum associated with a given magnetic ordering difficult. Meta-heuristic approaches~\cite{Casadei2012,Payne2018,Payne2019} that aim to mitigate the multiple-minima problem have been proposed and a systematic effort to explore the efficacy of such methods in the context of specific 4\textit{f-}electron systems such as KYbSe$_2$ is worth pursuing in future. 

The above DFT simulations of electronic structure in this work were carried out using the Vienna Ab Initio Simulation Package (VASP)~\cite{Kresse1993} version 6.2.1. which employs a planewave basis set in conjunction with PAW potentials~\cite{Kresse1999}.  A planewave cutoff of 400 eV was used. A 72 atom $\sqrt{3} \times 2\sqrt{3} \times 1$ magnetic supercell  was considered to model various low energy spin ordered configurations and a 4x2x1 k-point mesh was used to sample the corresponding Brillouin zone. Spin orbit coupling was included and an in-plane 120$^{\circ}$ ordered configuration was used to calculate the electronic DOS shown in Fig ~\ref{fig:dft}. For ionic positions, the experimentally determined geometry was used.

We also carried out first principles calculations of KYbSe$_2$ within the GGA+U+so approach \cite{perdew,Anisimov1991}, as implemented in the all-electron planewave density functional code WIEN2K \cite{blaha2001}. We find a saturation moment on the Yb site of 1.63~$\mu_B$, in good agreement with the approximate value of 1.5~$\mu_B$ found from the experimental measurements.   It is noteworthy that this former value is primarily orbital moment, with the Yb orbital moment, at 1.055~$\mu_B$, outstripping the Yb spin moment of 0.572~$\mu_B$ (there is an additional small component from the interstitial region, and Se spheres, of 0.12~$\mu_B$). 
The exact nature of exchange coupling, as well as the coupling to the lattice depicted by the substantial orbital moment, in such borderline quantum magnets \cite{pokharel,pandey} remains a matter of substantial debate and controversy.

\section{Data Availability}
 All plotted experimental data will be made publicly available.


\begin{thebibliography}{99}%
	\makeatletter
	\providecommand \@ifxundefined [1]{%
		\@ifx{#1\undefined}
	}%
	\providecommand \@ifnum [1]{%
		\ifnum #1\expandafter \@firstoftwo
		\else \expandafter \@secondoftwo
		\fi
	}%
	\providecommand \@ifx [1]{%
		\ifx #1\expandafter \@firstoftwo
		\else \expandafter \@secondoftwo
		\fi
	}%
	\providecommand \natexlab [1]{#1}%
	\providecommand \enquote  [1]{``#1''}%
	\providecommand \bibnamefont  [1]{#1}%
	\providecommand \bibfnamefont [1]{#1}%
	\providecommand \citenamefont [1]{#1}%
	\providecommand \href@noop [0]{\@secondoftwo}%
	\providecommand \href [0]{\begingroup \@sanitize@url \@href}%
	\providecommand \@href[1]{\@@startlink{#1}\@@href}%
	\providecommand \@@href[1]{\endgroup#1\@@endlink}%
	\providecommand \@sanitize@url [0]{\catcode `\\12\catcode `\$12\catcode
		`\&12\catcode `\#12\catcode `\^12\catcode `\_12\catcode `\%12\relax}%
	\providecommand \@@startlink[1]{}%
	\providecommand \@@endlink[0]{}%
	\providecommand \url  [0]{\begingroup\@sanitize@url \@url }%
	\providecommand \@url [1]{\endgroup\@href {#1}{\urlprefix }}%
	\providecommand \urlprefix  [0]{URL }%
	\providecommand \Eprint [0]{\href }%
	\providecommand \doibase [0]{http://dx.doi.org/}%
	\providecommand \selectlanguage [0]{\@gobble}%
	\providecommand \bibinfo  [0]{\@secondoftwo}%
	\providecommand \bibfield  [0]{\@secondoftwo}%
	\providecommand \translation [1]{[#1]}%
	\providecommand \BibitemOpen [0]{}%
	\providecommand \bibitemStop [0]{}%
	\providecommand \bibitemNoStop [0]{.\EOS\space}%
	\providecommand \EOS [0]{\spacefactor3000\relax}%
	\providecommand \BibitemShut  [1]{\csname bibitem#1\endcsname}%
	\let\auto@bib@innerbib\@empty
	\bibitem [{\citenamefont {Knolle}\ and\ \citenamefont
		{Moessner}(2019)}]{Knolle2019_review}%
	\BibitemOpen
	\bibfield  {author} {\bibinfo {author} {\bibfnamefont {J.}~\bibnamefont
			{Knolle}}\ and\ \bibinfo {author} {\bibfnamefont {R.}~\bibnamefont
			{Moessner}},\ }\href {\doibase 10.1146/annurev-conmatphys-031218-013401}
	{\bibfield  {journal} {\bibinfo  {journal} {Annual Review of Condensed Matter
				Physics}\ }\textbf {\bibinfo {volume} {10}},\ \bibinfo {pages} {451}
		(\bibinfo {year} {2019})}\BibitemShut {NoStop}%
	\bibitem [{\citenamefont {Broholm}\ \emph {et~al.}(2020)\citenamefont
		{Broholm}, \citenamefont {Cava}, \citenamefont {Kivelson}, \citenamefont
		{Nocera}, \citenamefont {Norman},\ and\ \citenamefont
		{Senthil}}]{broholm2019quantum}%
	\BibitemOpen
	\bibfield  {author} {\bibinfo {author} {\bibfnamefont {C.}~\bibnamefont
			{Broholm}}, \bibinfo {author} {\bibfnamefont {R.~J.}\ \bibnamefont {Cava}},
		\bibinfo {author} {\bibfnamefont {S.~A.}\ \bibnamefont {Kivelson}}, \bibinfo
		{author} {\bibfnamefont {D.~G.}\ \bibnamefont {Nocera}}, \bibinfo {author}
		{\bibfnamefont {M.~R.}\ \bibnamefont {Norman}}, \ and\ \bibinfo {author}
		{\bibfnamefont {T.}~\bibnamefont {Senthil}},\ }\href {\doibase
		10.1126/science.aay0668} {\bibfield  {journal} {\bibinfo  {journal}
			{Science}\ }\textbf {\bibinfo {volume} {367}} (\bibinfo {year} {2020}),\
		10.1126/science.aay0668}\BibitemShut {NoStop}%
	\bibitem [{\citenamefont {Savary}\ and\ \citenamefont
		{Balents}(2016)}]{Savary_2016review}%
	\BibitemOpen
	\bibfield  {author} {\bibinfo {author} {\bibfnamefont {L.}~\bibnamefont
			{Savary}}\ and\ \bibinfo {author} {\bibfnamefont {L.}~\bibnamefont
			{Balents}},\ }\href {\doibase 10.1088/0034-4885/80/1/016502} {\bibfield
		{journal} {\bibinfo  {journal} {Reports on Progress in Physics}\ }\textbf
		{\bibinfo {volume} {80}},\ \bibinfo {pages} {016502} (\bibinfo {year}
		{2016})}\BibitemShut {NoStop}%
	\bibitem [{\citenamefont {Zhou}\ \emph {et~al.}(2017)\citenamefont {Zhou},
		\citenamefont {Kanoda},\ and\ \citenamefont {Ng}}]{Zhou2017}%
	\BibitemOpen
	\bibfield  {author} {\bibinfo {author} {\bibfnamefont {Y.}~\bibnamefont
			{Zhou}}, \bibinfo {author} {\bibfnamefont {K.}~\bibnamefont {Kanoda}}, \ and\
		\bibinfo {author} {\bibfnamefont {T.-K.}\ \bibnamefont {Ng}},\ }\href
	{\doibase 10.1103/RevModPhys.89.025003} {\bibfield  {journal} {\bibinfo
			{journal} {Rev. Mod. Phys.}\ }\textbf {\bibinfo {volume} {89}},\ \bibinfo
		{pages} {025003} (\bibinfo {year} {2017})}\BibitemShut {NoStop}%
	\bibitem [{\citenamefont {Tokura}\ \emph {et~al.}(2017)\citenamefont {Tokura},
		\citenamefont {Kawasaki},\ and\ \citenamefont
		{Nagaosa}}]{tokura2017emergent}%
	\BibitemOpen
	\bibfield  {author} {\bibinfo {author} {\bibfnamefont {Y.}~\bibnamefont
			{Tokura}}, \bibinfo {author} {\bibfnamefont {M.}~\bibnamefont {Kawasaki}}, \
		and\ \bibinfo {author} {\bibfnamefont {N.}~\bibnamefont {Nagaosa}},\ }\href
	{https://doi.org/10.1038/nphys4274} {\bibfield  {journal} {\bibinfo
			{journal} {Nature Physics}\ }\textbf {\bibinfo {volume} {13}},\ \bibinfo
		{pages} {1056} (\bibinfo {year} {2017})}\BibitemShut {NoStop}%
	\bibitem [{\citenamefont {Yamashita}\ \emph {et~al.}(2008)\citenamefont
		{Yamashita}, \citenamefont {Nakazawa}, \citenamefont {Oguni}, \citenamefont
		{Oshima}, \citenamefont {Nojiri}, \citenamefont {Shimizu}, \citenamefont
		{Miyagawa},\ and\ \citenamefont {Kanoda}}]{Yamashita2008}%
	\BibitemOpen
	\bibfield  {author} {\bibinfo {author} {\bibfnamefont {S.}~\bibnamefont
			{Yamashita}}, \bibinfo {author} {\bibfnamefont {Y.}~\bibnamefont {Nakazawa}},
		\bibinfo {author} {\bibfnamefont {M.}~\bibnamefont {Oguni}}, \bibinfo
		{author} {\bibfnamefont {Y.}~\bibnamefont {Oshima}}, \bibinfo {author}
		{\bibfnamefont {H.}~\bibnamefont {Nojiri}}, \bibinfo {author} {\bibfnamefont
			{Y.}~\bibnamefont {Shimizu}}, \bibinfo {author} {\bibfnamefont
			{K.}~\bibnamefont {Miyagawa}}, \ and\ \bibinfo {author} {\bibfnamefont
			{K.}~\bibnamefont {Kanoda}},\ }\href {\doibase 10.1038/nphys942} {\bibfield
		{journal} {\bibinfo  {journal} {Nature Physics}\ }\textbf {\bibinfo {volume}
			{4}},\ \bibinfo {pages} {459} (\bibinfo {year} {2008})}\BibitemShut {NoStop}%
	\bibitem [{\citenamefont {Itou}\ \emph {et~al.}(2008)\citenamefont {Itou},
		\citenamefont {Oyamada}, \citenamefont {Maegawa}, \citenamefont {Tamura},\
		and\ \citenamefont {Kato}}]{Itou2008}%
	\BibitemOpen
	\bibfield  {author} {\bibinfo {author} {\bibfnamefont {T.}~\bibnamefont
			{Itou}}, \bibinfo {author} {\bibfnamefont {A.}~\bibnamefont {Oyamada}},
		\bibinfo {author} {\bibfnamefont {S.}~\bibnamefont {Maegawa}}, \bibinfo
		{author} {\bibfnamefont {M.}~\bibnamefont {Tamura}}, \ and\ \bibinfo {author}
		{\bibfnamefont {R.}~\bibnamefont {Kato}},\ }\href {\doibase
		10.1103/PhysRevB.77.104413} {\bibfield  {journal} {\bibinfo  {journal} {Phys.
				Rev. B}\ }\textbf {\bibinfo {volume} {77}},\ \bibinfo {pages} {104413}
		(\bibinfo {year} {2008})}\BibitemShut {NoStop}%
	\bibitem [{\citenamefont {Han}\ \emph {et~al.}(2012)\citenamefont {Han},
		\citenamefont {Helton}, \citenamefont {Chu}, \citenamefont {Nocera},
		\citenamefont {Rodriguez-Rivera}, \citenamefont {Broholm},\ and\
		\citenamefont {Lee}}]{han2012fractionalized}%
	\BibitemOpen
	\bibfield  {author} {\bibinfo {author} {\bibfnamefont {T.-H.}\ \bibnamefont
			{Han}}, \bibinfo {author} {\bibfnamefont {J.~S.}\ \bibnamefont {Helton}},
		\bibinfo {author} {\bibfnamefont {S.}~\bibnamefont {Chu}}, \bibinfo {author}
		{\bibfnamefont {D.~G.}\ \bibnamefont {Nocera}}, \bibinfo {author}
		{\bibfnamefont {J.~A.}\ \bibnamefont {Rodriguez-Rivera}}, \bibinfo {author}
		{\bibfnamefont {C.}~\bibnamefont {Broholm}}, \ and\ \bibinfo {author}
		{\bibfnamefont {Y.~S.}\ \bibnamefont {Lee}},\ }\href
	{https://doi.org/10.1038/nature11659} {\bibfield  {journal} {\bibinfo
			{journal} {Nature}\ }\textbf {\bibinfo {volume} {492}},\ \bibinfo {pages}
		{406} (\bibinfo {year} {2012})}\BibitemShut {NoStop}%
	\bibitem [{\citenamefont {Gaudet}\ \emph {et~al.}(2019)\citenamefont {Gaudet},
		\citenamefont {Smith}, \citenamefont {Dudemaine}, \citenamefont {Beare},
		\citenamefont {Buhariwalla}, \citenamefont {Butch}, \citenamefont {Stone},
		\citenamefont {Kolesnikov}, \citenamefont {Xu}, \citenamefont {Yahne},
		\citenamefont {Ross}, \citenamefont {Marjerrison}, \citenamefont {Garrett},
		\citenamefont {Luke}, \citenamefont {Bianchi},\ and\ \citenamefont
		{Gaulin}}]{Gaudet_2019}%
	\BibitemOpen
	\bibfield  {author} {\bibinfo {author} {\bibfnamefont {J.}~\bibnamefont
			{Gaudet}}, \bibinfo {author} {\bibfnamefont {E.~M.}\ \bibnamefont {Smith}},
		\bibinfo {author} {\bibfnamefont {J.}~\bibnamefont {Dudemaine}}, \bibinfo
		{author} {\bibfnamefont {J.}~\bibnamefont {Beare}}, \bibinfo {author}
		{\bibfnamefont {C.~R.~C.}\ \bibnamefont {Buhariwalla}}, \bibinfo {author}
		{\bibfnamefont {N.~P.}\ \bibnamefont {Butch}}, \bibinfo {author}
		{\bibfnamefont {M.~B.}\ \bibnamefont {Stone}}, \bibinfo {author}
		{\bibfnamefont {A.~I.}\ \bibnamefont {Kolesnikov}}, \bibinfo {author}
		{\bibfnamefont {G.}~\bibnamefont {Xu}}, \bibinfo {author} {\bibfnamefont
			{D.~R.}\ \bibnamefont {Yahne}}, \bibinfo {author} {\bibfnamefont {K.~A.}\
			\bibnamefont {Ross}}, \bibinfo {author} {\bibfnamefont {C.~A.}\ \bibnamefont
			{Marjerrison}}, \bibinfo {author} {\bibfnamefont {J.~D.}\ \bibnamefont
			{Garrett}}, \bibinfo {author} {\bibfnamefont {G.~M.}\ \bibnamefont {Luke}},
		\bibinfo {author} {\bibfnamefont {A.~D.}\ \bibnamefont {Bianchi}}, \ and\
		\bibinfo {author} {\bibfnamefont {B.~D.}\ \bibnamefont {Gaulin}},\ }\href
	{\doibase 10.1103/PhysRevLett.122.187201} {\bibfield  {journal} {\bibinfo
			{journal} {Phys. Rev. Lett.}\ }\textbf {\bibinfo {volume} {122}},\ \bibinfo
		{pages} {187201} (\bibinfo {year} {2019})}\BibitemShut {NoStop}%
	\bibitem [{\citenamefont {Gao}\ \emph {et~al.}(2019)\citenamefont {Gao},
		\citenamefont {Chen}, \citenamefont {Tam}, \citenamefont {Huang},
		\citenamefont {Sasmal}, \citenamefont {Adroja}, \citenamefont {Ye},
		\citenamefont {Cao}, \citenamefont {Sala}, \citenamefont {Stone},
		\citenamefont {Baines}, \citenamefont {Verezhak}, \citenamefont {Hu},
		\citenamefont {Chung}, \citenamefont {Xu}, \citenamefont {Cheong},
		\citenamefont {Nallaiyan}, \citenamefont {Spagna}, \citenamefont {Maple},
		\citenamefont {Nevidomskyy}, \citenamefont {Morosan}, \citenamefont {Chen},\
		and\ \citenamefont {Dai}}]{Gao2019}%
	\BibitemOpen
	\bibfield  {author} {\bibinfo {author} {\bibfnamefont {B.}~\bibnamefont
			{Gao}}, \bibinfo {author} {\bibfnamefont {T.}~\bibnamefont {Chen}}, \bibinfo
		{author} {\bibfnamefont {D.~W.}\ \bibnamefont {Tam}}, \bibinfo {author}
		{\bibfnamefont {C.-L.}\ \bibnamefont {Huang}}, \bibinfo {author}
		{\bibfnamefont {K.}~\bibnamefont {Sasmal}}, \bibinfo {author} {\bibfnamefont
			{D.~T.}\ \bibnamefont {Adroja}}, \bibinfo {author} {\bibfnamefont
			{F.}~\bibnamefont {Ye}}, \bibinfo {author} {\bibfnamefont {H.}~\bibnamefont
			{Cao}}, \bibinfo {author} {\bibfnamefont {G.}~\bibnamefont {Sala}}, \bibinfo
		{author} {\bibfnamefont {M.~B.}\ \bibnamefont {Stone}}, \bibinfo {author}
		{\bibfnamefont {C.}~\bibnamefont {Baines}}, \bibinfo {author} {\bibfnamefont
			{J.~A.~T.}\ \bibnamefont {Verezhak}}, \bibinfo {author} {\bibfnamefont
			{H.}~\bibnamefont {Hu}}, \bibinfo {author} {\bibfnamefont {J.-H.}\
			\bibnamefont {Chung}}, \bibinfo {author} {\bibfnamefont {X.}~\bibnamefont
			{Xu}}, \bibinfo {author} {\bibfnamefont {S.-W.}\ \bibnamefont {Cheong}},
		\bibinfo {author} {\bibfnamefont {M.}~\bibnamefont {Nallaiyan}}, \bibinfo
		{author} {\bibfnamefont {S.}~\bibnamefont {Spagna}}, \bibinfo {author}
		{\bibfnamefont {M.~B.}\ \bibnamefont {Maple}}, \bibinfo {author}
		{\bibfnamefont {A.~H.}\ \bibnamefont {Nevidomskyy}}, \bibinfo {author}
		{\bibfnamefont {E.}~\bibnamefont {Morosan}}, \bibinfo {author} {\bibfnamefont
			{G.}~\bibnamefont {Chen}}, \ and\ \bibinfo {author} {\bibfnamefont
			{P.}~\bibnamefont {Dai}},\ }\href {\doibase 10.1038/s41567-019-0577-6}
	{\bibfield  {journal} {\bibinfo  {journal} {Nature Physics}\ }\textbf
		{\bibinfo {volume} {15}},\ \bibinfo {pages} {1052} (\bibinfo {year}
		{2019})}\BibitemShut {NoStop}%
	\bibitem [{\citenamefont {Anderson}(1973)}]{Anderson1973}%
	\BibitemOpen
	\bibfield  {author} {\bibinfo {author} {\bibfnamefont {P.}~\bibnamefont
			{Anderson}},\ }\href {\doibase https://doi.org/10.1016/0025-5408(73)90167-0}
	{\bibfield  {journal} {\bibinfo  {journal} {Materials Research Bulletin}\
		}\textbf {\bibinfo {volume} {8}},\ \bibinfo {pages} {153 } (\bibinfo {year}
		{1973})}\BibitemShut {NoStop}%
	\bibitem [{\citenamefont {White}\ and\ \citenamefont
		{Chernyshev}(2007)}]{PhysRevLett.99.127004}%
	\BibitemOpen
	\bibfield  {author} {\bibinfo {author} {\bibfnamefont {S.~R.}\ \bibnamefont
			{White}}\ and\ \bibinfo {author} {\bibfnamefont {A.~L.}\ \bibnamefont
			{Chernyshev}},\ }\href {\doibase 10.1103/PhysRevLett.99.127004} {\bibfield
		{journal} {\bibinfo  {journal} {Phys. Rev. Lett.}\ }\textbf {\bibinfo
			{volume} {99}},\ \bibinfo {pages} {127004} (\bibinfo {year}
		{2007})}\BibitemShut {NoStop}%
	\bibitem [{\citenamefont {Zhu}\ and\ \citenamefont
		{White}(2015)}]{PhysRevB.92.041105}%
	\BibitemOpen
	\bibfield  {author} {\bibinfo {author} {\bibfnamefont {Z.}~\bibnamefont
			{Zhu}}\ and\ \bibinfo {author} {\bibfnamefont {S.~R.}\ \bibnamefont
			{White}},\ }\href {\doibase 10.1103/PhysRevB.92.041105} {\bibfield  {journal}
		{\bibinfo  {journal} {Phys. Rev. B}\ }\textbf {\bibinfo {volume} {92}},\
		\bibinfo {pages} {041105} (\bibinfo {year} {2015})}\BibitemShut {NoStop}%
	\bibitem [{\citenamefont {Hu}\ \emph {et~al.}(2015)\citenamefont {Hu},
		\citenamefont {Gong}, \citenamefont {Zhu},\ and\ \citenamefont
		{Sheng}}]{PhysRevB.92.140403}%
	\BibitemOpen
	\bibfield  {author} {\bibinfo {author} {\bibfnamefont {W.-J.}\ \bibnamefont
			{Hu}}, \bibinfo {author} {\bibfnamefont {S.-S.}\ \bibnamefont {Gong}},
		\bibinfo {author} {\bibfnamefont {W.}~\bibnamefont {Zhu}}, \ and\ \bibinfo
		{author} {\bibfnamefont {D.~N.}\ \bibnamefont {Sheng}},\ }\href {\doibase
		10.1103/PhysRevB.92.140403} {\bibfield  {journal} {\bibinfo  {journal} {Phys.
				Rev. B}\ }\textbf {\bibinfo {volume} {92}},\ \bibinfo {pages} {140403}
		(\bibinfo {year} {2015})}\BibitemShut {NoStop}%
	\bibitem [{\citenamefont {Iqbal}\ \emph {et~al.}(2016)\citenamefont {Iqbal},
		\citenamefont {Hu}, \citenamefont {Thomale}, \citenamefont {Poilblanc},\ and\
		\citenamefont {Becca}}]{PhysRevB.93.144411}%
	\BibitemOpen
	\bibfield  {author} {\bibinfo {author} {\bibfnamefont {Y.}~\bibnamefont
			{Iqbal}}, \bibinfo {author} {\bibfnamefont {W.-J.}\ \bibnamefont {Hu}},
		\bibinfo {author} {\bibfnamefont {R.}~\bibnamefont {Thomale}}, \bibinfo
		{author} {\bibfnamefont {D.}~\bibnamefont {Poilblanc}}, \ and\ \bibinfo
		{author} {\bibfnamefont {F.}~\bibnamefont {Becca}},\ }\href {\doibase
		10.1103/PhysRevB.93.144411} {\bibfield  {journal} {\bibinfo  {journal} {Phys.
				Rev. B}\ }\textbf {\bibinfo {volume} {93}},\ \bibinfo {pages} {144411}
		(\bibinfo {year} {2016})}\BibitemShut {NoStop}%
	\bibitem [{\citenamefont {Saadatmand}\ and\ \citenamefont
		{McCulloch}(2016)}]{PhysRevB.94.121111}%
	\BibitemOpen
	\bibfield  {author} {\bibinfo {author} {\bibfnamefont {S.~N.}\ \bibnamefont
			{Saadatmand}}\ and\ \bibinfo {author} {\bibfnamefont {I.~P.}\ \bibnamefont
			{McCulloch}},\ }\href {\doibase 10.1103/PhysRevB.94.121111} {\bibfield
		{journal} {\bibinfo  {journal} {Phys. Rev. B}\ }\textbf {\bibinfo {volume}
			{94}},\ \bibinfo {pages} {121111} (\bibinfo {year} {2016})}\BibitemShut
	{NoStop}%
	\bibitem [{\citenamefont {Wietek}\ and\ \citenamefont
		{L\"auchli}(2017)}]{PhysRevB.95.035141}%
	\BibitemOpen
	\bibfield  {author} {\bibinfo {author} {\bibfnamefont {A.}~\bibnamefont
			{Wietek}}\ and\ \bibinfo {author} {\bibfnamefont {A.~M.}\ \bibnamefont
			{L\"auchli}},\ }\href {\doibase 10.1103/PhysRevB.95.035141} {\bibfield
		{journal} {\bibinfo  {journal} {Phys. Rev. B}\ }\textbf {\bibinfo {volume}
			{95}},\ \bibinfo {pages} {035141} (\bibinfo {year} {2017})}\BibitemShut
	{NoStop}%
	\bibitem [{\citenamefont {Gong}\ \emph {et~al.}(2017)\citenamefont {Gong},
		\citenamefont {Zhu}, \citenamefont {Zhu}, \citenamefont {Sheng},\ and\
		\citenamefont {Yang}}]{PhysRevB.96.075116}%
	\BibitemOpen
	\bibfield  {author} {\bibinfo {author} {\bibfnamefont {S.-S.}\ \bibnamefont
			{Gong}}, \bibinfo {author} {\bibfnamefont {W.}~\bibnamefont {Zhu}}, \bibinfo
		{author} {\bibfnamefont {J.-X.}\ \bibnamefont {Zhu}}, \bibinfo {author}
		{\bibfnamefont {D.~N.}\ \bibnamefont {Sheng}}, \ and\ \bibinfo {author}
		{\bibfnamefont {K.}~\bibnamefont {Yang}},\ }\href {\doibase
		10.1103/PhysRevB.96.075116} {\bibfield  {journal} {\bibinfo  {journal} {Phys.
				Rev. B}\ }\textbf {\bibinfo {volume} {96}},\ \bibinfo {pages} {075116}
		(\bibinfo {year} {2017})}\BibitemShut {NoStop}%
	\bibitem [{\citenamefont {Hu}\ \emph {et~al.}(2019)\citenamefont {Hu},
		\citenamefont {Zhu}, \citenamefont {Eggert},\ and\ \citenamefont
		{He}}]{PhysRevLett.123.207203}%
	\BibitemOpen
	\bibfield  {author} {\bibinfo {author} {\bibfnamefont {S.}~\bibnamefont
			{Hu}}, \bibinfo {author} {\bibfnamefont {W.}~\bibnamefont {Zhu}}, \bibinfo
		{author} {\bibfnamefont {S.}~\bibnamefont {Eggert}}, \ and\ \bibinfo {author}
		{\bibfnamefont {Y.-C.}\ \bibnamefont {He}},\ }\href {\doibase
		10.1103/PhysRevLett.123.207203} {\bibfield  {journal} {\bibinfo  {journal}
			{Phys. Rev. Lett.}\ }\textbf {\bibinfo {volume} {123}},\ \bibinfo {pages}
		{207203} (\bibinfo {year} {2019})}\BibitemShut {NoStop}%
	\bibitem [{\citenamefont {Zhu}\ \emph {et~al.}(2018)\citenamefont {Zhu},
		\citenamefont {Maksimov}, \citenamefont {White},\ and\ \citenamefont
		{Chernyshev}}]{Zhu_2018}%
	\BibitemOpen
	\bibfield  {author} {\bibinfo {author} {\bibfnamefont {Z.}~\bibnamefont
			{Zhu}}, \bibinfo {author} {\bibfnamefont {P.~A.}\ \bibnamefont {Maksimov}},
		\bibinfo {author} {\bibfnamefont {S.~R.}\ \bibnamefont {White}}, \ and\
		\bibinfo {author} {\bibfnamefont {A.~L.}\ \bibnamefont {Chernyshev}},\ }\href
	{\doibase 10.1103/PhysRevLett.120.207203} {\bibfield  {journal} {\bibinfo
			{journal} {Phys. Rev. Lett.}\ }\textbf {\bibinfo {volume} {120}},\ \bibinfo
		{pages} {207203} (\bibinfo {year} {2018})}\BibitemShut {NoStop}%
	\bibitem [{\citenamefont {Ding}\ \emph {et~al.}(2019)\citenamefont {Ding},
		\citenamefont {Manuel}, \citenamefont {Bachus}, \citenamefont {Gru\ss{}ler},
		\citenamefont {Gegenwart}, \citenamefont {Singleton}, \citenamefont
		{Johnson}, \citenamefont {Walker}, \citenamefont {Adroja}, \citenamefont
		{Hillier},\ and\ \citenamefont {Tsirlin}}]{Ding_2019_NYO}%
	\BibitemOpen
	\bibfield  {author} {\bibinfo {author} {\bibfnamefont {L.}~\bibnamefont
			{Ding}}, \bibinfo {author} {\bibfnamefont {P.}~\bibnamefont {Manuel}},
		\bibinfo {author} {\bibfnamefont {S.}~\bibnamefont {Bachus}}, \bibinfo
		{author} {\bibfnamefont {F.}~\bibnamefont {Gru\ss{}ler}}, \bibinfo {author}
		{\bibfnamefont {P.}~\bibnamefont {Gegenwart}}, \bibinfo {author}
		{\bibfnamefont {J.}~\bibnamefont {Singleton}}, \bibinfo {author}
		{\bibfnamefont {R.~D.}\ \bibnamefont {Johnson}}, \bibinfo {author}
		{\bibfnamefont {H.~C.}\ \bibnamefont {Walker}}, \bibinfo {author}
		{\bibfnamefont {D.~T.}\ \bibnamefont {Adroja}}, \bibinfo {author}
		{\bibfnamefont {A.~D.}\ \bibnamefont {Hillier}}, \ and\ \bibinfo {author}
		{\bibfnamefont {A.~A.}\ \bibnamefont {Tsirlin}},\ }\href {\doibase
		10.1103/PhysRevB.100.144432} {\bibfield  {journal} {\bibinfo  {journal}
			{Phys. Rev. B}\ }\textbf {\bibinfo {volume} {100}},\ \bibinfo {pages}
		{144432} (\bibinfo {year} {2019})}\BibitemShut {NoStop}%
	\bibitem [{\citenamefont {Bordelon}\ \emph {et~al.}(2019)\citenamefont
		{Bordelon}, \citenamefont {Kenney}, \citenamefont {Liu}, \citenamefont
		{Hogan}, \citenamefont {Posthuma}, \citenamefont {Kavand}, \citenamefont
		{Lyu}, \citenamefont {Sherwin}, \citenamefont {Butch}, \citenamefont {Brown},
		\citenamefont {Graf}, \citenamefont {Balents},\ and\ \citenamefont
		{Wilson}}]{Bordelon2019}%
	\BibitemOpen
	\bibfield  {author} {\bibinfo {author} {\bibfnamefont {M.~M.}\ \bibnamefont
			{Bordelon}}, \bibinfo {author} {\bibfnamefont {E.}~\bibnamefont {Kenney}},
		\bibinfo {author} {\bibfnamefont {C.}~\bibnamefont {Liu}}, \bibinfo {author}
		{\bibfnamefont {T.}~\bibnamefont {Hogan}}, \bibinfo {author} {\bibfnamefont
			{L.}~\bibnamefont {Posthuma}}, \bibinfo {author} {\bibfnamefont
			{M.}~\bibnamefont {Kavand}}, \bibinfo {author} {\bibfnamefont
			{Y.}~\bibnamefont {Lyu}}, \bibinfo {author} {\bibfnamefont {M.}~\bibnamefont
			{Sherwin}}, \bibinfo {author} {\bibfnamefont {N.~P.}\ \bibnamefont {Butch}},
		\bibinfo {author} {\bibfnamefont {C.}~\bibnamefont {Brown}}, \bibinfo
		{author} {\bibfnamefont {M.~J.}\ \bibnamefont {Graf}}, \bibinfo {author}
		{\bibfnamefont {L.}~\bibnamefont {Balents}}, \ and\ \bibinfo {author}
		{\bibfnamefont {S.~D.}\ \bibnamefont {Wilson}},\ }\href {\doibase
		10.1038/s41567-019-0594-5} {\bibfield  {journal} {\bibinfo  {journal} {Nature
				Physics}\ }\textbf {\bibinfo {volume} {15}},\ \bibinfo {pages} {1058}
		(\bibinfo {year} {2019})}\BibitemShut {NoStop}%
	\bibitem [{\citenamefont {Ranjith}\ \emph
		{et~al.}(2019{\natexlab{a}})\citenamefont {Ranjith}, \citenamefont
		{Dmytriieva}, \citenamefont {Khim}, \citenamefont {Sichelschmidt},
		\citenamefont {Luther}, \citenamefont {Ehlers}, \citenamefont {Yasuoka},
		\citenamefont {Wosnitza}, \citenamefont {Tsirlin}, \citenamefont {K\"uhne},\
		and\ \citenamefont {Baenitz}}]{Ranjinth2019}%
	\BibitemOpen
	\bibfield  {author} {\bibinfo {author} {\bibfnamefont {K.~M.}\ \bibnamefont
			{Ranjith}}, \bibinfo {author} {\bibfnamefont {D.}~\bibnamefont {Dmytriieva}},
		\bibinfo {author} {\bibfnamefont {S.}~\bibnamefont {Khim}}, \bibinfo {author}
		{\bibfnamefont {J.}~\bibnamefont {Sichelschmidt}}, \bibinfo {author}
		{\bibfnamefont {S.}~\bibnamefont {Luther}}, \bibinfo {author} {\bibfnamefont
			{D.}~\bibnamefont {Ehlers}}, \bibinfo {author} {\bibfnamefont
			{H.}~\bibnamefont {Yasuoka}}, \bibinfo {author} {\bibfnamefont
			{J.}~\bibnamefont {Wosnitza}}, \bibinfo {author} {\bibfnamefont {A.~A.}\
			\bibnamefont {Tsirlin}}, \bibinfo {author} {\bibfnamefont {H.}~\bibnamefont
			{K\"uhne}}, \ and\ \bibinfo {author} {\bibfnamefont {M.}~\bibnamefont
			{Baenitz}},\ }\href {\doibase 10.1103/PhysRevB.99.180401} {\bibfield
		{journal} {\bibinfo  {journal} {Phys. Rev. B}\ }\textbf {\bibinfo {volume}
			{99}},\ \bibinfo {pages} {180401} (\bibinfo {year}
		{2019}{\natexlab{a}})}\BibitemShut {NoStop}%
	\bibitem [{\citenamefont {Baenitz}\ \emph {et~al.}(2018)\citenamefont
		{Baenitz}, \citenamefont {Schlender}, \citenamefont {Sichelschmidt},
		\citenamefont {Onykiienko}, \citenamefont {Zangeneh}, \citenamefont
		{Ranjith}, \citenamefont {Sarkar}, \citenamefont {Hozoi}, \citenamefont
		{Walker}, \citenamefont {Orain}, \citenamefont {Yasuoka}, \citenamefont
		{van~den Brink}, \citenamefont {Klauss}, \citenamefont {Inosov},\ and\
		\citenamefont {Doert}}]{Baenitz_2018}%
	\BibitemOpen
	\bibfield  {author} {\bibinfo {author} {\bibfnamefont {M.}~\bibnamefont
			{Baenitz}}, \bibinfo {author} {\bibfnamefont {P.}~\bibnamefont {Schlender}},
		\bibinfo {author} {\bibfnamefont {J.}~\bibnamefont {Sichelschmidt}}, \bibinfo
		{author} {\bibfnamefont {Y.~A.}\ \bibnamefont {Onykiienko}}, \bibinfo
		{author} {\bibfnamefont {Z.}~\bibnamefont {Zangeneh}}, \bibinfo {author}
		{\bibfnamefont {K.~M.}\ \bibnamefont {Ranjith}}, \bibinfo {author}
		{\bibfnamefont {R.}~\bibnamefont {Sarkar}}, \bibinfo {author} {\bibfnamefont
			{L.}~\bibnamefont {Hozoi}}, \bibinfo {author} {\bibfnamefont {H.~C.}\
			\bibnamefont {Walker}}, \bibinfo {author} {\bibfnamefont {J.-C.}\
			\bibnamefont {Orain}}, \bibinfo {author} {\bibfnamefont {H.}~\bibnamefont
			{Yasuoka}}, \bibinfo {author} {\bibfnamefont {J.}~\bibnamefont {van~den
				Brink}}, \bibinfo {author} {\bibfnamefont {H.~H.}\ \bibnamefont {Klauss}},
		\bibinfo {author} {\bibfnamefont {D.~S.}\ \bibnamefont {Inosov}}, \ and\
		\bibinfo {author} {\bibfnamefont {T.}~\bibnamefont {Doert}},\ }\href
	{\doibase 10.1103/PhysRevB.98.220409} {\bibfield  {journal} {\bibinfo
			{journal} {Phys. Rev. B}\ }\textbf {\bibinfo {volume} {98}},\ \bibinfo
		{pages} {220409} (\bibinfo {year} {2018})}\BibitemShut {NoStop}%
	\bibitem [{\citenamefont {Sarkar}\ \emph {et~al.}(2019)\citenamefont {Sarkar},
		\citenamefont {Schlender}, \citenamefont {Grinenko}, \citenamefont
		{Haeussler}, \citenamefont {Baker}, \citenamefont {Doert},\ and\
		\citenamefont {Klauss}}]{sarkar2019quantum}%
	\BibitemOpen
	\bibfield  {author} {\bibinfo {author} {\bibfnamefont {R.}~\bibnamefont
			{Sarkar}}, \bibinfo {author} {\bibfnamefont {P.}~\bibnamefont {Schlender}},
		\bibinfo {author} {\bibfnamefont {V.}~\bibnamefont {Grinenko}}, \bibinfo
		{author} {\bibfnamefont {E.}~\bibnamefont {Haeussler}}, \bibinfo {author}
		{\bibfnamefont {P.~J.}\ \bibnamefont {Baker}}, \bibinfo {author}
		{\bibfnamefont {T.}~\bibnamefont {Doert}}, \ and\ \bibinfo {author}
		{\bibfnamefont {H.-H.}\ \bibnamefont {Klauss}},\ }\href {\doibase
		10.1103/PhysRevB.100.241116} {\bibfield  {journal} {\bibinfo  {journal}
			{Phys. Rev. B}\ }\textbf {\bibinfo {volume} {100}},\ \bibinfo {pages}
		{241116} (\bibinfo {year} {2019})}\BibitemShut {NoStop}%
	\bibitem [{\citenamefont {Ranjith}\ \emph
		{et~al.}(2019{\natexlab{b}})\citenamefont {Ranjith}, \citenamefont {Luther},
		\citenamefont {Reimann}, \citenamefont {Schmidt}, \citenamefont {Schlender},
		\citenamefont {Sichelschmidt}, \citenamefont {Yasuoka}, \citenamefont
		{Strydom}, \citenamefont {Skourski}, \citenamefont {Wosnitza}, \citenamefont
		{K\"uhne}, \citenamefont {Doert},\ and\ \citenamefont
		{Baenitz}}]{Ranjith2019_2}%
	\BibitemOpen
	\bibfield  {author} {\bibinfo {author} {\bibfnamefont {K.~M.}\ \bibnamefont
			{Ranjith}}, \bibinfo {author} {\bibfnamefont {S.}~\bibnamefont {Luther}},
		\bibinfo {author} {\bibfnamefont {T.}~\bibnamefont {Reimann}}, \bibinfo
		{author} {\bibfnamefont {B.}~\bibnamefont {Schmidt}}, \bibinfo {author}
		{\bibfnamefont {P.}~\bibnamefont {Schlender}}, \bibinfo {author}
		{\bibfnamefont {J.}~\bibnamefont {Sichelschmidt}}, \bibinfo {author}
		{\bibfnamefont {H.}~\bibnamefont {Yasuoka}}, \bibinfo {author} {\bibfnamefont
			{A.~M.}\ \bibnamefont {Strydom}}, \bibinfo {author} {\bibfnamefont
			{Y.}~\bibnamefont {Skourski}}, \bibinfo {author} {\bibfnamefont
			{J.}~\bibnamefont {Wosnitza}}, \bibinfo {author} {\bibfnamefont
			{H.}~\bibnamefont {K\"uhne}}, \bibinfo {author} {\bibfnamefont
			{T.}~\bibnamefont {Doert}}, \ and\ \bibinfo {author} {\bibfnamefont
			{M.}~\bibnamefont {Baenitz}},\ }\href {\doibase 10.1103/PhysRevB.100.224417}
	{\bibfield  {journal} {\bibinfo  {journal} {Phys. Rev. B}\ }\textbf {\bibinfo
			{volume} {100}},\ \bibinfo {pages} {224417} (\bibinfo {year}
		{2019}{\natexlab{b}})}\BibitemShut {NoStop}%
	\bibitem [{\citenamefont {Dai}\ \emph {et~al.}(2021)\citenamefont {Dai},
		\citenamefont {Zhang}, \citenamefont {Xie}, \citenamefont {Duan},
		\citenamefont {Gao}, \citenamefont {Zhu}, \citenamefont {Feng}, \citenamefont
		{Tao}, \citenamefont {Huang}, \citenamefont {Cao}, \citenamefont
		{Podlesnyak}, \citenamefont {Granroth}, \citenamefont {Everett},
		\citenamefont {Neuefeind}, \citenamefont {Voneshen}, \citenamefont {Wang},
		\citenamefont {Tan}, \citenamefont {Morosan}, \citenamefont {Wang},
		\citenamefont {Lin}, \citenamefont {Shu}, \citenamefont {Chen}, \citenamefont
		{Guo}, \citenamefont {Lu},\ and\ \citenamefont {Dai}}]{Dai_2021}%
	\BibitemOpen
	\bibfield  {author} {\bibinfo {author} {\bibfnamefont {P.-L.}\ \bibnamefont
			{Dai}}, \bibinfo {author} {\bibfnamefont {G.}~\bibnamefont {Zhang}}, \bibinfo
		{author} {\bibfnamefont {Y.}~\bibnamefont {Xie}}, \bibinfo {author}
		{\bibfnamefont {C.}~\bibnamefont {Duan}}, \bibinfo {author} {\bibfnamefont
			{Y.}~\bibnamefont {Gao}}, \bibinfo {author} {\bibfnamefont {Z.}~\bibnamefont
			{Zhu}}, \bibinfo {author} {\bibfnamefont {E.}~\bibnamefont {Feng}}, \bibinfo
		{author} {\bibfnamefont {Z.}~\bibnamefont {Tao}}, \bibinfo {author}
		{\bibfnamefont {C.-L.}\ \bibnamefont {Huang}}, \bibinfo {author}
		{\bibfnamefont {H.}~\bibnamefont {Cao}}, \bibinfo {author} {\bibfnamefont
			{A.}~\bibnamefont {Podlesnyak}}, \bibinfo {author} {\bibfnamefont {G.~E.}\
			\bibnamefont {Granroth}}, \bibinfo {author} {\bibfnamefont {M.~S.}\
			\bibnamefont {Everett}}, \bibinfo {author} {\bibfnamefont {J.~C.}\
			\bibnamefont {Neuefeind}}, \bibinfo {author} {\bibfnamefont {D.}~\bibnamefont
			{Voneshen}}, \bibinfo {author} {\bibfnamefont {S.}~\bibnamefont {Wang}},
		\bibinfo {author} {\bibfnamefont {G.}~\bibnamefont {Tan}}, \bibinfo {author}
		{\bibfnamefont {E.}~\bibnamefont {Morosan}}, \bibinfo {author} {\bibfnamefont
			{X.}~\bibnamefont {Wang}}, \bibinfo {author} {\bibfnamefont {H.-Q.}\
			\bibnamefont {Lin}}, \bibinfo {author} {\bibfnamefont {L.}~\bibnamefont
			{Shu}}, \bibinfo {author} {\bibfnamefont {G.}~\bibnamefont {Chen}}, \bibinfo
		{author} {\bibfnamefont {Y.}~\bibnamefont {Guo}}, \bibinfo {author}
		{\bibfnamefont {X.}~\bibnamefont {Lu}}, \ and\ \bibinfo {author}
		{\bibfnamefont {P.}~\bibnamefont {Dai}},\ }\href {\doibase
		10.1103/PhysRevX.11.021044} {\bibfield  {journal} {\bibinfo  {journal} {Phys.
				Rev. X}\ }\textbf {\bibinfo {volume} {11}},\ \bibinfo {pages} {021044}
		(\bibinfo {year} {2021})}\BibitemShut {NoStop}%
	\bibitem [{\citenamefont {Xie}\ \emph {et~al.}(2021)\citenamefont {Xie},
		\citenamefont {Xing}, \citenamefont {Nikitin}, \citenamefont {Nishimoto},
		\citenamefont {Brando}, \citenamefont {Khanenko}, \citenamefont
		{Sichelschmidt}, \citenamefont {Sanjeewa}, \citenamefont {Sefat},\ and\
		\citenamefont {Podlesnyak}}]{xie2021field}%
	\BibitemOpen
	\bibfield  {author} {\bibinfo {author} {\bibfnamefont {T.}~\bibnamefont
			{Xie}}, \bibinfo {author} {\bibfnamefont {J.}~\bibnamefont {Xing}}, \bibinfo
		{author} {\bibfnamefont {S.}~\bibnamefont {Nikitin}}, \bibinfo {author}
		{\bibfnamefont {S.}~\bibnamefont {Nishimoto}}, \bibinfo {author}
		{\bibfnamefont {M.}~\bibnamefont {Brando}}, \bibinfo {author} {\bibfnamefont
			{P.}~\bibnamefont {Khanenko}}, \bibinfo {author} {\bibfnamefont
			{J.}~\bibnamefont {Sichelschmidt}}, \bibinfo {author} {\bibfnamefont
			{L.}~\bibnamefont {Sanjeewa}}, \bibinfo {author} {\bibfnamefont {A.~S.}\
			\bibnamefont {Sefat}}, \ and\ \bibinfo {author} {\bibfnamefont
			{A.}~\bibnamefont {Podlesnyak}},\ }\href {https://arxiv.org/abs/2106.12451}
	{\bibfield  {journal} {\bibinfo  {journal} {arXiv preprint arXiv:2106.12451}\
		} (\bibinfo {year} {2021})}\BibitemShut {NoStop}%
	\bibitem [{\citenamefont {Zhang}\ \emph
		{et~al.}(2019{\natexlab{a}})\citenamefont {Zhang}, \citenamefont {Changlani},
		\citenamefont {Plumb}, \citenamefont {Tchernyshyov},\ and\ \citenamefont
		{Moessner}}]{Zhang_2019}%
	\BibitemOpen
	\bibfield  {author} {\bibinfo {author} {\bibfnamefont {S.}~\bibnamefont
			{Zhang}}, \bibinfo {author} {\bibfnamefont {H.~J.}\ \bibnamefont
			{Changlani}}, \bibinfo {author} {\bibfnamefont {K.~W.}\ \bibnamefont
			{Plumb}}, \bibinfo {author} {\bibfnamefont {O.}~\bibnamefont {Tchernyshyov}},
		\ and\ \bibinfo {author} {\bibfnamefont {R.}~\bibnamefont {Moessner}},\
	}\href {\doibase 10.1103/PhysRevLett.122.167203} {\bibfield  {journal}
		{\bibinfo  {journal} {Phys. Rev. Lett.}\ }\textbf {\bibinfo {volume} {122}},\
		\bibinfo {pages} {167203} (\bibinfo {year} {2019}{\natexlab{a}})}\BibitemShut
	{NoStop}%
	\bibitem [{\citenamefont {Zhu}\ \emph {et~al.}(2017)\citenamefont {Zhu},
		\citenamefont {Maksimov}, \citenamefont {White},\ and\ \citenamefont
		{Chernyshev}}]{Zhu_2017_YMGO}%
	\BibitemOpen
	\bibfield  {author} {\bibinfo {author} {\bibfnamefont {Z.}~\bibnamefont
			{Zhu}}, \bibinfo {author} {\bibfnamefont {P.~A.}\ \bibnamefont {Maksimov}},
		\bibinfo {author} {\bibfnamefont {S.~R.}\ \bibnamefont {White}}, \ and\
		\bibinfo {author} {\bibfnamefont {A.~L.}\ \bibnamefont {Chernyshev}},\ }\href
	{\doibase 10.1103/PhysRevLett.119.157201} {\bibfield  {journal} {\bibinfo
			{journal} {Phys. Rev. Lett.}\ }\textbf {\bibinfo {volume} {119}},\ \bibinfo
		{pages} {157201} (\bibinfo {year} {2017})}\BibitemShut {NoStop}%
	\bibitem [{\citenamefont {Xing}\ \emph {et~al.}(2021)\citenamefont {Xing},
		\citenamefont {Sanjeewa}, \citenamefont {May},\ and\ \citenamefont
		{Sefat}}]{xing2021_KYS}%
	\BibitemOpen
	\bibfield  {author} {\bibinfo {author} {\bibfnamefont {J.}~\bibnamefont
			{Xing}}, \bibinfo {author} {\bibfnamefont {L.~D.}\ \bibnamefont {Sanjeewa}},
		\bibinfo {author} {\bibfnamefont {A.~F.}\ \bibnamefont {May}}, \ and\
		\bibinfo {author} {\bibfnamefont {A.~S.}\ \bibnamefont {Sefat}},\ }\href
	{\doibase 10.1063/5.0071161} {\bibfield  {journal} {\bibinfo  {journal} {APL
				Materials}\ }\textbf {\bibinfo {volume} {9}},\ \bibinfo {pages} {111104}
		(\bibinfo {year} {2021})}\BibitemShut {NoStop}%
	\bibitem [{\citenamefont {Scheie}\ \emph {et~al.}(2021)\citenamefont {Scheie},
		\citenamefont {Laurell}, \citenamefont {Samarakoon}, \citenamefont {Lake},
		\citenamefont {Nagler}, \citenamefont {Granroth}, \citenamefont {Okamoto},
		\citenamefont {Alvarez},\ and\ \citenamefont
		{Tennant}}]{scheie2021witnessing}%
	\BibitemOpen
	\bibfield  {author} {\bibinfo {author} {\bibfnamefont {A.}~\bibnamefont
			{Scheie}}, \bibinfo {author} {\bibfnamefont {P.}~\bibnamefont {Laurell}},
		\bibinfo {author} {\bibfnamefont {A.~M.}\ \bibnamefont {Samarakoon}},
		\bibinfo {author} {\bibfnamefont {B.}~\bibnamefont {Lake}}, \bibinfo {author}
		{\bibfnamefont {S.~E.}\ \bibnamefont {Nagler}}, \bibinfo {author}
		{\bibfnamefont {G.~E.}\ \bibnamefont {Granroth}}, \bibinfo {author}
		{\bibfnamefont {S.}~\bibnamefont {Okamoto}}, \bibinfo {author} {\bibfnamefont
			{G.}~\bibnamefont {Alvarez}}, \ and\ \bibinfo {author} {\bibfnamefont
			{D.~A.}\ \bibnamefont {Tennant}},\ }\href {\doibase
		10.1103/PhysRevB.103.224434} {\bibfield  {journal} {\bibinfo  {journal}
			{Phys. Rev. B}\ }\textbf {\bibinfo {volume} {103}},\ \bibinfo {pages}
		{224434} (\bibinfo {year} {2021})}\BibitemShut {NoStop}%
	\bibitem [{\citenamefont {Ehlers}\ \emph {et~al.}(2011)\citenamefont {Ehlers},
		\citenamefont {Podlesnyak}, \citenamefont {Niedziela}, \citenamefont
		{Iverson},\ and\ \citenamefont {Sokol}}]{CNCS}%
	\BibitemOpen
	\bibfield  {author} {\bibinfo {author} {\bibfnamefont {G.}~\bibnamefont
			{Ehlers}}, \bibinfo {author} {\bibfnamefont {A.~A.}\ \bibnamefont
			{Podlesnyak}}, \bibinfo {author} {\bibfnamefont {J.~L.}\ \bibnamefont
			{Niedziela}}, \bibinfo {author} {\bibfnamefont {E.~B.}\ \bibnamefont
			{Iverson}}, \ and\ \bibinfo {author} {\bibfnamefont {P.~E.}\ \bibnamefont
			{Sokol}},\ }\href {\doibase 10.1063/1.3626935} {\bibfield  {journal}
		{\bibinfo  {journal} {Review of Scientific Instruments}\ }\textbf {\bibinfo
			{volume} {82}},\ \bibinfo {pages} {085108} (\bibinfo {year}
		{2011})}\BibitemShut {NoStop}%
	\bibitem [{\citenamefont {Mason}\ \emph {et~al.}(2006)\citenamefont {Mason},
		\citenamefont {Abernathy}, \citenamefont {Anderson}, \citenamefont {Ankner},
		\citenamefont {Egami}, \citenamefont {Ehlers}, \citenamefont {Ekkebus},
		\citenamefont {Granroth}, \citenamefont {Hagen}, \citenamefont {Herwig},
		\citenamefont {Hodges}, \citenamefont {Hoffmann}, \citenamefont {Horak},
		\citenamefont {Horton}, \citenamefont {Klose}, \citenamefont {Larese},
		\citenamefont {Mesecar}, \citenamefont {Myles}, \citenamefont {Neuefeind},
		\citenamefont {Ohl}, \citenamefont {Tulk}, \citenamefont {Wang},\ and\
		\citenamefont {Zhao}}]{mason2006spallation}%
	\BibitemOpen
	\bibfield  {author} {\bibinfo {author} {\bibfnamefont {T.~E.}\ \bibnamefont
			{Mason}}, \bibinfo {author} {\bibfnamefont {D.}~\bibnamefont {Abernathy}},
		\bibinfo {author} {\bibfnamefont {I.}~\bibnamefont {Anderson}}, \bibinfo
		{author} {\bibfnamefont {J.}~\bibnamefont {Ankner}}, \bibinfo {author}
		{\bibfnamefont {T.}~\bibnamefont {Egami}}, \bibinfo {author} {\bibfnamefont
			{G.}~\bibnamefont {Ehlers}}, \bibinfo {author} {\bibfnamefont
			{A.}~\bibnamefont {Ekkebus}}, \bibinfo {author} {\bibfnamefont
			{G.}~\bibnamefont {Granroth}}, \bibinfo {author} {\bibfnamefont
			{M.}~\bibnamefont {Hagen}}, \bibinfo {author} {\bibfnamefont
			{K.}~\bibnamefont {Herwig}}, \bibinfo {author} {\bibfnamefont
			{J.}~\bibnamefont {Hodges}}, \bibinfo {author} {\bibfnamefont
			{C.}~\bibnamefont {Hoffmann}}, \bibinfo {author} {\bibfnamefont
			{C.}~\bibnamefont {Horak}}, \bibinfo {author} {\bibfnamefont
			{L.}~\bibnamefont {Horton}}, \bibinfo {author} {\bibfnamefont
			{F.}~\bibnamefont {Klose}}, \bibinfo {author} {\bibfnamefont
			{J.}~\bibnamefont {Larese}}, \bibinfo {author} {\bibfnamefont
			{A.}~\bibnamefont {Mesecar}}, \bibinfo {author} {\bibfnamefont
			{D.}~\bibnamefont {Myles}}, \bibinfo {author} {\bibfnamefont
			{J.}~\bibnamefont {Neuefeind}}, \bibinfo {author} {\bibfnamefont
			{M.}~\bibnamefont {Ohl}}, \bibinfo {author} {\bibfnamefont {C.}~\bibnamefont
			{Tulk}}, \bibinfo {author} {\bibfnamefont {X.-L.}\ \bibnamefont {Wang}}, \
		and\ \bibinfo {author} {\bibfnamefont {J.}~\bibnamefont {Zhao}},\ }\href
	{https://doi.org/10.1016/j.physb.2006.05.281} {\bibfield  {journal} {\bibinfo
			{journal} {Physica B: Condensed Matter}\ }\textbf {\bibinfo {volume}
			{385}},\ \bibinfo {pages} {955} (\bibinfo {year} {2006})}\BibitemShut
	{NoStop}%
	\bibitem [{\citenamefont {Macdougal}\ \emph {et~al.}(2020)\citenamefont
		{Macdougal}, \citenamefont {Williams}, \citenamefont {Prabhakaran},
		\citenamefont {Bewley}, \citenamefont {Voneshen},\ and\ \citenamefont
		{Coldea}}]{Macdougal_2020}%
	\BibitemOpen
	\bibfield  {author} {\bibinfo {author} {\bibfnamefont {D.}~\bibnamefont
			{Macdougal}}, \bibinfo {author} {\bibfnamefont {S.}~\bibnamefont {Williams}},
		\bibinfo {author} {\bibfnamefont {D.}~\bibnamefont {Prabhakaran}}, \bibinfo
		{author} {\bibfnamefont {R.~I.}\ \bibnamefont {Bewley}}, \bibinfo {author}
		{\bibfnamefont {D.~J.}\ \bibnamefont {Voneshen}}, \ and\ \bibinfo {author}
		{\bibfnamefont {R.}~\bibnamefont {Coldea}},\ }\href {\doibase
		10.1103/PhysRevB.102.064421} {\bibfield  {journal} {\bibinfo  {journal}
			{Phys. Rev. B}\ }\textbf {\bibinfo {volume} {102}},\ \bibinfo {pages}
		{064421} (\bibinfo {year} {2020})}\BibitemShut {NoStop}%
	\bibitem [{\citenamefont {Zhou}\ \emph {et~al.}(2012)\citenamefont {Zhou},
		\citenamefont {Xu}, \citenamefont {Hallas}, \citenamefont {Silverstein},
		\citenamefont {Wiebe}, \citenamefont {Umegaki}, \citenamefont {Yan},
		\citenamefont {Murphy}, \citenamefont {Park}, \citenamefont {Qiu},
		\citenamefont {Copley}, \citenamefont {Gardner},\ and\ \citenamefont
		{Takano}}]{Zhou_2012}%
	\BibitemOpen
	\bibfield  {author} {\bibinfo {author} {\bibfnamefont {H.~D.}\ \bibnamefont
			{Zhou}}, \bibinfo {author} {\bibfnamefont {C.}~\bibnamefont {Xu}}, \bibinfo
		{author} {\bibfnamefont {A.~M.}\ \bibnamefont {Hallas}}, \bibinfo {author}
		{\bibfnamefont {H.~J.}\ \bibnamefont {Silverstein}}, \bibinfo {author}
		{\bibfnamefont {C.~R.}\ \bibnamefont {Wiebe}}, \bibinfo {author}
		{\bibfnamefont {I.}~\bibnamefont {Umegaki}}, \bibinfo {author} {\bibfnamefont
			{J.~Q.}\ \bibnamefont {Yan}}, \bibinfo {author} {\bibfnamefont {T.~P.}\
			\bibnamefont {Murphy}}, \bibinfo {author} {\bibfnamefont {J.-H.}\
			\bibnamefont {Park}}, \bibinfo {author} {\bibfnamefont {Y.}~\bibnamefont
			{Qiu}}, \bibinfo {author} {\bibfnamefont {J.~R.~D.}\ \bibnamefont {Copley}},
		\bibinfo {author} {\bibfnamefont {J.~S.}\ \bibnamefont {Gardner}}, \ and\
		\bibinfo {author} {\bibfnamefont {Y.}~\bibnamefont {Takano}},\ }\href
	{\doibase 10.1103/PhysRevLett.109.267206} {\bibfield  {journal} {\bibinfo
			{journal} {Phys. Rev. Lett.}\ }\textbf {\bibinfo {volume} {109}},\ \bibinfo
		{pages} {267206} (\bibinfo {year} {2012})}\BibitemShut {NoStop}%
	\bibitem [{\citenamefont {Ito}\ \emph {et~al.}(2017)\citenamefont {Ito},
		\citenamefont {Kurita}, \citenamefont {Tanaka}, \citenamefont
		{Ohira-Kawamura}, \citenamefont {Nakajima}, \citenamefont {Itoh},
		\citenamefont {Kuwahara},\ and\ \citenamefont {Kakurai}}]{Ito2017}%
	\BibitemOpen
	\bibfield  {author} {\bibinfo {author} {\bibfnamefont {S.}~\bibnamefont
			{Ito}}, \bibinfo {author} {\bibfnamefont {N.}~\bibnamefont {Kurita}},
		\bibinfo {author} {\bibfnamefont {H.}~\bibnamefont {Tanaka}}, \bibinfo
		{author} {\bibfnamefont {S.}~\bibnamefont {Ohira-Kawamura}}, \bibinfo
		{author} {\bibfnamefont {K.}~\bibnamefont {Nakajima}}, \bibinfo {author}
		{\bibfnamefont {S.}~\bibnamefont {Itoh}}, \bibinfo {author} {\bibfnamefont
			{K.}~\bibnamefont {Kuwahara}}, \ and\ \bibinfo {author} {\bibfnamefont
			{K.}~\bibnamefont {Kakurai}},\ }\href {\doibase 10.1038/s41467-017-00316-x}
	{\bibfield  {journal} {\bibinfo  {journal} {Nature Communications}\ }\textbf
		{\bibinfo {volume} {8}},\ \bibinfo {pages} {235} (\bibinfo {year}
		{2017})}\BibitemShut {NoStop}%
	\bibitem [{\citenamefont {Ma}\ \emph {et~al.}(2016)\citenamefont {Ma},
		\citenamefont {Kamiya}, \citenamefont {Hong}, \citenamefont {Cao},
		\citenamefont {Ehlers}, \citenamefont {Tian}, \citenamefont {Batista},
		\citenamefont {Dun}, \citenamefont {Zhou},\ and\ \citenamefont
		{Matsuda}}]{Ma_2016}%
	\BibitemOpen
	\bibfield  {author} {\bibinfo {author} {\bibfnamefont {J.}~\bibnamefont
			{Ma}}, \bibinfo {author} {\bibfnamefont {Y.}~\bibnamefont {Kamiya}}, \bibinfo
		{author} {\bibfnamefont {T.}~\bibnamefont {Hong}}, \bibinfo {author}
		{\bibfnamefont {H.~B.}\ \bibnamefont {Cao}}, \bibinfo {author} {\bibfnamefont
			{G.}~\bibnamefont {Ehlers}}, \bibinfo {author} {\bibfnamefont
			{W.}~\bibnamefont {Tian}}, \bibinfo {author} {\bibfnamefont {C.~D.}\
			\bibnamefont {Batista}}, \bibinfo {author} {\bibfnamefont {Z.~L.}\
			\bibnamefont {Dun}}, \bibinfo {author} {\bibfnamefont {H.~D.}\ \bibnamefont
			{Zhou}}, \ and\ \bibinfo {author} {\bibfnamefont {M.}~\bibnamefont
			{Matsuda}},\ }\href {\doibase 10.1103/PhysRevLett.116.087201} {\bibfield
		{journal} {\bibinfo  {journal} {Phys. Rev. Lett.}\ }\textbf {\bibinfo
			{volume} {116}},\ \bibinfo {pages} {087201} (\bibinfo {year}
		{2016})}\BibitemShut {NoStop}%
	\bibitem [{\citenamefont {Zheng}\ \emph {et~al.}(2006)\citenamefont {Zheng},
		\citenamefont {Fj\ae{}restad}, \citenamefont {Singh}, \citenamefont
		{McKenzie},\ and\ \citenamefont {Coldea}}]{Zheng_2006}%
	\BibitemOpen
	\bibfield  {author} {\bibinfo {author} {\bibfnamefont {W.}~\bibnamefont
			{Zheng}}, \bibinfo {author} {\bibfnamefont {J.~O.}\ \bibnamefont
			{Fj\ae{}restad}}, \bibinfo {author} {\bibfnamefont {R.~R.~P.}\ \bibnamefont
			{Singh}}, \bibinfo {author} {\bibfnamefont {R.~H.}\ \bibnamefont {McKenzie}},
		\ and\ \bibinfo {author} {\bibfnamefont {R.}~\bibnamefont {Coldea}},\ }\href
	{\doibase 10.1103/PhysRevB.74.224420} {\bibfield  {journal} {\bibinfo
			{journal} {Phys. Rev. B}\ }\textbf {\bibinfo {volume} {74}},\ \bibinfo
		{pages} {224420} (\bibinfo {year} {2006})}\BibitemShut {NoStop}%
	\bibitem [{\citenamefont {Starykh}\ \emph {et~al.}(2006)\citenamefont
		{Starykh}, \citenamefont {Chubukov},\ and\ \citenamefont
		{Abanov}}]{Starykh_2006}%
	\BibitemOpen
	\bibfield  {author} {\bibinfo {author} {\bibfnamefont {O.~A.}\ \bibnamefont
			{Starykh}}, \bibinfo {author} {\bibfnamefont {A.~V.}\ \bibnamefont
			{Chubukov}}, \ and\ \bibinfo {author} {\bibfnamefont {A.~G.}\ \bibnamefont
			{Abanov}},\ }\href {\doibase 10.1103/PhysRevB.74.180403} {\bibfield
		{journal} {\bibinfo  {journal} {Phys. Rev. B}\ }\textbf {\bibinfo {volume}
			{74}},\ \bibinfo {pages} {180403} (\bibinfo {year} {2006})}\BibitemShut
	{NoStop}%
	\bibitem [{\citenamefont {Chernyshev}\ and\ \citenamefont
		{Zhitomirsky}(2009)}]{Chernyshev_2009}%
	\BibitemOpen
	\bibfield  {author} {\bibinfo {author} {\bibfnamefont {A.~L.}\ \bibnamefont
			{Chernyshev}}\ and\ \bibinfo {author} {\bibfnamefont {M.~E.}\ \bibnamefont
			{Zhitomirsky}},\ }\href {\doibase 10.1103/PhysRevB.79.144416} {\bibfield
		{journal} {\bibinfo  {journal} {Phys. Rev. B}\ }\textbf {\bibinfo {volume}
			{79}},\ \bibinfo {pages} {144416} (\bibinfo {year} {2009})}\BibitemShut
	{NoStop}%
	\bibitem [{\citenamefont {Laurell}\ \emph {et~al.}(2021)\citenamefont
		{Laurell}, \citenamefont {Scheie}, \citenamefont {Mukherjee}, \citenamefont
		{Koza}, \citenamefont {Enderle}, \citenamefont {Tylczynski}, \citenamefont
		{Okamoto}, \citenamefont {Coldea}, \citenamefont {Tennant},\ and\
		\citenamefont {Alvarez}}]{laurell2020dynamics}%
	\BibitemOpen
	\bibfield  {author} {\bibinfo {author} {\bibfnamefont {P.}~\bibnamefont
			{Laurell}}, \bibinfo {author} {\bibfnamefont {A.}~\bibnamefont {Scheie}},
		\bibinfo {author} {\bibfnamefont {C.~J.}\ \bibnamefont {Mukherjee}}, \bibinfo
		{author} {\bibfnamefont {M.~M.}\ \bibnamefont {Koza}}, \bibinfo {author}
		{\bibfnamefont {M.}~\bibnamefont {Enderle}}, \bibinfo {author} {\bibfnamefont
			{Z.}~\bibnamefont {Tylczynski}}, \bibinfo {author} {\bibfnamefont
			{S.}~\bibnamefont {Okamoto}}, \bibinfo {author} {\bibfnamefont
			{R.}~\bibnamefont {Coldea}}, \bibinfo {author} {\bibfnamefont {D.~A.}\
			\bibnamefont {Tennant}}, \ and\ \bibinfo {author} {\bibfnamefont
			{G.}~\bibnamefont {Alvarez}},\ }\href {\doibase
		10.1103/PhysRevLett.127.037201} {\bibfield  {journal} {\bibinfo  {journal}
			{Phys. Rev. Lett.}\ }\textbf {\bibinfo {volume} {127}},\ \bibinfo {pages}
		{037201} (\bibinfo {year} {2021})}\BibitemShut {NoStop}%
	\bibitem [{\citenamefont {Lake}\ \emph {et~al.}(2013)\citenamefont {Lake},
		\citenamefont {Tennant}, \citenamefont {Caux}, \citenamefont {Barthel},
		\citenamefont {Schollw\"ock}, \citenamefont {Nagler},\ and\ \citenamefont
		{Frost}}]{Lake2013}%
	\BibitemOpen
	\bibfield  {author} {\bibinfo {author} {\bibfnamefont {B.}~\bibnamefont
			{Lake}}, \bibinfo {author} {\bibfnamefont {D.~A.}\ \bibnamefont {Tennant}},
		\bibinfo {author} {\bibfnamefont {J.-S.}\ \bibnamefont {Caux}}, \bibinfo
		{author} {\bibfnamefont {T.}~\bibnamefont {Barthel}}, \bibinfo {author}
		{\bibfnamefont {U.}~\bibnamefont {Schollw\"ock}}, \bibinfo {author}
		{\bibfnamefont {S.~E.}\ \bibnamefont {Nagler}}, \ and\ \bibinfo {author}
		{\bibfnamefont {C.~D.}\ \bibnamefont {Frost}},\ }\href {\doibase
		10.1103/PhysRevLett.111.137205} {\bibfield  {journal} {\bibinfo  {journal}
			{Phys. Rev. Lett.}\ }\textbf {\bibinfo {volume} {111}},\ \bibinfo {pages}
		{137205} (\bibinfo {year} {2013})}\BibitemShut {NoStop}%
	\bibitem [{\citenamefont {Plumb}\ \emph {et~al.}(2019)\citenamefont {Plumb},
		\citenamefont {Changlani}, \citenamefont {Scheie}, \citenamefont {Zhang},
		\citenamefont {Krizan}, \citenamefont {Rodriguez-Rivera}, \citenamefont
		{Qiu}, \citenamefont {Winn}, \citenamefont {Cava},\ and\ \citenamefont
		{Broholm}}]{plumb2019continuum}%
	\BibitemOpen
	\bibfield  {author} {\bibinfo {author} {\bibfnamefont {K.}~\bibnamefont
			{Plumb}}, \bibinfo {author} {\bibfnamefont {H.~J.}\ \bibnamefont
			{Changlani}}, \bibinfo {author} {\bibfnamefont {A.}~\bibnamefont {Scheie}},
		\bibinfo {author} {\bibfnamefont {S.}~\bibnamefont {Zhang}}, \bibinfo
		{author} {\bibfnamefont {J.}~\bibnamefont {Krizan}}, \bibinfo {author}
		{\bibfnamefont {J.}~\bibnamefont {Rodriguez-Rivera}}, \bibinfo {author}
		{\bibfnamefont {Y.}~\bibnamefont {Qiu}}, \bibinfo {author} {\bibfnamefont
			{B.}~\bibnamefont {Winn}}, \bibinfo {author} {\bibfnamefont {R.}~\bibnamefont
			{Cava}}, \ and\ \bibinfo {author} {\bibfnamefont {C.~L.}\ \bibnamefont
			{Broholm}},\ }\href {https://doi.org/10.1038/s41567-018-0317-3} {\bibfield
		{journal} {\bibinfo  {journal} {Nature Physics}\ }\textbf {\bibinfo {volume}
			{15}},\ \bibinfo {pages} {54} (\bibinfo {year} {2019})}\BibitemShut {NoStop}%
	\bibitem [{\citenamefont {Shen}\ \emph {et~al.}(2016)\citenamefont {Shen},
		\citenamefont {Li}, \citenamefont {Wo}, \citenamefont {Li}, \citenamefont
		{Shen}, \citenamefont {Pan}, \citenamefont {Wang}, \citenamefont {Walker},
		\citenamefont {Steffens}, \citenamefont {Boehm}, \citenamefont {Hao},
		\citenamefont {Quintero-Castro}, \citenamefont {Harriger}, \citenamefont
		{Frontzek}, \citenamefont {Hao}, \citenamefont {Meng}, \citenamefont {Zhang},
		\citenamefont {Chen},\ and\ \citenamefont {Zhao}}]{Shen2016}%
	\BibitemOpen
	\bibfield  {author} {\bibinfo {author} {\bibfnamefont {Y.}~\bibnamefont
			{Shen}}, \bibinfo {author} {\bibfnamefont {Y.-D.}\ \bibnamefont {Li}},
		\bibinfo {author} {\bibfnamefont {H.}~\bibnamefont {Wo}}, \bibinfo {author}
		{\bibfnamefont {Y.}~\bibnamefont {Li}}, \bibinfo {author} {\bibfnamefont
			{S.}~\bibnamefont {Shen}}, \bibinfo {author} {\bibfnamefont {B.}~\bibnamefont
			{Pan}}, \bibinfo {author} {\bibfnamefont {Q.}~\bibnamefont {Wang}}, \bibinfo
		{author} {\bibfnamefont {H.~C.}\ \bibnamefont {Walker}}, \bibinfo {author}
		{\bibfnamefont {P.}~\bibnamefont {Steffens}}, \bibinfo {author}
		{\bibfnamefont {M.}~\bibnamefont {Boehm}}, \bibinfo {author} {\bibfnamefont
			{Y.}~\bibnamefont {Hao}}, \bibinfo {author} {\bibfnamefont {D.~L.}\
			\bibnamefont {Quintero-Castro}}, \bibinfo {author} {\bibfnamefont {L.~W.}\
			\bibnamefont {Harriger}}, \bibinfo {author} {\bibfnamefont {M.~D.}\
			\bibnamefont {Frontzek}}, \bibinfo {author} {\bibfnamefont {L.}~\bibnamefont
			{Hao}}, \bibinfo {author} {\bibfnamefont {S.}~\bibnamefont {Meng}}, \bibinfo
		{author} {\bibfnamefont {Q.}~\bibnamefont {Zhang}}, \bibinfo {author}
		{\bibfnamefont {G.}~\bibnamefont {Chen}}, \ and\ \bibinfo {author}
		{\bibfnamefont {J.}~\bibnamefont {Zhao}},\ }\href {\doibase
		10.1038/nature20614} {\bibfield  {journal} {\bibinfo  {journal} {Nature}\
		}\textbf {\bibinfo {volume} {540}},\ \bibinfo {pages} {559} (\bibinfo {year}
		{2016})}\BibitemShut {NoStop}%
	\bibitem [{\citenamefont {Paddison}\ \emph {et~al.}(2017)\citenamefont
		{Paddison}, \citenamefont {Daum}, \citenamefont {Dun}, \citenamefont
		{Ehlers}, \citenamefont {Liu}, \citenamefont {Stone}, \citenamefont {Zhou},\
		and\ \citenamefont {Mourigal}}]{Paddison2017}%
	\BibitemOpen
	\bibfield  {author} {\bibinfo {author} {\bibfnamefont {J.~A.~M.}\
			\bibnamefont {Paddison}}, \bibinfo {author} {\bibfnamefont {M.}~\bibnamefont
			{Daum}}, \bibinfo {author} {\bibfnamefont {Z.}~\bibnamefont {Dun}}, \bibinfo
		{author} {\bibfnamefont {G.}~\bibnamefont {Ehlers}}, \bibinfo {author}
		{\bibfnamefont {Y.}~\bibnamefont {Liu}}, \bibinfo {author} {\bibfnamefont
			{M.}~\bibnamefont {Stone}}, \bibinfo {author} {\bibfnamefont
			{H.}~\bibnamefont {Zhou}}, \ and\ \bibinfo {author} {\bibfnamefont
			{M.}~\bibnamefont {Mourigal}},\ }\href {\doibase 10.1038/nphys3971}
	{\bibfield  {journal} {\bibinfo  {journal} {Nature Physics}\ }\textbf
		{\bibinfo {volume} {13}},\ \bibinfo {pages} {117} (\bibinfo {year}
		{2017})}\BibitemShut {NoStop}%
	\bibitem [{\citenamefont {Abernathy}\ \emph {et~al.}(2012)\citenamefont
		{Abernathy}, \citenamefont {Stone}, \citenamefont {Loguillo}, \citenamefont
		{Lucas}, \citenamefont {Delaire}, \citenamefont {Tang}, \citenamefont {Lin},\
		and\ \citenamefont {Fultz}}]{ARCS}%
	\BibitemOpen
	\bibfield  {author} {\bibinfo {author} {\bibfnamefont {D.~L.}\ \bibnamefont
			{Abernathy}}, \bibinfo {author} {\bibfnamefont {M.~B.}\ \bibnamefont
			{Stone}}, \bibinfo {author} {\bibfnamefont {M.~J.}\ \bibnamefont {Loguillo}},
		\bibinfo {author} {\bibfnamefont {M.~S.}\ \bibnamefont {Lucas}}, \bibinfo
		{author} {\bibfnamefont {O.}~\bibnamefont {Delaire}}, \bibinfo {author}
		{\bibfnamefont {X.}~\bibnamefont {Tang}}, \bibinfo {author} {\bibfnamefont
			{J.~Y.~Y.}\ \bibnamefont {Lin}}, \ and\ \bibinfo {author} {\bibfnamefont
			{B.}~\bibnamefont {Fultz}},\ }\href {\doibase 10.1063/1.3680104} {\bibfield
		{journal} {\bibinfo  {journal} {Review of Scientific Instruments}\ }\textbf
		{\bibinfo {volume} {83}},\ \bibinfo {pages} {015114} (\bibinfo {year}
		{2012})}\BibitemShut {NoStop}%
	\bibitem [{\citenamefont {Scheie}(2021)}]{PyCrystalField}%
	\BibitemOpen
	\bibfield  {author} {\bibinfo {author} {\bibfnamefont {A.}~\bibnamefont
			{Scheie}},\ }\href {https://doi.org/10.1107/S160057672001554X} {\bibfield
		{journal} {\bibinfo  {journal} {Journal of Applied Crystallography}\ }\textbf
		{\bibinfo {volume} {54}} (\bibinfo {year} {2021})}\BibitemShut {NoStop}%
	\bibitem [{\citenamefont {Wootters}(1998)}]{Wootters_1998}%
	\BibitemOpen
	\bibfield  {author} {\bibinfo {author} {\bibfnamefont {W.~K.}\ \bibnamefont
			{Wootters}},\ }\href {\doibase 10.1103/PhysRevLett.80.2245} {\bibfield
		{journal} {\bibinfo  {journal} {Phys. Rev. Lett.}\ }\textbf {\bibinfo
			{volume} {80}},\ \bibinfo {pages} {2245} (\bibinfo {year}
		{1998})}\BibitemShut {NoStop}%
	\bibitem [{\citenamefont {Coffman}\ \emph {et~al.}(2000)\citenamefont
		{Coffman}, \citenamefont {Kundu},\ and\ \citenamefont
		{Wootters}}]{PhysRevA.61.052306}%
	\BibitemOpen
	\bibfield  {author} {\bibinfo {author} {\bibfnamefont {V.}~\bibnamefont
			{Coffman}}, \bibinfo {author} {\bibfnamefont {J.}~\bibnamefont {Kundu}}, \
		and\ \bibinfo {author} {\bibfnamefont {W.~K.}\ \bibnamefont {Wootters}},\
	}\href {\doibase 10.1103/PhysRevA.61.052306} {\bibfield  {journal} {\bibinfo
			{journal} {Phys. Rev. A}\ }\textbf {\bibinfo {volume} {61}},\ \bibinfo
		{pages} {052306} (\bibinfo {year} {2000})}\BibitemShut {NoStop}%
	\bibitem [{\citenamefont {Roscilde}\ \emph {et~al.}(2004)\citenamefont
		{Roscilde}, \citenamefont {Verrucchi}, \citenamefont {Fubini}, \citenamefont
		{Haas},\ and\ \citenamefont {Tognetti}}]{Roscilde_2004}%
	\BibitemOpen
	\bibfield  {author} {\bibinfo {author} {\bibfnamefont {T.}~\bibnamefont
			{Roscilde}}, \bibinfo {author} {\bibfnamefont {P.}~\bibnamefont {Verrucchi}},
		\bibinfo {author} {\bibfnamefont {A.}~\bibnamefont {Fubini}}, \bibinfo
		{author} {\bibfnamefont {S.}~\bibnamefont {Haas}}, \ and\ \bibinfo {author}
		{\bibfnamefont {V.}~\bibnamefont {Tognetti}},\ }\href {\doibase
		10.1103/PhysRevLett.93.167203} {\bibfield  {journal} {\bibinfo  {journal}
			{Phys. Rev. Lett.}\ }\textbf {\bibinfo {volume} {93}},\ \bibinfo {pages}
		{167203} (\bibinfo {year} {2004})}\BibitemShut {NoStop}%
	\bibitem [{\citenamefont {Amico}\ \emph {et~al.}(2006)\citenamefont {Amico},
		\citenamefont {Baroni}, \citenamefont {Fubini}, \citenamefont {Patan\`e},
		\citenamefont {Tognetti},\ and\ \citenamefont {Verrucchi}}]{Amico_2006}%
	\BibitemOpen
	\bibfield  {author} {\bibinfo {author} {\bibfnamefont {L.}~\bibnamefont
			{Amico}}, \bibinfo {author} {\bibfnamefont {F.}~\bibnamefont {Baroni}},
		\bibinfo {author} {\bibfnamefont {A.}~\bibnamefont {Fubini}}, \bibinfo
		{author} {\bibfnamefont {D.}~\bibnamefont {Patan\`e}}, \bibinfo {author}
		{\bibfnamefont {V.}~\bibnamefont {Tognetti}}, \ and\ \bibinfo {author}
		{\bibfnamefont {P.}~\bibnamefont {Verrucchi}},\ }\href {\doibase
		10.1103/PhysRevA.74.022322} {\bibfield  {journal} {\bibinfo  {journal} {Phys.
				Rev. A}\ }\textbf {\bibinfo {volume} {74}},\ \bibinfo {pages} {022322}
		(\bibinfo {year} {2006})}\BibitemShut {NoStop}%
	\bibitem [{\citenamefont {Hauke}\ \emph {et~al.}(2016)\citenamefont {Hauke},
		\citenamefont {Heyl}, \citenamefont {Tagliacozzo},\ and\ \citenamefont
		{Zoller}}]{Hauke2016}%
	\BibitemOpen
	\bibfield  {author} {\bibinfo {author} {\bibfnamefont {P.}~\bibnamefont
			{Hauke}}, \bibinfo {author} {\bibfnamefont {M.}~\bibnamefont {Heyl}},
		\bibinfo {author} {\bibfnamefont {L.}~\bibnamefont {Tagliacozzo}}, \ and\
		\bibinfo {author} {\bibfnamefont {P.}~\bibnamefont {Zoller}},\ }\href
	{\doibase 10.1038/nphys3700} {\bibfield  {journal} {\bibinfo  {journal} {Nat.
				Phys.}\ }\textbf {\bibinfo {volume} {12}},\ \bibinfo {pages} {778} (\bibinfo
		{year} {2016})}\BibitemShut {NoStop}%
	\bibitem [{\citenamefont {Osborne}\ and\ \citenamefont
		{Verstraete}(2006)}]{PhysRevLett.96.220503}%
	\BibitemOpen
	\bibfield  {author} {\bibinfo {author} {\bibfnamefont {T.~J.}\ \bibnamefont
			{Osborne}}\ and\ \bibinfo {author} {\bibfnamefont {F.}~\bibnamefont
			{Verstraete}},\ }\href {\doibase 10.1103/PhysRevLett.96.220503} {\bibfield
		{journal} {\bibinfo  {journal} {Phys. Rev. Lett.}\ }\textbf {\bibinfo
			{volume} {96}},\ \bibinfo {pages} {220503} (\bibinfo {year}
		{2006})}\BibitemShut {NoStop}%
	\bibitem [{\citenamefont {Baskaran}\ \emph {et~al.}(2007)\citenamefont
		{Baskaran}, \citenamefont {Mandal},\ and\ \citenamefont
		{Shankar}}]{PhysRevLett.98.247201}%
	\BibitemOpen
	\bibfield  {author} {\bibinfo {author} {\bibfnamefont {G.}~\bibnamefont
			{Baskaran}}, \bibinfo {author} {\bibfnamefont {S.}~\bibnamefont {Mandal}}, \
		and\ \bibinfo {author} {\bibfnamefont {R.}~\bibnamefont {Shankar}},\ }\href
	{\doibase 10.1103/PhysRevLett.98.247201} {\bibfield  {journal} {\bibinfo
			{journal} {Phys. Rev. Lett.}\ }\textbf {\bibinfo {volume} {98}},\ \bibinfo
		{pages} {247201} (\bibinfo {year} {2007})}\BibitemShut {NoStop}%
	\bibitem [{\citenamefont {Paddison}(2020)}]{Paddison_2020}%
	\BibitemOpen
	\bibfield  {author} {\bibinfo {author} {\bibfnamefont {J.~A.~M.}\
			\bibnamefont {Paddison}},\ }\href {\doibase 10.1103/PhysRevLett.125.247202}
	{\bibfield  {journal} {\bibinfo  {journal} {Phys. Rev. Lett.}\ }\textbf
		{\bibinfo {volume} {125}},\ \bibinfo {pages} {247202} (\bibinfo {year}
		{2020})}\BibitemShut {NoStop}%
	\bibitem [{\citenamefont {Huberman}\ \emph {et~al.}(2008)\citenamefont
		{Huberman}, \citenamefont {Tennant}, \citenamefont {Cowley}, \citenamefont
		{Coldea},\ and\ \citenamefont {Frost}}]{Huberman_2008}%
	\BibitemOpen
	\bibfield  {author} {\bibinfo {author} {\bibfnamefont {T.}~\bibnamefont
			{Huberman}}, \bibinfo {author} {\bibfnamefont {D.~A.}\ \bibnamefont
			{Tennant}}, \bibinfo {author} {\bibfnamefont {R.~A.}\ \bibnamefont {Cowley}},
		\bibinfo {author} {\bibfnamefont {R.}~\bibnamefont {Coldea}}, \ and\ \bibinfo
		{author} {\bibfnamefont {C.~D.}\ \bibnamefont {Frost}},\ }\href {\doibase
		10.1088/1742-5468/2008/05/p05017} {\bibfield  {journal} {\bibinfo  {journal}
			{Journal of Statistical Mechanics: Theory and Experiment}\ }\textbf {\bibinfo
			{volume} {2008}},\ \bibinfo {pages} {P05017} (\bibinfo {year}
		{2008})}\BibitemShut {NoStop}%
	\bibitem [{\citenamefont {Arovas}\ and\ \citenamefont
		{Auerbach}(1988)}]{Arovas1988}%
	\BibitemOpen
	\bibfield  {author} {\bibinfo {author} {\bibfnamefont {D.~P.}\ \bibnamefont
			{Arovas}}\ and\ \bibinfo {author} {\bibfnamefont {A.}~\bibnamefont
			{Auerbach}},\ }\href {\doibase 10.1103/PhysRevB.38.316} {\bibfield  {journal}
		{\bibinfo  {journal} {Phys. Rev. B}\ }\textbf {\bibinfo {volume} {38}},\
		\bibinfo {pages} {316} (\bibinfo {year} {1988})}\BibitemShut {NoStop}%
	\bibitem [{\citenamefont {Auerbach}(1994)}]{Auerbach1994}%
	\BibitemOpen
	\bibfield  {author} {\bibinfo {author} {\bibfnamefont {A.}~\bibnamefont
			{Auerbach}},\ }\href@noop {} {\emph {\bibinfo {title} {Interacting electrons
				and quantum magnetism}}}\ (\bibinfo  {publisher} {Springer-Verlag},\ \bibinfo
	{address} {New York},\ \bibinfo {year} {1994})\BibitemShut {NoStop}%
	\bibitem [{\citenamefont {Ghioldi}\ \emph {et~al.}(2018)\citenamefont
		{Ghioldi}, \citenamefont {Gonzalez}, \citenamefont {Zhang}, \citenamefont
		{Kamiya}, \citenamefont {Manuel}, \citenamefont {Trumper},\ and\
		\citenamefont {Batista}}]{Ghioldi_2018}%
	\BibitemOpen
	\bibfield  {author} {\bibinfo {author} {\bibfnamefont {E.~A.}\ \bibnamefont
			{Ghioldi}}, \bibinfo {author} {\bibfnamefont {M.~G.}\ \bibnamefont
			{Gonzalez}}, \bibinfo {author} {\bibfnamefont {S.-S.}\ \bibnamefont {Zhang}},
		\bibinfo {author} {\bibfnamefont {Y.}~\bibnamefont {Kamiya}}, \bibinfo
		{author} {\bibfnamefont {L.~O.}\ \bibnamefont {Manuel}}, \bibinfo {author}
		{\bibfnamefont {A.~E.}\ \bibnamefont {Trumper}}, \ and\ \bibinfo {author}
		{\bibfnamefont {C.~D.}\ \bibnamefont {Batista}},\ }\href {\doibase
		10.1103/PhysRevB.98.184403} {\bibfield  {journal} {\bibinfo  {journal} {Phys.
				Rev. B}\ }\textbf {\bibinfo {volume} {98}},\ \bibinfo {pages} {184403}
		(\bibinfo {year} {2018})}\BibitemShut {NoStop}%
	\bibitem [{\citenamefont {Ferrari}\ and\ \citenamefont
		{Becca}(2019)}]{Ferrari_2019}%
	\BibitemOpen
	\bibfield  {author} {\bibinfo {author} {\bibfnamefont {F.}~\bibnamefont
			{Ferrari}}\ and\ \bibinfo {author} {\bibfnamefont {F.}~\bibnamefont
			{Becca}},\ }\href {\doibase 10.1103/PhysRevX.9.031026} {\bibfield  {journal}
		{\bibinfo  {journal} {Phys. Rev. X}\ }\textbf {\bibinfo {volume} {9}},\
		\bibinfo {pages} {031026} (\bibinfo {year} {2019})}\BibitemShut {NoStop}%
	\bibitem [{\citenamefont {Lake}\ \emph {et~al.}(2005)\citenamefont {Lake},
		\citenamefont {Tennant}, \citenamefont {Frost},\ and\ \citenamefont
		{Nagler}}]{Lake2005}%
	\BibitemOpen
	\bibfield  {author} {\bibinfo {author} {\bibfnamefont {B.}~\bibnamefont
			{Lake}}, \bibinfo {author} {\bibfnamefont {D.~A.}\ \bibnamefont {Tennant}},
		\bibinfo {author} {\bibfnamefont {C.~D.}\ \bibnamefont {Frost}}, \ and\
		\bibinfo {author} {\bibfnamefont {S.~E.}\ \bibnamefont {Nagler}},\ }\href
	{\doibase 10.1038/nmat1327} {\bibfield  {journal} {\bibinfo  {journal} {Nat.
				Mater.}\ }\textbf {\bibinfo {volume} {4}},\ \bibinfo {pages} {329} (\bibinfo
		{year} {2005})}\BibitemShut {NoStop}%
	\bibitem [{\citenamefont {Schr{\"o}der}\ \emph {et~al.}(2000)\citenamefont
		{Schr{\"o}der}, \citenamefont {Aeppli}, \citenamefont {Coldea}, \citenamefont
		{Adams}, \citenamefont {Stockert}, \citenamefont {L{\"o}hneysen},
		\citenamefont {Bucher}, \citenamefont {Ramazashvili},\ and\ \citenamefont
		{Coleman}}]{schroder2000onset}%
	\BibitemOpen
	\bibfield  {author} {\bibinfo {author} {\bibfnamefont {A.}~\bibnamefont
			{Schr{\"o}der}}, \bibinfo {author} {\bibfnamefont {G.}~\bibnamefont
			{Aeppli}}, \bibinfo {author} {\bibfnamefont {R.}~\bibnamefont {Coldea}},
		\bibinfo {author} {\bibfnamefont {M.}~\bibnamefont {Adams}}, \bibinfo
		{author} {\bibfnamefont {O.}~\bibnamefont {Stockert}}, \bibinfo {author}
		{\bibfnamefont {H.}~\bibnamefont {L{\"o}hneysen}}, \bibinfo {author}
		{\bibfnamefont {E.}~\bibnamefont {Bucher}}, \bibinfo {author} {\bibfnamefont
			{R.}~\bibnamefont {Ramazashvili}}, \ and\ \bibinfo {author} {\bibfnamefont
			{P.}~\bibnamefont {Coleman}},\ }\href {\doibase 10.1038/35030039} {\bibfield
		{journal} {\bibinfo  {journal} {Nature}\ }\textbf {\bibinfo {volume} {407}},\
		\bibinfo {pages} {351} (\bibinfo {year} {2000})}\BibitemShut {NoStop}%
	\bibitem [{\citenamefont {Chakravarty}\ \emph {et~al.}(1989)\citenamefont
		{Chakravarty}, \citenamefont {Halperin},\ and\ \citenamefont
		{Nelson}}]{Chakravarty_1989}%
	\BibitemOpen
	\bibfield  {author} {\bibinfo {author} {\bibfnamefont {S.}~\bibnamefont
			{Chakravarty}}, \bibinfo {author} {\bibfnamefont {B.~I.}\ \bibnamefont
			{Halperin}}, \ and\ \bibinfo {author} {\bibfnamefont {D.~R.}\ \bibnamefont
			{Nelson}},\ }\href {\doibase 10.1103/PhysRevB.39.2344} {\bibfield  {journal}
		{\bibinfo  {journal} {Phys. Rev. B}\ }\textbf {\bibinfo {volume} {39}},\
		\bibinfo {pages} {2344} (\bibinfo {year} {1989})}\BibitemShut {NoStop}%
	\bibitem [{\citenamefont {Sachdev}\ and\ \citenamefont
		{Ye}(1992)}]{Sachdev_1992}%
	\BibitemOpen
	\bibfield  {author} {\bibinfo {author} {\bibfnamefont {S.}~\bibnamefont
			{Sachdev}}\ and\ \bibinfo {author} {\bibfnamefont {J.}~\bibnamefont {Ye}},\
	}\href {\doibase 10.1103/PhysRevLett.69.2411} {\bibfield  {journal} {\bibinfo
			{journal} {Phys. Rev. Lett.}\ }\textbf {\bibinfo {volume} {69}},\ \bibinfo
		{pages} {2411} (\bibinfo {year} {1992})}\BibitemShut {NoStop}%
	\bibitem [{\citenamefont {Sachdev}(1992)}]{Sachdev92}%
	\BibitemOpen
	\bibfield  {author} {\bibinfo {author} {\bibfnamefont {S.}~\bibnamefont
			{Sachdev}},\ }\href {\doibase 10.1103/PhysRevB.45.12377} {\bibfield
		{journal} {\bibinfo  {journal} {Phys. Rev. B}\ }\textbf {\bibinfo {volume}
			{45}},\ \bibinfo {pages} {12377} (\bibinfo {year} {1992})}\BibitemShut
	{NoStop}%
	\bibitem [{\citenamefont {Wang}\ and\ \citenamefont
		{Vishwanath}(2006)}]{Wang06}%
	\BibitemOpen
	\bibfield  {author} {\bibinfo {author} {\bibfnamefont {F.}~\bibnamefont
			{Wang}}\ and\ \bibinfo {author} {\bibfnamefont {A.}~\bibnamefont
			{Vishwanath}},\ }\href {\doibase 10.1103/PhysRevB.74.174423} {\bibfield
		{journal} {\bibinfo  {journal} {Phys. Rev. B}\ }\textbf {\bibinfo {volume}
			{74}},\ \bibinfo {pages} {174423} (\bibinfo {year} {2006})}\BibitemShut
	{NoStop}%
	\bibitem [{\citenamefont {Azaria}\ \emph {et~al.}(1990)\citenamefont {Azaria},
		\citenamefont {Delamotte},\ and\ \citenamefont {Jolicoeur}}]{Azaria90}%
	\BibitemOpen
	\bibfield  {author} {\bibinfo {author} {\bibfnamefont {P.}~\bibnamefont
			{Azaria}}, \bibinfo {author} {\bibfnamefont {B.}~\bibnamefont {Delamotte}}, \
		and\ \bibinfo {author} {\bibfnamefont {T.}~\bibnamefont {Jolicoeur}},\ }\href
	{\doibase 10.1103/PhysRevLett.64.3175} {\bibfield  {journal} {\bibinfo
			{journal} {Phys. Rev. Lett.}\ }\textbf {\bibinfo {volume} {64}},\ \bibinfo
		{pages} {3175} (\bibinfo {year} {1990})}\BibitemShut {NoStop}%
	\bibitem [{\citenamefont {Chubukov}\ \emph {et~al.}(1994)\citenamefont
		{Chubukov}, \citenamefont {Sachdev},\ and\ \citenamefont
		{Senthil}}]{Chubukov94}%
	\BibitemOpen
	\bibfield  {author} {\bibinfo {author} {\bibfnamefont {A.~V.}\ \bibnamefont
			{Chubukov}}, \bibinfo {author} {\bibfnamefont {S.}~\bibnamefont {Sachdev}}, \
		and\ \bibinfo {author} {\bibfnamefont {T.}~\bibnamefont {Senthil}},\ }\href
	{\doibase https://doi.org/10.1016/0550-3213(94)90023-X} {\bibfield  {journal}
		{\bibinfo  {journal} {Nuclear Physics B}\ }\textbf {\bibinfo {volume}
			{426}},\ \bibinfo {pages} {601} (\bibinfo {year} {1994})}\BibitemShut
	{NoStop}%
	\bibitem [{\citenamefont {Jia}\ \emph {et~al.}(2020)\citenamefont {Jia},
		\citenamefont {Gong}, \citenamefont {Liu}, \citenamefont {Zhao},
		\citenamefont {Dong}, \citenamefont {Dai}, \citenamefont {Li}, \citenamefont
		{Lei}, \citenamefont {Yu}, \citenamefont {Zhang},\ and\ \citenamefont
		{Jin}}]{Jia_2020}%
	\BibitemOpen
	\bibfield  {author} {\bibinfo {author} {\bibfnamefont {Y.-T.}\ \bibnamefont
			{Jia}}, \bibinfo {author} {\bibfnamefont {C.-S.}\ \bibnamefont {Gong}},
		\bibinfo {author} {\bibfnamefont {Y.-X.}\ \bibnamefont {Liu}}, \bibinfo
		{author} {\bibfnamefont {J.-F.}\ \bibnamefont {Zhao}}, \bibinfo {author}
		{\bibfnamefont {C.}~\bibnamefont {Dong}}, \bibinfo {author} {\bibfnamefont
			{G.-Y.}\ \bibnamefont {Dai}}, \bibinfo {author} {\bibfnamefont {X.-D.}\
			\bibnamefont {Li}}, \bibinfo {author} {\bibfnamefont {H.-C.}\ \bibnamefont
			{Lei}}, \bibinfo {author} {\bibfnamefont {R.-Z.}\ \bibnamefont {Yu}},
		\bibinfo {author} {\bibfnamefont {G.-M.}\ \bibnamefont {Zhang}}, \ and\
		\bibinfo {author} {\bibfnamefont {C.-Q.}\ \bibnamefont {Jin}},\ }\href
	{\doibase 10.1088/0256-307x/37/9/097404} {\bibfield  {journal} {\bibinfo
			{journal} {Chinese Physics Letters}\ }\textbf {\bibinfo {volume} {37}},\
		\bibinfo {pages} {097404} (\bibinfo {year} {2020})}\BibitemShut {NoStop}%
	\bibitem [{\citenamefont {Zhang}\ \emph {et~al.}(2020)\citenamefont {Zhang},
		\citenamefont {Yin}, \citenamefont {Ma}, \citenamefont {Liu}, \citenamefont
		{Li}, \citenamefont {Jin}, \citenamefont {Ji}, \citenamefont {Wang},
		\citenamefont {Wang}, \citenamefont {Yu} \emph {et~al.}}]{zhang2020pressure}%
	\BibitemOpen
	\bibfield  {author} {\bibinfo {author} {\bibfnamefont {Z.}~\bibnamefont
			{Zhang}}, \bibinfo {author} {\bibfnamefont {Y.}~\bibnamefont {Yin}}, \bibinfo
		{author} {\bibfnamefont {X.}~\bibnamefont {Ma}}, \bibinfo {author}
		{\bibfnamefont {W.}~\bibnamefont {Liu}}, \bibinfo {author} {\bibfnamefont
			{J.}~\bibnamefont {Li}}, \bibinfo {author} {\bibfnamefont {F.}~\bibnamefont
			{Jin}}, \bibinfo {author} {\bibfnamefont {J.}~\bibnamefont {Ji}}, \bibinfo
		{author} {\bibfnamefont {Y.}~\bibnamefont {Wang}}, \bibinfo {author}
		{\bibfnamefont {X.}~\bibnamefont {Wang}}, \bibinfo {author} {\bibfnamefont
			{X.}~\bibnamefont {Yu}},  \emph {et~al.},\ }\href
	{https://arxiv.org/abs/2003.11479} {\bibfield  {journal} {\bibinfo  {journal}
			{arXiv preprint arXiv:2003.11479}\ } (\bibinfo {year} {2020})}\BibitemShut
	{NoStop}%
	\bibitem [{\citenamefont {Brown}(1998)}]{BrownFF}%
	\BibitemOpen
	\bibfield  {author} {\bibinfo {author} {\bibfnamefont {P.~J.}\ \bibnamefont
			{Brown}},\ }\href {https://www.ill.eu/sites/ccsl/ffacts/} {\enquote {\bibinfo
			{title} {Magnetic form factors},}\ }\bibinfo {howpublished} {The Cambridge
		Crystallographic Subroutine Library} (\bibinfo {year} {1998})\BibitemShut
	{NoStop}%
	\bibitem [{\citenamefont {James}\ and\ \citenamefont
		{Roos}(1975)}]{James_1975}%
	\BibitemOpen
	\bibfield  {author} {\bibinfo {author} {\bibfnamefont {F.}~\bibnamefont
			{James}}\ and\ \bibinfo {author} {\bibfnamefont {M.}~\bibnamefont {Roos}},\
	}\href {\doibase https://doi.org/10.1016/0010-4655(75)90039-9} {\bibfield
		{journal} {\bibinfo  {journal} {Comp. Phys. Commun.}\ }\textbf {\bibinfo
			{volume} {10}},\ \bibinfo {pages} {343 } (\bibinfo {year}
		{1975})}\BibitemShut {NoStop}%
	\bibitem [{\citenamefont {Zhang}\ \emph
		{et~al.}(2019{\natexlab{b}})\citenamefont {Zhang}, \citenamefont {Ghioldi},
		\citenamefont {Kamiya}, \citenamefont {Manuel}, \citenamefont {Trumper},\
		and\ \citenamefont {Batista}}]{Zhang19}%
	\BibitemOpen
	\bibfield  {author} {\bibinfo {author} {\bibfnamefont {S.-S.}\ \bibnamefont
			{Zhang}}, \bibinfo {author} {\bibfnamefont {E.~A.}\ \bibnamefont {Ghioldi}},
		\bibinfo {author} {\bibfnamefont {Y.}~\bibnamefont {Kamiya}}, \bibinfo
		{author} {\bibfnamefont {L.~O.}\ \bibnamefont {Manuel}}, \bibinfo {author}
		{\bibfnamefont {A.~E.}\ \bibnamefont {Trumper}}, \ and\ \bibinfo {author}
		{\bibfnamefont {C.~D.}\ \bibnamefont {Batista}},\ }\href {\doibase
		10.1103/PhysRevB.100.104431} {\bibfield  {journal} {\bibinfo  {journal}
			{Phys. Rev. B}\ }\textbf {\bibinfo {volume} {100}},\ \bibinfo {pages}
		{104431} (\bibinfo {year} {2019}{\natexlab{b}})}\BibitemShut {NoStop}%
	\bibitem [{\citenamefont {Szasz}\ \emph {et~al.}(2020)\citenamefont {Szasz},
		\citenamefont {Motruk}, \citenamefont {Zaletel},\ and\ \citenamefont
		{Moore}}]{Szasz20}%
	\BibitemOpen
	\bibfield  {author} {\bibinfo {author} {\bibfnamefont {A.}~\bibnamefont
			{Szasz}}, \bibinfo {author} {\bibfnamefont {J.}~\bibnamefont {Motruk}},
		\bibinfo {author} {\bibfnamefont {M.~P.}\ \bibnamefont {Zaletel}}, \ and\
		\bibinfo {author} {\bibfnamefont {J.~E.}\ \bibnamefont {Moore}},\ }\href
	{\doibase 10.1103/PhysRevX.10.021042} {\bibfield  {journal} {\bibinfo
			{journal} {Phys. Rev. X}\ }\textbf {\bibinfo {volume} {10}},\ \bibinfo
		{pages} {021042} (\bibinfo {year} {2020})}\BibitemShut {NoStop}%
	\bibitem [{\citenamefont {Schollwock}(2011)}]{schollwock2011}%
	\BibitemOpen
	\bibfield  {author} {\bibinfo {author} {\bibfnamefont {U.}~\bibnamefont
			{Schollwock}},\ }\href {\doibase https://doi.org/10.1016/j.aop.2010.09.012}
	{\bibfield  {journal} {\bibinfo  {journal} {Annals of Physics}\ }\textbf
		{\bibinfo {volume} {326}},\ \bibinfo {pages} {96 } (\bibinfo {year}
		{2011})},\ \bibinfo {note} {january 2011 Special Issue}\BibitemShut {NoStop}%
	\bibitem [{\citenamefont {Vanderstraeten}\ \emph {et~al.}(2019)\citenamefont
		{Vanderstraeten}, \citenamefont {Haegeman},\ and\ \citenamefont
		{Verstraete}}]{Vanderstraeten2019}%
	\BibitemOpen
	\bibfield  {author} {\bibinfo {author} {\bibfnamefont {L.}~\bibnamefont
			{Vanderstraeten}}, \bibinfo {author} {\bibfnamefont {J.}~\bibnamefont
			{Haegeman}}, \ and\ \bibinfo {author} {\bibfnamefont {F.}~\bibnamefont
			{Verstraete}},\ }\href {\doibase 10.21468/SciPostPhysLectNotes.7} {\bibfield
		{journal} {\bibinfo  {journal} {SciPost Phys. Lect. Notes}\ ,\ \bibinfo
			{pages} {7}} (\bibinfo {year} {2019})}\BibitemShut {NoStop}%
	\bibitem [{\citenamefont {Fishman}\ \emph {et~al.}(2020)\citenamefont
		{Fishman}, \citenamefont {White},\ and\ \citenamefont
		{Stoudenmire}}]{itensor}%
	\BibitemOpen
	\bibfield  {author} {\bibinfo {author} {\bibfnamefont {M.}~\bibnamefont
			{Fishman}}, \bibinfo {author} {\bibfnamefont {S.~R.}\ \bibnamefont {White}},
		\ and\ \bibinfo {author} {\bibfnamefont {E.~M.}\ \bibnamefont
			{Stoudenmire}},\ }\href {https://arxiv.org/abs/2007.14822v1} {\enquote
		{\bibinfo {title} {The \mbox{ITensor} software library for tensor network
				calculations},}\ } (\bibinfo {year} {2020}),\ \Eprint
	{http://arxiv.org/abs/2007.14822} {arXiv:2007.14822} \BibitemShut {NoStop}%
	\bibitem [{\citenamefont {Hohenberg}(1964)}]{Hohenberg1964}%
	\BibitemOpen
	\bibfield  {author} {\bibinfo {author} {\bibfnamefont {P.}~\bibnamefont
			{Hohenberg}},\ }\href {\doibase 10.1103/PhysRev.136.B864} {\bibfield
		{journal} {\bibinfo  {journal} {Physical Review}\ }\textbf {\bibinfo {volume}
			{136}},\ \bibinfo {pages} {B864} (\bibinfo {year} {1964})}\BibitemShut
	{NoStop}%
	\bibitem [{\citenamefont {Kohn}\ and\ \citenamefont {Sham}(1965)}]{Kohn1965}%
	\BibitemOpen
	\bibfield  {author} {\bibinfo {author} {\bibfnamefont {W.}~\bibnamefont
			{Kohn}}\ and\ \bibinfo {author} {\bibfnamefont {L.~J.}\ \bibnamefont
			{Sham}},\ }\href {\doibase 10.1103/PhysRev.140.A1133} {\bibfield  {journal}
		{\bibinfo  {journal} {Physical Review}\ }\textbf {\bibinfo {volume} {140}},\
		\bibinfo {pages} {A1133} (\bibinfo {year} {1965})}\BibitemShut {NoStop}%
	\bibitem [{\citenamefont {Cohen}\ \emph {et~al.}(2008)\citenamefont {Cohen},
		\citenamefont {Mori-S{\'{a}}nchez},\ and\ \citenamefont {Yang}}]{Cohen2008}%
	\BibitemOpen
	\bibfield  {author} {\bibinfo {author} {\bibfnamefont {A.~J.}\ \bibnamefont
			{Cohen}}, \bibinfo {author} {\bibfnamefont {P.}~\bibnamefont
			{Mori-S{\'{a}}nchez}}, \ and\ \bibinfo {author} {\bibfnamefont
			{W.}~\bibnamefont {Yang}},\ }\href {\doibase 10.1126/science.1158722}
	{\enquote {\bibinfo {title} {{Insights into current limitations of density
					functional theory}},}\ } (\bibinfo {year} {2008})\BibitemShut {NoStop}%
	\bibitem [{\citenamefont {Duan}\ \emph {et~al.}(2018)\citenamefont {Duan},
		\citenamefont {Wu}, \citenamefont {Chen}, \citenamefont {Zhang},
		\citenamefont {Liu}, \citenamefont {Yuan},\ and\ \citenamefont
		{Cao}}]{Duan2018}%
	\BibitemOpen
	\bibfield  {author} {\bibinfo {author} {\bibfnamefont {X.}~\bibnamefont
			{Duan}}, \bibinfo {author} {\bibfnamefont {F.}~\bibnamefont {Wu}}, \bibinfo
		{author} {\bibfnamefont {J.}~\bibnamefont {Chen}}, \bibinfo {author}
		{\bibfnamefont {P.}~\bibnamefont {Zhang}}, \bibinfo {author} {\bibfnamefont
			{Y.}~\bibnamefont {Liu}}, \bibinfo {author} {\bibfnamefont {H.}~\bibnamefont
			{Yuan}}, \ and\ \bibinfo {author} {\bibfnamefont {C.}~\bibnamefont {Cao}},\
	}\href {\doibase 10.1038/s42005-018-0074-8} {\bibfield  {journal} {\bibinfo
			{journal} {Communications Physics 2018 1:1}\ }\textbf {\bibinfo {volume}
			{1}},\ \bibinfo {pages} {1} (\bibinfo {year} {2018})}\BibitemShut {NoStop}%
	\bibitem [{\citenamefont {Seidl}\ \emph {et~al.}(1996)\citenamefont {Seidl},
		\citenamefont {G{\"{o}}rling}, \citenamefont {Vogl}, \citenamefont
		{Majewski},\ and\ \citenamefont {Levy}}]{Seidl1996}%
	\BibitemOpen
	\bibfield  {author} {\bibinfo {author} {\bibfnamefont {a.}~\bibnamefont
			{Seidl}}, \bibinfo {author} {\bibfnamefont {a.}~\bibnamefont
			{G{\"{o}}rling}}, \bibinfo {author} {\bibfnamefont {P.}~\bibnamefont {Vogl}},
		\bibinfo {author} {\bibfnamefont {J.}~\bibnamefont {Majewski}}, \ and\
		\bibinfo {author} {\bibfnamefont {M.}~\bibnamefont {Levy}},\ }\href
	{http://www.ncbi.nlm.nih.gov/pubmed/9983927} {\bibfield  {journal} {\bibinfo
			{journal} {Physical review. B, Condensed matter}\ }\textbf {\bibinfo {volume}
			{53}},\ \bibinfo {pages} {3764} (\bibinfo {year} {1996})}\BibitemShut
	{NoStop}%
	\bibitem [{\citenamefont {Perdew}\ \emph {et~al.}(2017)\citenamefont {Perdew},
		\citenamefont {Yang}, \citenamefont {Burke}, \citenamefont {Yang},
		\citenamefont {Gross}, \citenamefont {Scheffler}, \citenamefont {Scuseria},
		\citenamefont {Henderson}, \citenamefont {Zhang}, \citenamefont {Ruzsinszky},
		\citenamefont {Peng}, \citenamefont {Sun}, \citenamefont {Trushin},\ and\
		\citenamefont {G{\"{o}}rling}}]{Perdew2017}%
	\BibitemOpen
	\bibfield  {author} {\bibinfo {author} {\bibfnamefont {J.~P.}\ \bibnamefont
			{Perdew}}, \bibinfo {author} {\bibfnamefont {W.}~\bibnamefont {Yang}},
		\bibinfo {author} {\bibfnamefont {K.}~\bibnamefont {Burke}}, \bibinfo
		{author} {\bibfnamefont {Z.}~\bibnamefont {Yang}}, \bibinfo {author}
		{\bibfnamefont {E.~K.}\ \bibnamefont {Gross}}, \bibinfo {author}
		{\bibfnamefont {M.}~\bibnamefont {Scheffler}}, \bibinfo {author}
		{\bibfnamefont {G.~E.}\ \bibnamefont {Scuseria}}, \bibinfo {author}
		{\bibfnamefont {T.~M.}\ \bibnamefont {Henderson}}, \bibinfo {author}
		{\bibfnamefont {I.~Y.}\ \bibnamefont {Zhang}}, \bibinfo {author}
		{\bibfnamefont {A.}~\bibnamefont {Ruzsinszky}}, \bibinfo {author}
		{\bibfnamefont {H.}~\bibnamefont {Peng}}, \bibinfo {author} {\bibfnamefont
			{J.}~\bibnamefont {Sun}}, \bibinfo {author} {\bibfnamefont {E.}~\bibnamefont
			{Trushin}}, \ and\ \bibinfo {author} {\bibfnamefont {A.}~\bibnamefont
			{G{\"{o}}rling}},\ }\href {\doibase 10.1073/pnas.1621352114} {\bibfield
		{journal} {\bibinfo  {journal} {Proceedings of the National Academy of
				Sciences of the United States of America}\ }\textbf {\bibinfo {volume}
			{114}},\ \bibinfo {pages} {2801} (\bibinfo {year} {2017})}\BibitemShut
	{NoStop}%
	\bibitem [{\citenamefont {Perdew}(1981)}]{Perdew1981}%
	\BibitemOpen
	\bibfield  {author} {\bibinfo {author} {\bibfnamefont {J.~P.}\ \bibnamefont
			{Perdew}},\ }\href {\doibase 10.1103/PhysRevB.23.5048} {\bibfield  {journal}
		{\bibinfo  {journal} {Physical Review B}\ }\textbf {\bibinfo {volume} {23}},\
		\bibinfo {pages} {5048} (\bibinfo {year} {1981})}\BibitemShut {NoStop}%
	\bibitem [{\citenamefont {Sun}\ \emph {et~al.}(2015)\citenamefont {Sun},
		\citenamefont {Ruzsinszky},\ and\ \citenamefont {Perdew}}]{Sun2015a}%
	\BibitemOpen
	\bibfield  {author} {\bibinfo {author} {\bibfnamefont {J.}~\bibnamefont
			{Sun}}, \bibinfo {author} {\bibfnamefont {A.}~\bibnamefont {Ruzsinszky}}, \
		and\ \bibinfo {author} {\bibfnamefont {J.}~\bibnamefont {Perdew}},\ }\href
	{\doibase 10.1103/PhysRevLett.115.036402} {\bibfield  {journal} {\bibinfo
			{journal} {Physical Review Letters}\ }\textbf {\bibinfo {volume} {115}},\
		\bibinfo {pages} {036402} (\bibinfo {year} {2015})}\BibitemShut {NoStop}%
	\bibitem [{\citenamefont {Anisimov}\ \emph {et~al.}(1997)\citenamefont
		{Anisimov}, \citenamefont {Aryasetiawan},\ and\ \citenamefont
		{Lichtenstein}}]{Anisimov1997}%
	\BibitemOpen
	\bibfield  {author} {\bibinfo {author} {\bibfnamefont {V.~I.}\ \bibnamefont
			{Anisimov}}, \bibinfo {author} {\bibfnamefont {F.}~\bibnamefont
			{Aryasetiawan}}, \ and\ \bibinfo {author} {\bibfnamefont {A.~I.}\
			\bibnamefont {Lichtenstein}},\ }\href {\doibase 10.1088/0953-8984/9/4/002}
	{\bibfield  {journal} {\bibinfo  {journal} {Journal of Physics: Condensed
				Matter}\ }\textbf {\bibinfo {volume} {9}},\ \bibinfo {pages} {767} (\bibinfo
		{year} {1997})}\BibitemShut {NoStop}%
	\bibitem [{\citenamefont {Heyd}\ \emph {et~al.}(2003)\citenamefont {Heyd},
		\citenamefont {Scuseria},\ and\ \citenamefont {Ernzerhof}}]{Heyd2003}%
	\BibitemOpen
	\bibfield  {author} {\bibinfo {author} {\bibfnamefont {J.}~\bibnamefont
			{Heyd}}, \bibinfo {author} {\bibfnamefont {G.~E.}\ \bibnamefont {Scuseria}},
		\ and\ \bibinfo {author} {\bibfnamefont {M.}~\bibnamefont {Ernzerhof}},\
	}\href {\doibase 10.1063/1.1564060} {\bibfield  {journal} {\bibinfo
			{journal} {Journal of Chemical Physics}\ }\textbf {\bibinfo {volume} {118}},\
		\bibinfo {pages} {8207} (\bibinfo {year} {2003})}\BibitemShut {NoStop}%
	\bibitem [{\citenamefont {Deilynazar}\ \emph {et~al.}(2015)\citenamefont
		{Deilynazar}, \citenamefont {Khorasani}, \citenamefont {Alaei},\ and\
		\citenamefont {{Javad Hashemifar}}}]{Deilynazar2015}%
	\BibitemOpen
	\bibfield  {author} {\bibinfo {author} {\bibfnamefont {N.}~\bibnamefont
			{Deilynazar}}, \bibinfo {author} {\bibfnamefont {E.}~\bibnamefont
			{Khorasani}}, \bibinfo {author} {\bibfnamefont {M.}~\bibnamefont {Alaei}}, \
		and\ \bibinfo {author} {\bibfnamefont {S.}~\bibnamefont {{Javad
					Hashemifar}}},\ }\href {\doibase 10.1016/j.jmmm.2015.05.042} {\bibfield
		{journal} {\bibinfo  {journal} {Journal of Magnetism and Magnetic Materials}\
		}\textbf {\bibinfo {volume} {393}},\ \bibinfo {pages} {127} (\bibinfo {year}
		{2015})},\ \Eprint {http://arxiv.org/abs/1502.01814} {arXiv:1502.01814}
	\BibitemShut {NoStop}%
	\bibitem [{\citenamefont {Payne}\ \emph {et~al.}(2019)\citenamefont {Payne},
		\citenamefont {Aveda{\~{n}}o-Franco}, \citenamefont {He}, \citenamefont
		{Bousquet},\ and\ \citenamefont {Romero}}]{Payne2019}%
	\BibitemOpen
	\bibfield  {author} {\bibinfo {author} {\bibfnamefont {A.}~\bibnamefont
			{Payne}}, \bibinfo {author} {\bibfnamefont {G.}~\bibnamefont
			{Aveda{\~{n}}o-Franco}}, \bibinfo {author} {\bibfnamefont {X.}~\bibnamefont
			{He}}, \bibinfo {author} {\bibfnamefont {E.}~\bibnamefont {Bousquet}}, \ and\
		\bibinfo {author} {\bibfnamefont {A.~H.}\ \bibnamefont {Romero}},\ }\href
	{\doibase 10.1039/c9cp03618k} {\bibfield  {journal} {\bibinfo  {journal}
			{Physical Chemistry Chemical Physics}\ }\textbf {\bibinfo {volume} {21}},\
		\bibinfo {pages} {21932} (\bibinfo {year} {2019})}\BibitemShut {NoStop}%
	\bibitem [{\citenamefont {Casadei}\ \emph {et~al.}(2012)\citenamefont
		{Casadei}, \citenamefont {Ren}, \citenamefont {Rinke}, \citenamefont
		{Rubio},\ and\ \citenamefont {Scheffler}}]{Casadei2012}%
	\BibitemOpen
	\bibfield  {author} {\bibinfo {author} {\bibfnamefont {M.}~\bibnamefont
			{Casadei}}, \bibinfo {author} {\bibfnamefont {X.}~\bibnamefont {Ren}},
		\bibinfo {author} {\bibfnamefont {P.}~\bibnamefont {Rinke}}, \bibinfo
		{author} {\bibfnamefont {A.}~\bibnamefont {Rubio}}, \ and\ \bibinfo {author}
		{\bibfnamefont {M.}~\bibnamefont {Scheffler}},\ }\href {\doibase
		10.1103/PhysRevLett.109.146402} {\bibfield  {journal} {\bibinfo  {journal}
			{Physical Review Letters}\ }\textbf {\bibinfo {volume} {109}},\ \bibinfo
		{pages} {1} (\bibinfo {year} {2012})}\BibitemShut {NoStop}%
	\bibitem [{\citenamefont {Payne}\ \emph {et~al.}(2018)\citenamefont {Payne},
		\citenamefont {Avenda{\~{n}}o-Franco}, \citenamefont {Bousquet},\ and\
		\citenamefont {Romero}}]{Payne2018}%
	\BibitemOpen
	\bibfield  {author} {\bibinfo {author} {\bibfnamefont {A.}~\bibnamefont
			{Payne}}, \bibinfo {author} {\bibfnamefont {G.}~\bibnamefont
			{Avenda{\~{n}}o-Franco}}, \bibinfo {author} {\bibfnamefont {E.}~\bibnamefont
			{Bousquet}}, \ and\ \bibinfo {author} {\bibfnamefont {A.~H.}\ \bibnamefont
			{Romero}},\ }\href {\doibase 10.1021/acs.jctc.8b00404} {\bibfield  {journal}
		{\bibinfo  {journal} {Journal of Chemical Theory and Computation}\ }\textbf
		{\bibinfo {volume} {14}},\ \bibinfo {pages} {4455} (\bibinfo {year}
		{2018})}\BibitemShut {NoStop}%
	\bibitem [{\citenamefont {Kresse}\ and\ \citenamefont
		{Hafner}(1993)}]{Kresse1993}%
	\BibitemOpen
	\bibfield  {author} {\bibinfo {author} {\bibfnamefont {G.}~\bibnamefont
			{Kresse}}\ and\ \bibinfo {author} {\bibfnamefont {J.}~\bibnamefont
			{Hafner}},\ }\href {\doibase 10.1103/PhysRevB.47.558} {\bibfield  {journal}
		{\bibinfo  {journal} {Physical Review B}\ }\textbf {\bibinfo {volume} {47}},\
		\bibinfo {pages} {558} (\bibinfo {year} {1993})}\BibitemShut {NoStop}%
	\bibitem [{\citenamefont {Kresse}\ and\ \citenamefont
		{Joubert}(1999)}]{Kresse1999}%
	\BibitemOpen
	\bibfield  {author} {\bibinfo {author} {\bibfnamefont {G.}~\bibnamefont
			{Kresse}}\ and\ \bibinfo {author} {\bibfnamefont {D.}~\bibnamefont
			{Joubert}},\ }\href {\doibase 10.1103/PhysRevB.59.1758} {\bibfield  {journal}
		{\bibinfo  {journal} {Physical Review B}\ }\textbf {\bibinfo {volume} {59}},\
		\bibinfo {pages} {1758} (\bibinfo {year} {1999})}\BibitemShut {NoStop}%
	\bibitem [{\citenamefont {Perdew}\ \emph {et~al.}(1996)\citenamefont {Perdew},
		\citenamefont {Burke},\ and\ \citenamefont {Ernzerhof}}]{perdew}%
	\BibitemOpen
	\bibfield  {author} {\bibinfo {author} {\bibfnamefont {J.~P.}\ \bibnamefont
			{Perdew}}, \bibinfo {author} {\bibfnamefont {K.}~\bibnamefont {Burke}}, \
		and\ \bibinfo {author} {\bibfnamefont {M.}~\bibnamefont {Ernzerhof}},\ }\href
	{\doibase 10.1103/PhysRevLett.77.3865} {\bibfield  {journal} {\bibinfo
			{journal} {Phys. Rev. Lett.}\ }\textbf {\bibinfo {volume} {77}},\ \bibinfo
		{pages} {3865} (\bibinfo {year} {1996})}\BibitemShut {NoStop}%
	\bibitem [{\citenamefont {Anisimov}\ \emph {et~al.}(1991)\citenamefont
		{Anisimov}, \citenamefont {Zaanen},\ and\ \citenamefont
		{Andersen}}]{Anisimov1991}%
	\BibitemOpen
	\bibfield  {author} {\bibinfo {author} {\bibfnamefont {V.~I.}\ \bibnamefont
			{Anisimov}}, \bibinfo {author} {\bibfnamefont {J.}~\bibnamefont {Zaanen}}, \
		and\ \bibinfo {author} {\bibfnamefont {O.~K.}\ \bibnamefont {Andersen}},\
	}\href {\doibase 10.1103/PhysRevB.44.943} {\bibfield  {journal} {\bibinfo
			{journal} {Phys. Rev. B}\ }\textbf {\bibinfo {volume} {44}},\ \bibinfo
		{pages} {943} (\bibinfo {year} {1991})}\BibitemShut {NoStop}%
	\bibitem [{\citenamefont {Blaha}\ \emph {et~al.}(2001)\citenamefont {Blaha},
		\citenamefont {Schwarz}, \citenamefont {Madsen}, \citenamefont {Kvasnicka},
		\citenamefont {Luitz} \emph {et~al.}}]{blaha2001}%
	\BibitemOpen
	\bibfield  {author} {\bibinfo {author} {\bibfnamefont {P.}~\bibnamefont
			{Blaha}}, \bibinfo {author} {\bibfnamefont {K.}~\bibnamefont {Schwarz}},
		\bibinfo {author} {\bibfnamefont {G.~K.}\ \bibnamefont {Madsen}}, \bibinfo
		{author} {\bibfnamefont {D.}~\bibnamefont {Kvasnicka}}, \bibinfo {author}
		{\bibfnamefont {J.}~\bibnamefont {Luitz}},  \emph {et~al.},\ }\href@noop {}
	{\bibfield  {journal} {\bibinfo  {journal} {An augmented plane wave+ local
				orbitals program for calculating crystal properties}\ }\textbf {\bibinfo
			{volume} {60}} (\bibinfo {year} {2001})}\BibitemShut {NoStop}%
	\bibitem [{\citenamefont {Pokharel}\ \emph {et~al.}(2018)\citenamefont
		{Pokharel}, \citenamefont {May}, \citenamefont {Parker}, \citenamefont
		{Calder}, \citenamefont {Ehlers}, \citenamefont {Huq}, \citenamefont
		{Kimber}, \citenamefont {Arachchige}, \citenamefont {Poudel}, \citenamefont
		{McGuire}, \citenamefont {Mandrus},\ and\ \citenamefont
		{Christianson}}]{pokharel}%
	\BibitemOpen
	\bibfield  {author} {\bibinfo {author} {\bibfnamefont {G.}~\bibnamefont
			{Pokharel}}, \bibinfo {author} {\bibfnamefont {A.~F.}\ \bibnamefont {May}},
		\bibinfo {author} {\bibfnamefont {D.~S.}\ \bibnamefont {Parker}}, \bibinfo
		{author} {\bibfnamefont {S.}~\bibnamefont {Calder}}, \bibinfo {author}
		{\bibfnamefont {G.}~\bibnamefont {Ehlers}}, \bibinfo {author} {\bibfnamefont
			{A.}~\bibnamefont {Huq}}, \bibinfo {author} {\bibfnamefont {S.~A.~J.}\
			\bibnamefont {Kimber}}, \bibinfo {author} {\bibfnamefont {H.~S.}\
			\bibnamefont {Arachchige}}, \bibinfo {author} {\bibfnamefont
			{L.}~\bibnamefont {Poudel}}, \bibinfo {author} {\bibfnamefont {M.~A.}\
			\bibnamefont {McGuire}}, \bibinfo {author} {\bibfnamefont {D.}~\bibnamefont
			{Mandrus}}, \ and\ \bibinfo {author} {\bibfnamefont {A.~D.}\ \bibnamefont
			{Christianson}},\ }\href {\doibase 10.1103/PhysRevB.97.134117} {\bibfield
		{journal} {\bibinfo  {journal} {Phys. Rev. B}\ }\textbf {\bibinfo {volume}
			{97}},\ \bibinfo {pages} {134117} (\bibinfo {year} {2018})}\BibitemShut
	{NoStop}%
	\bibitem [{\citenamefont {Pandey}\ and\ \citenamefont {Parker}(2018)}]{pandey}%
	\BibitemOpen
	\bibfield  {author} {\bibinfo {author} {\bibfnamefont {T.}~\bibnamefont
			{Pandey}}\ and\ \bibinfo {author} {\bibfnamefont {D.~S.}\ \bibnamefont
			{Parker}},\ }\href {\doibase 10.1103/PhysRevApplied.10.034038} {\bibfield
		{journal} {\bibinfo  {journal} {Phys. Rev. Applied}\ }\textbf {\bibinfo
			{volume} {10}},\ \bibinfo {pages} {034038} (\bibinfo {year}
		{2018})}\BibitemShut {NoStop}%
\end{thebibliography}

\begin{thebibliography}{28}%
	\makeatletter
	\providecommand \@ifxundefined [1]{%
		\@ifx{#1\undefined}
	}%
	\providecommand \@ifnum [1]{%
		\ifnum #1\expandafter \@firstoftwo
		\else \expandafter \@secondoftwo
		\fi
	}%
	\providecommand \@ifx [1]{%
		\ifx #1\expandafter \@firstoftwo
		\else \expandafter \@secondoftwo
		\fi
	}%
	\providecommand \natexlab [1]{#1}%
	\providecommand \enquote  [1]{``#1''}%
	\providecommand \bibnamefont  [1]{#1}%
	\providecommand \bibfnamefont [1]{#1}%
	\providecommand \citenamefont [1]{#1}%
	\providecommand \href@noop [0]{\@secondoftwo}%
	\providecommand \href [0]{\begingroup \@sanitize@url \@href}%
	\providecommand \@href[1]{\@@startlink{#1}\@@href}%
	\providecommand \@@href[1]{\endgroup#1\@@endlink}%
	\providecommand \@sanitize@url [0]{\catcode `\\12\catcode `\$12\catcode
		`\&12\catcode `\#12\catcode `\^12\catcode `\_12\catcode `\%12\relax}%
	\providecommand \@@startlink[1]{}%
	\providecommand \@@endlink[0]{}%
	\providecommand \url  [0]{\begingroup\@sanitize@url \@url }%
	\providecommand \@url [1]{\endgroup\@href {#1}{\urlprefix }}%
	\providecommand \urlprefix  [0]{URL }%
	\providecommand \Eprint [0]{\href }%
	\providecommand \doibase [0]{http://dx.doi.org/}%
	\providecommand \selectlanguage [0]{\@gobble}%
	\providecommand \bibinfo  [0]{\@secondoftwo}%
	\providecommand \bibfield  [0]{\@secondoftwo}%
	\providecommand \translation [1]{[#1]}%
	\providecommand \BibitemOpen [0]{}%
	\providecommand \bibitemStop [0]{}%
	\providecommand \bibitemNoStop [0]{.\EOS\space}%
	\providecommand \EOS [0]{\spacefactor3000\relax}%
	\providecommand \BibitemShut  [1]{\csname bibitem#1\endcsname}%
	\let\auto@bib@innerbib\@empty
	\bibitem [{\citenamefont {Bleaney}(1963)}]{Bleaney1963}%
	\BibitemOpen
	\bibfield  {author} {\bibinfo {author} {\bibfnamefont {B.}~\bibnamefont
			{Bleaney}},\ }\href {\doibase 10.1063/1.1729355} {\bibfield  {journal}
		{\bibinfo  {journal} {Journal of Applied Physics}\ }\textbf {\bibinfo
			{volume} {34}},\ \bibinfo {pages} {1024} (\bibinfo {year}
		{1963})}\BibitemShut {NoStop}%
	\bibitem [{\citenamefont {Bruker}(2015)}]{Apex3}%
	\BibitemOpen
	\bibfield  {author} {\bibinfo {author} {\bibnamefont {Bruker}},\ }\href@noop
	{} {\enquote {\bibinfo {title} {Apex3},}\ } (\bibinfo {year}
	{2015})\BibitemShut {NoStop}%
	\bibitem [{\citenamefont {Sheldrick}(2008)}]{sheldrick2008short}%
	\BibitemOpen
	\bibfield  {author} {\bibinfo {author} {\bibfnamefont {G.~M.}\ \bibnamefont
			{Sheldrick}},\ }\href {\doibase 10.1107/S0108767307043930} {\bibfield
		{journal} {\bibinfo  {journal} {Acta Crystallographica Section A: Foundations
				of Crystallography}\ }\textbf {\bibinfo {volume} {64}},\ \bibinfo {pages}
		{112} (\bibinfo {year} {2008})}\BibitemShut {NoStop}%
	\bibitem [{\citenamefont {Spek}(2009)}]{spek2009structure}%
	\BibitemOpen
	\bibfield  {author} {\bibinfo {author} {\bibfnamefont {A.~L.}\ \bibnamefont
			{Spek}},\ }\href {\doibase 10.1107/S090744490804362X} {\bibfield  {journal}
		{\bibinfo  {journal} {Acta Crystallographica Section D: Biological
				Crystallography}\ }\textbf {\bibinfo {volume} {65}},\ \bibinfo {pages} {148}
		(\bibinfo {year} {2009})}\BibitemShut {NoStop}%
	\bibitem [{\citenamefont {Dai}\ \emph {et~al.}(2021)\citenamefont {Dai},
		\citenamefont {Zhang}, \citenamefont {Xie}, \citenamefont {Duan},
		\citenamefont {Gao}, \citenamefont {Zhu}, \citenamefont {Feng}, \citenamefont
		{Tao}, \citenamefont {Huang}, \citenamefont {Cao}, \citenamefont
		{Podlesnyak}, \citenamefont {Granroth}, \citenamefont {Everett},
		\citenamefont {Neuefeind}, \citenamefont {Voneshen}, \citenamefont {Wang},
		\citenamefont {Tan}, \citenamefont {Morosan}, \citenamefont {Wang},
		\citenamefont {Lin}, \citenamefont {Shu}, \citenamefont {Chen}, \citenamefont
		{Guo}, \citenamefont {Lu},\ and\ \citenamefont {Dai}}]{Dai_2021}%
	\BibitemOpen
	\bibfield  {author} {\bibinfo {author} {\bibfnamefont {P.-L.}\ \bibnamefont
			{Dai}}, \bibinfo {author} {\bibfnamefont {G.}~\bibnamefont {Zhang}}, \bibinfo
		{author} {\bibfnamefont {Y.}~\bibnamefont {Xie}}, \bibinfo {author}
		{\bibfnamefont {C.}~\bibnamefont {Duan}}, \bibinfo {author} {\bibfnamefont
			{Y.}~\bibnamefont {Gao}}, \bibinfo {author} {\bibfnamefont {Z.}~\bibnamefont
			{Zhu}}, \bibinfo {author} {\bibfnamefont {E.}~\bibnamefont {Feng}}, \bibinfo
		{author} {\bibfnamefont {Z.}~\bibnamefont {Tao}}, \bibinfo {author}
		{\bibfnamefont {C.-L.}\ \bibnamefont {Huang}}, \bibinfo {author}
		{\bibfnamefont {H.}~\bibnamefont {Cao}}, \bibinfo {author} {\bibfnamefont
			{A.}~\bibnamefont {Podlesnyak}}, \bibinfo {author} {\bibfnamefont {G.~E.}\
			\bibnamefont {Granroth}}, \bibinfo {author} {\bibfnamefont {M.~S.}\
			\bibnamefont {Everett}}, \bibinfo {author} {\bibfnamefont {J.~C.}\
			\bibnamefont {Neuefeind}}, \bibinfo {author} {\bibfnamefont {D.}~\bibnamefont
			{Voneshen}}, \bibinfo {author} {\bibfnamefont {S.}~\bibnamefont {Wang}},
		\bibinfo {author} {\bibfnamefont {G.}~\bibnamefont {Tan}}, \bibinfo {author}
		{\bibfnamefont {E.}~\bibnamefont {Morosan}}, \bibinfo {author} {\bibfnamefont
			{X.}~\bibnamefont {Wang}}, \bibinfo {author} {\bibfnamefont {H.-Q.}\
			\bibnamefont {Lin}}, \bibinfo {author} {\bibfnamefont {L.}~\bibnamefont
			{Shu}}, \bibinfo {author} {\bibfnamefont {G.}~\bibnamefont {Chen}}, \bibinfo
		{author} {\bibfnamefont {Y.}~\bibnamefont {Guo}}, \bibinfo {author}
		{\bibfnamefont {X.}~\bibnamefont {Lu}}, \ and\ \bibinfo {author}
		{\bibfnamefont {P.}~\bibnamefont {Dai}},\ }\href {\doibase
		10.1103/PhysRevX.11.021044} {\bibfield  {journal} {\bibinfo  {journal} {Phys.
				Rev. X}\ }\textbf {\bibinfo {volume} {11}},\ \bibinfo {pages} {021044}
		(\bibinfo {year} {2021})}\BibitemShut {NoStop}%
	\bibitem [{\citenamefont {Capriotti}\ \emph {et~al.}(1999)\citenamefont
		{Capriotti}, \citenamefont {Trumper},\ and\ \citenamefont
		{Sorella}}]{Capriotti_1999}%
	\BibitemOpen
	\bibfield  {author} {\bibinfo {author} {\bibfnamefont {L.}~\bibnamefont
			{Capriotti}}, \bibinfo {author} {\bibfnamefont {A.~E.}\ \bibnamefont
			{Trumper}}, \ and\ \bibinfo {author} {\bibfnamefont {S.}~\bibnamefont
			{Sorella}},\ }\href {\doibase 10.1103/PhysRevLett.82.3899} {\bibfield
		{journal} {\bibinfo  {journal} {Phys. Rev. Lett.}\ }\textbf {\bibinfo
			{volume} {82}},\ \bibinfo {pages} {3899} (\bibinfo {year}
		{1999})}\BibitemShut {NoStop}%
	\bibitem [{\citenamefont {White}\ and\ \citenamefont
		{Chernyshev}(2007)}]{PhysRevLett.99.127004}%
	\BibitemOpen
	\bibfield  {author} {\bibinfo {author} {\bibfnamefont {S.~R.}\ \bibnamefont
			{White}}\ and\ \bibinfo {author} {\bibfnamefont {A.~L.}\ \bibnamefont
			{Chernyshev}},\ }\href {\doibase 10.1103/PhysRevLett.99.127004} {\bibfield
		{journal} {\bibinfo  {journal} {Phys. Rev. Lett.}\ }\textbf {\bibinfo
			{volume} {99}},\ \bibinfo {pages} {127004} (\bibinfo {year}
		{2007})}\BibitemShut {NoStop}%
	\bibitem [{\citenamefont {Zheng}\ \emph {et~al.}(2006)\citenamefont {Zheng},
		\citenamefont {Fj\ae{}restad}, \citenamefont {Singh}, \citenamefont
		{McKenzie},\ and\ \citenamefont {Coldea}}]{Zheng_2006}%
	\BibitemOpen
	\bibfield  {author} {\bibinfo {author} {\bibfnamefont {W.}~\bibnamefont
			{Zheng}}, \bibinfo {author} {\bibfnamefont {J.~O.}\ \bibnamefont
			{Fj\ae{}restad}}, \bibinfo {author} {\bibfnamefont {R.~R.~P.}\ \bibnamefont
			{Singh}}, \bibinfo {author} {\bibfnamefont {R.~H.}\ \bibnamefont {McKenzie}},
		\ and\ \bibinfo {author} {\bibfnamefont {R.}~\bibnamefont {Coldea}},\ }\href
	{\doibase 10.1103/PhysRevB.74.224420} {\bibfield  {journal} {\bibinfo
			{journal} {Phys. Rev. B}\ }\textbf {\bibinfo {volume} {74}},\ \bibinfo
		{pages} {224420} (\bibinfo {year} {2006})}\BibitemShut {NoStop}%
	\bibitem [{\citenamefont {Ghioldi}\ \emph {et~al.}(2018)\citenamefont
		{Ghioldi}, \citenamefont {Gonzalez}, \citenamefont {Zhang}, \citenamefont
		{Kamiya}, \citenamefont {Manuel}, \citenamefont {Trumper},\ and\
		\citenamefont {Batista}}]{Ghioldi_2018}%
	\BibitemOpen
	\bibfield  {author} {\bibinfo {author} {\bibfnamefont {E.~A.}\ \bibnamefont
			{Ghioldi}}, \bibinfo {author} {\bibfnamefont {M.~G.}\ \bibnamefont
			{Gonzalez}}, \bibinfo {author} {\bibfnamefont {S.-S.}\ \bibnamefont {Zhang}},
		\bibinfo {author} {\bibfnamefont {Y.}~\bibnamefont {Kamiya}}, \bibinfo
		{author} {\bibfnamefont {L.~O.}\ \bibnamefont {Manuel}}, \bibinfo {author}
		{\bibfnamefont {A.~E.}\ \bibnamefont {Trumper}}, \ and\ \bibinfo {author}
		{\bibfnamefont {C.~D.}\ \bibnamefont {Batista}},\ }\href {\doibase
		10.1103/PhysRevB.98.184403} {\bibfield  {journal} {\bibinfo  {journal} {Phys.
				Rev. B}\ }\textbf {\bibinfo {volume} {98}},\ \bibinfo {pages} {184403}
		(\bibinfo {year} {2018})}\BibitemShut {NoStop}%
	\bibitem [{\citenamefont {Amico}\ \emph {et~al.}(2004)\citenamefont {Amico},
		\citenamefont {Osterloh}, \citenamefont {Plastina}, \citenamefont {Fazio},\
		and\ \citenamefont {Massimo~Palma}}]{PhysRevA.69.022304}%
	\BibitemOpen
	\bibfield  {author} {\bibinfo {author} {\bibfnamefont {L.}~\bibnamefont
			{Amico}}, \bibinfo {author} {\bibfnamefont {A.}~\bibnamefont {Osterloh}},
		\bibinfo {author} {\bibfnamefont {F.}~\bibnamefont {Plastina}}, \bibinfo
		{author} {\bibfnamefont {R.}~\bibnamefont {Fazio}}, \ and\ \bibinfo {author}
		{\bibfnamefont {G.}~\bibnamefont {Massimo~Palma}},\ }\href {\doibase
		10.1103/PhysRevA.69.022304} {\bibfield  {journal} {\bibinfo  {journal} {Phys.
				Rev. A}\ }\textbf {\bibinfo {volume} {69}},\ \bibinfo {pages} {022304}
		(\bibinfo {year} {2004})}\BibitemShut {NoStop}%
	\bibitem [{\citenamefont {Hauke}\ \emph {et~al.}(2016)\citenamefont {Hauke},
		\citenamefont {Heyl}, \citenamefont {Tagliacozzo},\ and\ \citenamefont
		{Zoller}}]{Hauke2016}%
	\BibitemOpen
	\bibfield  {author} {\bibinfo {author} {\bibfnamefont {P.}~\bibnamefont
			{Hauke}}, \bibinfo {author} {\bibfnamefont {M.}~\bibnamefont {Heyl}},
		\bibinfo {author} {\bibfnamefont {L.}~\bibnamefont {Tagliacozzo}}, \ and\
		\bibinfo {author} {\bibfnamefont {P.}~\bibnamefont {Zoller}},\ }\href
	{\doibase 10.1038/nphys3700} {\bibfield  {journal} {\bibinfo  {journal} {Nat.
				Phys.}\ }\textbf {\bibinfo {volume} {12}},\ \bibinfo {pages} {778} (\bibinfo
		{year} {2016})}\BibitemShut {NoStop}%
	\bibitem [{\citenamefont {Scheie}\ \emph {et~al.}(2021)\citenamefont {Scheie},
		\citenamefont {Laurell}, \citenamefont {Samarakoon}, \citenamefont {Lake},
		\citenamefont {Nagler}, \citenamefont {Granroth}, \citenamefont {Okamoto},
		\citenamefont {Alvarez},\ and\ \citenamefont
		{Tennant}}]{scheie2021witnessing}%
	\BibitemOpen
	\bibfield  {author} {\bibinfo {author} {\bibfnamefont {A.}~\bibnamefont
			{Scheie}}, \bibinfo {author} {\bibfnamefont {P.}~\bibnamefont {Laurell}},
		\bibinfo {author} {\bibfnamefont {A.~M.}\ \bibnamefont {Samarakoon}},
		\bibinfo {author} {\bibfnamefont {B.}~\bibnamefont {Lake}}, \bibinfo {author}
		{\bibfnamefont {S.~E.}\ \bibnamefont {Nagler}}, \bibinfo {author}
		{\bibfnamefont {G.~E.}\ \bibnamefont {Granroth}}, \bibinfo {author}
		{\bibfnamefont {S.}~\bibnamefont {Okamoto}}, \bibinfo {author} {\bibfnamefont
			{G.}~\bibnamefont {Alvarez}}, \ and\ \bibinfo {author} {\bibfnamefont
			{D.~A.}\ \bibnamefont {Tennant}},\ }\href {\doibase
		10.1103/PhysRevB.103.224434} {\bibfield  {journal} {\bibinfo  {journal}
			{Phys. Rev. B}\ }\textbf {\bibinfo {volume} {103}},\ \bibinfo {pages}
		{224434} (\bibinfo {year} {2021})}\BibitemShut {NoStop}%
	\bibitem [{\citenamefont {Scheie}\ \emph {et~al.}(2023)\citenamefont {Scheie},
		\citenamefont {Laurell}, \citenamefont {Samarakoon}, \citenamefont {Lake},
		\citenamefont {Nagler}, \citenamefont {Granroth}, \citenamefont {Okamoto},
		\citenamefont {Alvarez},\ and\ \citenamefont
		{Tennant}}]{PhysRevB.107.059902}%
	\BibitemOpen
	\bibfield  {author} {\bibinfo {author} {\bibfnamefont {A.}~\bibnamefont
			{Scheie}}, \bibinfo {author} {\bibfnamefont {P.}~\bibnamefont {Laurell}},
		\bibinfo {author} {\bibfnamefont {A.~M.}\ \bibnamefont {Samarakoon}},
		\bibinfo {author} {\bibfnamefont {B.}~\bibnamefont {Lake}}, \bibinfo {author}
		{\bibfnamefont {S.~E.}\ \bibnamefont {Nagler}}, \bibinfo {author}
		{\bibfnamefont {G.~E.}\ \bibnamefont {Granroth}}, \bibinfo {author}
		{\bibfnamefont {S.}~\bibnamefont {Okamoto}}, \bibinfo {author} {\bibfnamefont
			{G.}~\bibnamefont {Alvarez}}, \ and\ \bibinfo {author} {\bibfnamefont
			{D.~A.}\ \bibnamefont {Tennant}},\ }\href {\doibase
		10.1103/PhysRevB.107.059902} {\bibfield  {journal} {\bibinfo  {journal}
			{Phys. Rev. B}\ }\textbf {\bibinfo {volume} {107}},\ \bibinfo {pages}
		{059902} (\bibinfo {year} {2023})}\BibitemShut {NoStop}%
	\bibitem [{\citenamefont {Ranjith}\ \emph {et~al.}(2019)\citenamefont
		{Ranjith}, \citenamefont {Luther}, \citenamefont {Reimann}, \citenamefont
		{Schmidt}, \citenamefont {Schlender}, \citenamefont {Sichelschmidt},
		\citenamefont {Yasuoka}, \citenamefont {Strydom}, \citenamefont {Skourski},
		\citenamefont {Wosnitza}, \citenamefont {K\"uhne}, \citenamefont {Doert},\
		and\ \citenamefont {Baenitz}}]{Ranjith2019_2}%
	\BibitemOpen
	\bibfield  {author} {\bibinfo {author} {\bibfnamefont {K.~M.}\ \bibnamefont
			{Ranjith}}, \bibinfo {author} {\bibfnamefont {S.}~\bibnamefont {Luther}},
		\bibinfo {author} {\bibfnamefont {T.}~\bibnamefont {Reimann}}, \bibinfo
		{author} {\bibfnamefont {B.}~\bibnamefont {Schmidt}}, \bibinfo {author}
		{\bibfnamefont {P.}~\bibnamefont {Schlender}}, \bibinfo {author}
		{\bibfnamefont {J.}~\bibnamefont {Sichelschmidt}}, \bibinfo {author}
		{\bibfnamefont {H.}~\bibnamefont {Yasuoka}}, \bibinfo {author} {\bibfnamefont
			{A.~M.}\ \bibnamefont {Strydom}}, \bibinfo {author} {\bibfnamefont
			{Y.}~\bibnamefont {Skourski}}, \bibinfo {author} {\bibfnamefont
			{J.}~\bibnamefont {Wosnitza}}, \bibinfo {author} {\bibfnamefont
			{H.}~\bibnamefont {K\"uhne}}, \bibinfo {author} {\bibfnamefont
			{T.}~\bibnamefont {Doert}}, \ and\ \bibinfo {author} {\bibfnamefont
			{M.}~\bibnamefont {Baenitz}},\ }\href {\doibase 10.1103/PhysRevB.100.224417}
	{\bibfield  {journal} {\bibinfo  {journal} {Phys. Rev. B}\ }\textbf {\bibinfo
			{volume} {100}},\ \bibinfo {pages} {224417} (\bibinfo {year}
		{2019})}\BibitemShut {NoStop}%
	\bibitem [{\citenamefont {Hutchings}(1964)}]{Hutchings1964}%
	\BibitemOpen
	\bibfield  {author} {\bibinfo {author} {\bibfnamefont {M.}~\bibnamefont
			{Hutchings}},\ }\href {\doibase
		https://doi.org/10.1016/S0081-1947(08)60517-2} {\ \bibinfo {series} {Solid
			State Physics},\ \textbf {\bibinfo {volume} {16}},\ \bibinfo {pages} {227 }
		(\bibinfo {year} {1964})}\BibitemShut {NoStop}%
	\bibitem [{\citenamefont {Aczel}\ \emph {et~al.}(2012)\citenamefont {Aczel},
		\citenamefont {Granroth}, \citenamefont {MacDougall}, \citenamefont {Buyers},
		\citenamefont {Abernathy}, \citenamefont {Samolyuk}, \citenamefont {Stocks},\
		and\ \citenamefont {Nagler}}]{aczel2012quantum}%
	\BibitemOpen
	\bibfield  {author} {\bibinfo {author} {\bibfnamefont {A.~A.}\ \bibnamefont
			{Aczel}}, \bibinfo {author} {\bibfnamefont {G.~E.}\ \bibnamefont {Granroth}},
		\bibinfo {author} {\bibfnamefont {G.~J.}\ \bibnamefont {MacDougall}},
		\bibinfo {author} {\bibfnamefont {W.}~\bibnamefont {Buyers}}, \bibinfo
		{author} {\bibfnamefont {D.~L.}\ \bibnamefont {Abernathy}}, \bibinfo {author}
		{\bibfnamefont {G.~D.}\ \bibnamefont {Samolyuk}}, \bibinfo {author}
		{\bibfnamefont {G.~M.}\ \bibnamefont {Stocks}}, \ and\ \bibinfo {author}
		{\bibfnamefont {S.~E.}\ \bibnamefont {Nagler}},\ }\href
	{https://doi.org/10.1038/ncomms2117} {\bibfield  {journal} {\bibinfo
			{journal} {Nature communications}\ }\textbf {\bibinfo {volume} {3}},\
		\bibinfo {pages} {1} (\bibinfo {year} {2012})}\BibitemShut {NoStop}%
	\bibitem [{\citenamefont {Zhang}\ \emph {et~al.}(2021)\citenamefont {Zhang},
		\citenamefont {Ma}, \citenamefont {Li}, \citenamefont {Wang}, \citenamefont
		{Adroja}, \citenamefont {Perring}, \citenamefont {Liu}, \citenamefont {Jin},
		\citenamefont {Ji}, \citenamefont {Wang}, \citenamefont {Kamiya},
		\citenamefont {Wang}, \citenamefont {Ma},\ and\ \citenamefont
		{Zhang}}]{Zhang_2021_NYS}%
	\BibitemOpen
	\bibfield  {author} {\bibinfo {author} {\bibfnamefont {Z.}~\bibnamefont
			{Zhang}}, \bibinfo {author} {\bibfnamefont {X.}~\bibnamefont {Ma}}, \bibinfo
		{author} {\bibfnamefont {J.}~\bibnamefont {Li}}, \bibinfo {author}
		{\bibfnamefont {G.}~\bibnamefont {Wang}}, \bibinfo {author} {\bibfnamefont
			{D.~T.}\ \bibnamefont {Adroja}}, \bibinfo {author} {\bibfnamefont {T.~P.}\
			\bibnamefont {Perring}}, \bibinfo {author} {\bibfnamefont {W.}~\bibnamefont
			{Liu}}, \bibinfo {author} {\bibfnamefont {F.}~\bibnamefont {Jin}}, \bibinfo
		{author} {\bibfnamefont {J.}~\bibnamefont {Ji}}, \bibinfo {author}
		{\bibfnamefont {Y.}~\bibnamefont {Wang}}, \bibinfo {author} {\bibfnamefont
			{Y.}~\bibnamefont {Kamiya}}, \bibinfo {author} {\bibfnamefont
			{X.}~\bibnamefont {Wang}}, \bibinfo {author} {\bibfnamefont {J.}~\bibnamefont
			{Ma}}, \ and\ \bibinfo {author} {\bibfnamefont {Q.}~\bibnamefont {Zhang}},\
	}\href {\doibase 10.1103/PhysRevB.103.035144} {\bibfield  {journal} {\bibinfo
			{journal} {Phys. Rev. B}\ }\textbf {\bibinfo {volume} {103}},\ \bibinfo
		{pages} {035144} (\bibinfo {year} {2021})}\BibitemShut {NoStop}%
	\bibitem [{\citenamefont {Scheie}(2021{\natexlab{a}})}]{PyCrystalField}%
	\BibitemOpen
	\bibfield  {author} {\bibinfo {author} {\bibfnamefont {A.}~\bibnamefont
			{Scheie}},\ }\href {https://doi.org/10.1107/S160057672001554X} {\bibfield
		{journal} {\bibinfo  {journal} {Journal of Applied Crystallography}\ }\textbf
		{\bibinfo {volume} {54}} (\bibinfo {year} {2021}{\natexlab{a}})}\BibitemShut
	{NoStop}%
	\bibitem [{\citenamefont {Powell}(1964)}]{PowellsMethod}%
	\BibitemOpen
	\bibfield  {author} {\bibinfo {author} {\bibfnamefont {M.~J.~D.}\
			\bibnamefont {Powell}},\ }\href {\doibase 10.1093/comjnl/7.2.155} {\bibfield
		{journal} {\bibinfo  {journal} {The Computer Journal}\ }\textbf {\bibinfo
			{volume} {7}},\ \bibinfo {pages} {155} (\bibinfo {year} {1964})}\BibitemShut
	{NoStop}%
	\bibitem [{\citenamefont {Virtanen}\ \emph {et~al.}(2020)\citenamefont
		{Virtanen}, \citenamefont {Gommers}, \citenamefont {Oliphant}, \citenamefont
		{Haberland}, \citenamefont {Reddy}, \citenamefont {Cournapeau}, \citenamefont
		{Burovski}, \citenamefont {Peterson}, \citenamefont {Weckesser},
		\citenamefont {Bright} \emph {et~al.}}]{virtanen2020scipy}%
	\BibitemOpen
	\bibfield  {author} {\bibinfo {author} {\bibfnamefont {P.}~\bibnamefont
			{Virtanen}}, \bibinfo {author} {\bibfnamefont {R.}~\bibnamefont {Gommers}},
		\bibinfo {author} {\bibfnamefont {T.~E.}\ \bibnamefont {Oliphant}}, \bibinfo
		{author} {\bibfnamefont {M.}~\bibnamefont {Haberland}}, \bibinfo {author}
		{\bibfnamefont {T.}~\bibnamefont {Reddy}}, \bibinfo {author} {\bibfnamefont
			{D.}~\bibnamefont {Cournapeau}}, \bibinfo {author} {\bibfnamefont
			{E.}~\bibnamefont {Burovski}}, \bibinfo {author} {\bibfnamefont
			{P.}~\bibnamefont {Peterson}}, \bibinfo {author} {\bibfnamefont
			{W.}~\bibnamefont {Weckesser}}, \bibinfo {author} {\bibfnamefont
			{J.}~\bibnamefont {Bright}},  \emph {et~al.},\ }\href {\doibase
		10.1038/s41592-019-0686-2} {\bibfield  {journal} {\bibinfo  {journal} {Nature
				Methods}\ }\textbf {\bibinfo {volume} {17}},\ \bibinfo {pages} {261}
		(\bibinfo {year} {2020})}\BibitemShut {NoStop}%
	\bibitem [{\citenamefont {Scheie}(2021{\natexlab{b}})}]{scheie2021quantifying}%
	\BibitemOpen
	\bibfield  {author} {\bibinfo {author} {\bibfnamefont {A.}~\bibnamefont
			{Scheie}},\ }\href {https://arxiv.org/abs/2107.14164} {\bibfield  {journal}
		{\bibinfo  {journal} {arXiv preprint arXiv:2107.14164}\ } (\bibinfo {year}
		{2021}{\natexlab{b}})}\BibitemShut {NoStop}%
	\bibitem [{\citenamefont {Pedregosa}\ \emph {et~al.}(2011)\citenamefont
		{Pedregosa}, \citenamefont {Varoquaux}, \citenamefont {Gramfort},
		\citenamefont {Michel}, \citenamefont {Thirion}, \citenamefont {Grisel},
		\citenamefont {Blondel}, \citenamefont {Prettenhofer}, \citenamefont {Weiss},
		\citenamefont {Dubourg}, \citenamefont {Vanderplas}, \citenamefont {Passos},
		\citenamefont {Cournapeau}, \citenamefont {Brucher}, \citenamefont {Perrot},\
		and\ \citenamefont {{{\'E}}douard Duchesnay}}]{scikit}%
	\BibitemOpen
	\bibfield  {author} {\bibinfo {author} {\bibfnamefont {F.}~\bibnamefont
			{Pedregosa}}, \bibinfo {author} {\bibfnamefont {G.}~\bibnamefont
			{Varoquaux}}, \bibinfo {author} {\bibfnamefont {A.}~\bibnamefont {Gramfort}},
		\bibinfo {author} {\bibfnamefont {V.}~\bibnamefont {Michel}}, \bibinfo
		{author} {\bibfnamefont {B.}~\bibnamefont {Thirion}}, \bibinfo {author}
		{\bibfnamefont {O.}~\bibnamefont {Grisel}}, \bibinfo {author} {\bibfnamefont
			{M.}~\bibnamefont {Blondel}}, \bibinfo {author} {\bibfnamefont
			{P.}~\bibnamefont {Prettenhofer}}, \bibinfo {author} {\bibfnamefont
			{R.}~\bibnamefont {Weiss}}, \bibinfo {author} {\bibfnamefont
			{V.}~\bibnamefont {Dubourg}}, \bibinfo {author} {\bibfnamefont
			{J.}~\bibnamefont {Vanderplas}}, \bibinfo {author} {\bibfnamefont
			{A.}~\bibnamefont {Passos}}, \bibinfo {author} {\bibfnamefont
			{D.}~\bibnamefont {Cournapeau}}, \bibinfo {author} {\bibfnamefont
			{M.}~\bibnamefont {Brucher}}, \bibinfo {author} {\bibfnamefont
			{M.}~\bibnamefont {Perrot}}, \ and\ \bibinfo {author} {\bibnamefont
			{{{\'E}}douard Duchesnay}},\ }\href
	{http://jmlr.org/papers/v12/pedregosa11a.html} {\bibfield  {journal}
		{\bibinfo  {journal} {Journal of Machine Learning Research}\ }\textbf
		{\bibinfo {volume} {12}},\ \bibinfo {pages} {2825} (\bibinfo {year}
		{2011})}\BibitemShut {NoStop}%
	\bibitem [{\citenamefont {Scheie}\ \emph {et~al.}(2020)\citenamefont {Scheie},
		\citenamefont {Garlea}, \citenamefont {Sanjeewa}, \citenamefont {Xing},\ and\
		\citenamefont {Sefat}}]{Scheie_2020}%
	\BibitemOpen
	\bibfield  {author} {\bibinfo {author} {\bibfnamefont {A.}~\bibnamefont
			{Scheie}}, \bibinfo {author} {\bibfnamefont {V.~O.}\ \bibnamefont {Garlea}},
		\bibinfo {author} {\bibfnamefont {L.~D.}\ \bibnamefont {Sanjeewa}}, \bibinfo
		{author} {\bibfnamefont {J.}~\bibnamefont {Xing}}, \ and\ \bibinfo {author}
		{\bibfnamefont {A.~S.}\ \bibnamefont {Sefat}},\ }\href {\doibase
		10.1103/PhysRevB.101.144432} {\bibfield  {journal} {\bibinfo  {journal}
			{Phys. Rev. B}\ }\textbf {\bibinfo {volume} {101}},\ \bibinfo {pages}
		{144432} (\bibinfo {year} {2020})}\BibitemShut {NoStop}%
	\bibitem [{\citenamefont {Xing}\ \emph {et~al.}(2021)\citenamefont {Xing},
		\citenamefont {Sanjeewa}, \citenamefont {May},\ and\ \citenamefont
		{Sefat}}]{xing2021_KYS}%
	\BibitemOpen
	\bibfield  {author} {\bibinfo {author} {\bibfnamefont {J.}~\bibnamefont
			{Xing}}, \bibinfo {author} {\bibfnamefont {L.~D.}\ \bibnamefont {Sanjeewa}},
		\bibinfo {author} {\bibfnamefont {A.~F.}\ \bibnamefont {May}}, \ and\
		\bibinfo {author} {\bibfnamefont {A.~S.}\ \bibnamefont {Sefat}},\ }\href
	{\doibase 10.1063/5.0071161} {\bibfield  {journal} {\bibinfo  {journal} {APL
				Materials}\ }\textbf {\bibinfo {volume} {9}},\ \bibinfo {pages} {111104}
		(\bibinfo {year} {2021})}\BibitemShut {NoStop}%
	\bibitem [{\citenamefont {Paddison}(2020)}]{Paddison_2020}%
	\BibitemOpen
	\bibfield  {author} {\bibinfo {author} {\bibfnamefont {J.~A.~M.}\
			\bibnamefont {Paddison}},\ }\href {\doibase 10.1103/PhysRevLett.125.247202}
	{\bibfield  {journal} {\bibinfo  {journal} {Phys. Rev. Lett.}\ }\textbf
		{\bibinfo {volume} {125}},\ \bibinfo {pages} {247202} (\bibinfo {year}
		{2020})}\BibitemShut {NoStop}%
	\bibitem [{\citenamefont {Brout}\ and\ \citenamefont
		{Thomas}(1967)}]{Brout_1967}%
	\BibitemOpen
	\bibfield  {author} {\bibinfo {author} {\bibfnamefont {R.}~\bibnamefont
			{Brout}}\ and\ \bibinfo {author} {\bibfnamefont {H.}~\bibnamefont {Thomas}},\
	}\href {\doibase 10.1103/PhysicsPhysiqueFizika.3.317} {\bibfield  {journal}
		{\bibinfo  {journal} {Physics Physique Fizika}\ }\textbf {\bibinfo {volume}
			{3}},\ \bibinfo {pages} {317} (\bibinfo {year} {1967})}\BibitemShut {NoStop}%
	\bibitem [{\citenamefont {Hohlwein}\ \emph {et~al.}(2003)\citenamefont
		{Hohlwein}, \citenamefont {Hoffmann},\ and\ \citenamefont
		{Schneider}}]{Hohlwein_2003}%
	\BibitemOpen
	\bibfield  {author} {\bibinfo {author} {\bibfnamefont {D.}~\bibnamefont
			{Hohlwein}}, \bibinfo {author} {\bibfnamefont {J.-U.}\ \bibnamefont
			{Hoffmann}}, \ and\ \bibinfo {author} {\bibfnamefont {R.}~\bibnamefont
			{Schneider}},\ }\href {\doibase 10.1103/PhysRevB.68.140408} {\bibfield
		{journal} {\bibinfo  {journal} {Phys. Rev. B}\ }\textbf {\bibinfo {volume}
			{68}},\ \bibinfo {pages} {140408} (\bibinfo {year} {2003})}\BibitemShut
	{NoStop}%
	\bibitem [{\citenamefont {Wysin}(2000)}]{Wysin_2000}%
	\BibitemOpen
	\bibfield  {author} {\bibinfo {author} {\bibfnamefont {G.~M.}\ \bibnamefont
			{Wysin}},\ }\href {\doibase 10.1103/PhysRevB.62.3251} {\bibfield  {journal}
		{\bibinfo  {journal} {Phys. Rev. B}\ }\textbf {\bibinfo {volume} {62}},\
		\bibinfo {pages} {3251} (\bibinfo {year} {2000})}\BibitemShut {NoStop}%
\end{thebibliography}

\section{Acknowledgments}

This research used resources at the Spallation Neutron Source and High Flux Isotope Reactor, DOE Office of Science User Facilities operated by the Oak Ridge National Laboratory. The work by D.A. Tennant is supported by the Quantum Science Center (QSC), a National Quantum Information Science Research Center of the U.S. Department of Energy (DOE). The work of J.A.M. Paddison (magnetic diffuse scattering fits) was supported by the U.S. Department of Energy, Office of Science, Basic Energy Sciences, Materials Sciences and Engineering Division. J. Xing and A. Sefat were supported by U.S. Department of Energy, Basic Energy Sciences, Materials Science and Engineering Division. L.O.M and A.E.T were supported by CONICET under Grant No. 364 (PIP2015).
This research used resources at the Missouri University Research Reactor and the Department of Chemistry in University of Missouri.
S.L., A.W., and R.M. were supported by the U.S. Department of Energy, Office of Science, National Quantum Information Science Research Centers, and Quantum Science Center.

N.E. Sherman, M. Dupont, J.E. Moore, (C.D. Pemmaraju, T.P. Devereaux) were supported by the U.S. Department of Energy, Office of Science, Office of Basic Energy Sciences, Materials Sciences and Engineering Division under Contract No. DE-AC02-05-CH11231 (DE-AC02-76SF00515) through the Theory Institute for Materials and Energy Spectroscopy (TIMES). J.E. Moore acknowledges additional support by a Simons Investigatorship.

This research used the Lawrencium computational cluster resource provided by the IT Division at the Lawrence Berkeley National Laboratory (supported by the Director, Office of Science, Office of Basic Energy Sciences, of the U.S. Department of Energy under Contract No. DE-AC02-05CH11231). This research also used resources of the National Energy Research Scientific Computing Center (NERSC), a U.S. Department of Energy Office of Science User Facility operated under Contract No. DE-AC02-05CH11231.

\section{Author contributions}

A.O. Scheie and D.A. Tennant conceived and coordinated the project. J. Xing, L.D. Sanjeewa, and A. Sefat synthesized and characterized single crystal KYbSe$_2$ samples for experiments. A.O. Scheie, D. Abernathy, D.M. Pajerowski, and T.J. Williams performed the neutron experiments, and A.O. Scheie analyzed the neutron data and calculated entanglement witnesses. J.A.M. Paddison performed ORF fits. E.A. Ghioldi, S-S. Zhang, L.O. Manuel, A.E. Trumper, and C.D. Batista carried out Schwinger boson calculations. N.E. Sherman, M. Dupont, and J.E. Moore carried out tensor network calculations of dynamical structure factors. C.D. Pemmaraju, T.P. Devereaux, and D.S. Parker carried out DFT calculations. S. Lee, A.J. Woods, and R. Movshovich performed heat capacity measurements.  A.O. Scheie, N.E. Sherman, M. Dupont, J.E. Moore, C.D. Batista, and D.A. Tennant wrote the manuscript with input from all co-authors.\\

\section{Competing financial interests}

The authors declare no competing financial interests.

\section{Corresponding authors}

Correspondence and requests for materials should be addressed to A. O. Scheie and C. D. Batista.



\newpage

\quad

\newpage

\section*{Supplemental Information for Witnessing quantum criticality and entanglement in the triangular antiferromagnet $\rm KYbSe_2$}

\renewcommand{\thefigure}{S\arabic{figure}}
\renewcommand{\thetable}{S\arabic{table}}
\renewcommand{\theequation}{S.\arabic{equation}}
\renewcommand{\thepage}{S\arabic{page}}  
\setcounter{figure}{0}
\setcounter{page}{1}
\setcounter{equation}{0}

{
	\section{Low temperature magnetic order}
	
	Diffuse scattering Hamiltonian fits predicted that KYbSe$_2$ is within the 120$^{\circ}$ ordered phase, so that it should order magnetically at the lowest temperatures.
	To test this, we measured the specific heat of KYbSe$_2$ using  quasi-adiabatic method in a dilution refrigerator. Heater was mounted on one side of the sapphire stage, with one large single sample (1.19 mg)  mounted on the other side with GE varnish. Ruthenium oxide resistance thermometer was glued on the top of the samples. A heat pulse is delivered to heat capacity stage, and the temperature of the thermometer is measured as a function of time.
	The results are shown in Fig. \ref{flo:magneticOrder}(a). A clear kink is visible at 290~mK consistent with an ordering transition. 
	
	To evaluate any possible sample-dependent properties, we remeasured heat capacity of a collection of ~11 small pieces (2.33 mg) [see the grey data in Fig. \ref{flo:magneticOrder}(a)], which shows a somewhat broadened transition. Neither sample had a measurable amount of crystalline disorder, suggesting that an appreciable amount of crystalline disorder may suppress the transition entirely.
	
	More clear evidence of static magnetism comes from the upturn at the lowest temperatures. We fit this with a nuclear Schottky anomaly using the hyperfine parameters for Yb$^{3+}$ in Ref. \cite{Bleaney1963}. As shown by the colored lines in  Fig. \ref{flo:magneticOrder}(a), the Yb moment size is well-constrained by the data, giving a fitted $\mu = (0.579 \pm 0.010) \> {\rm \mu_B}$. This method of determining the magnetic moment has the advantage of (a) being a local probe and insensitive to the particular type of magnetic order, and (b) is a fit extrapolated to zero temperature, and thus is an estimate of the $T=0$ ordered moment.
	
	\begin{figure}
		\centering\includegraphics[width=0.47\textwidth]{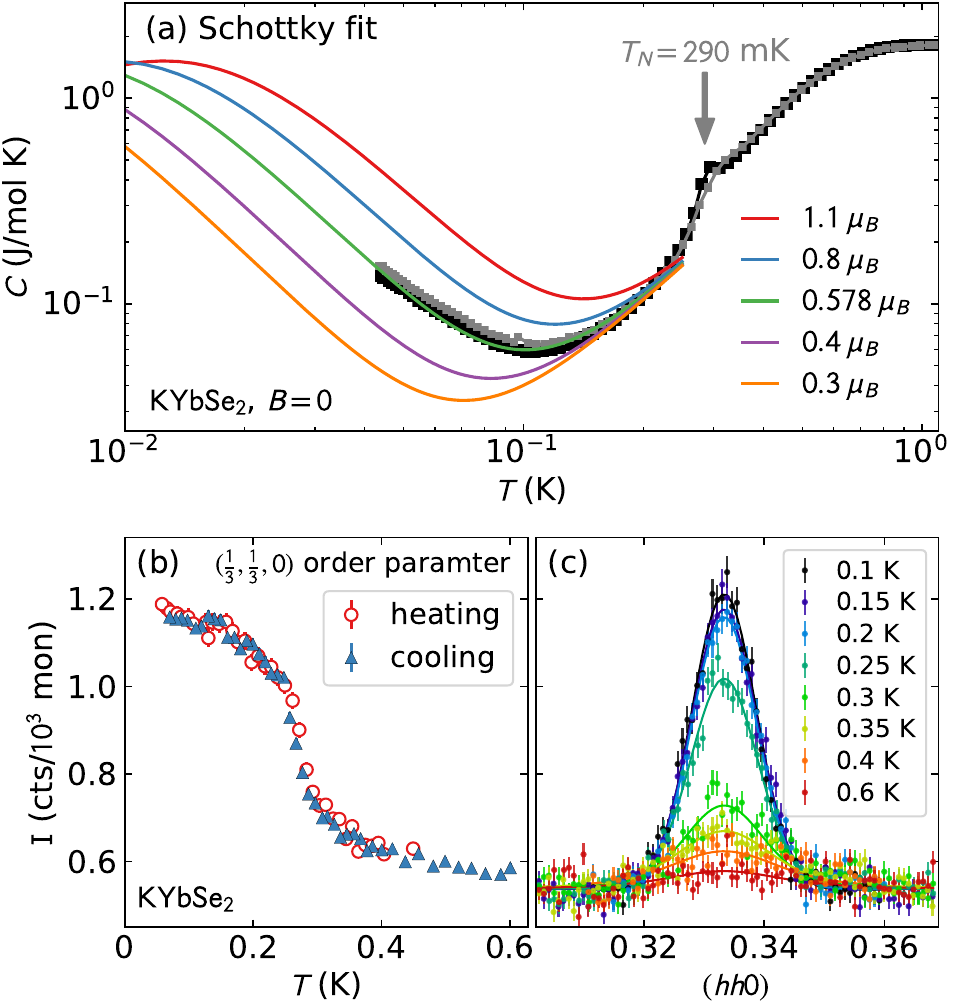}
		\caption{Evidence of magnetic order in KYbSe$_2$.  (a) Zero field heat capacity of a large (black) and small (grey) sample, showing a kink at 290 mK. An upturn at low temperatures is fitted with a nuclear Schottky anomaly, which uniquely constrains the local ordered Yb$^{3+}$ moment as shown by the colored curves.}
		\label{flo:magneticOrder}
	\end{figure}
	
	We also measured the neutron diffraction of KYbSe$_2$ using the CTAX spectrometer at the Oak Ridge National Laboratory HFIR reactor, using the same sample and sample mount as used in the CNCS experiment, but mounted in a dilution refrigerator. We measured with $E_i=E_f=4.8$~meV neutrons focusing on the $Q=(1/3,1/3,0)$ point, the wavevector associated with 120$^{\circ}$ order. This data is shown in Fig. \ref{flo:magneticOrder}(b), and shows a clear onset of elastic scattering at around 300~mK. Panel (c) shows cuts along the $(hh0)$ direction at several different temperatures, showing the emergence of the Bragg intensity. The steepest part of the order parameter curve is at 290~mK, confirming that the bump observed in heat capacity is indeed the transition to 120$^{\circ}$ magnetic order. Additionally, the agreement between the heat capacity and neutron order parameter curves shows that the sample mount used for the neutron experiments provides adequate thermal equilibration down to at least 290 mK.
	In summary, both heat capacity and neutron diffraction confirm that the KYbSe$_2$ $J_2/J_1$ ratio is indeed within the 120$^{\circ}$ ordered phase.

	\section{Crystal characterization}
	
	To investigate the quality of the KYbSe$_2$ crystals used in this experiment, we measured the single crystal X-ray diffraction using a Bruker Quest D8 single-crystal X-ray diffractometer. The data were collected at room temperature utilizing a Mo K$\alpha$ radiation, $\lambda = 0.71073 \AA$. The crystal diffraction images were collected using $\phi$ and $\omega$-scans. The diffractometer was equipped with an Incoatec I$\rm \mu$S source using the APEXIII software suite for data setup, collection, and processing \cite{Apex3}. The structure was resolved using intrinsic phasing and full-matrix least square methods with refinement on F2. Structure refinements were performed using the SHELXTL software suite \cite{sheldrick2008short}. All atoms were first refined with isotropic displacement parameters and then they were refined anisotropically. The final refinement was confirmed with \textit{CheckCif} \cite{spek2009structure}. A refinement with no site mixing fits the data extremely well.
	
	If we allow site mixing between K and Yb, similar to what was observed in NaYbSe$_2$ \cite{Dai_2021}, the K-Yb site mixing in KYbSe$_2$ refines to $(0.2 \pm 0.3)$\%, where the error bar indicates one standard deviation uncertainty as calculated by reduced $\chi^2$ contour. This is a full order of magnitude less site mixing than the 3\% site disorder found in NaYbSe$_2$ \cite{Dai_2021}. If we force the refined model to have a site-mixing greater than 1\%, as shown in Fig. \ref{flo:XRD}, we find a worse $R$-value and a visibly worse fit. Thus, to within uncertainty, KYbSe$_2$ has no K-Yb site mixing and can be considered as an ideal 2D triangular lattice.

	\begin{figure}
		\centering\includegraphics[width=0.42\textwidth]{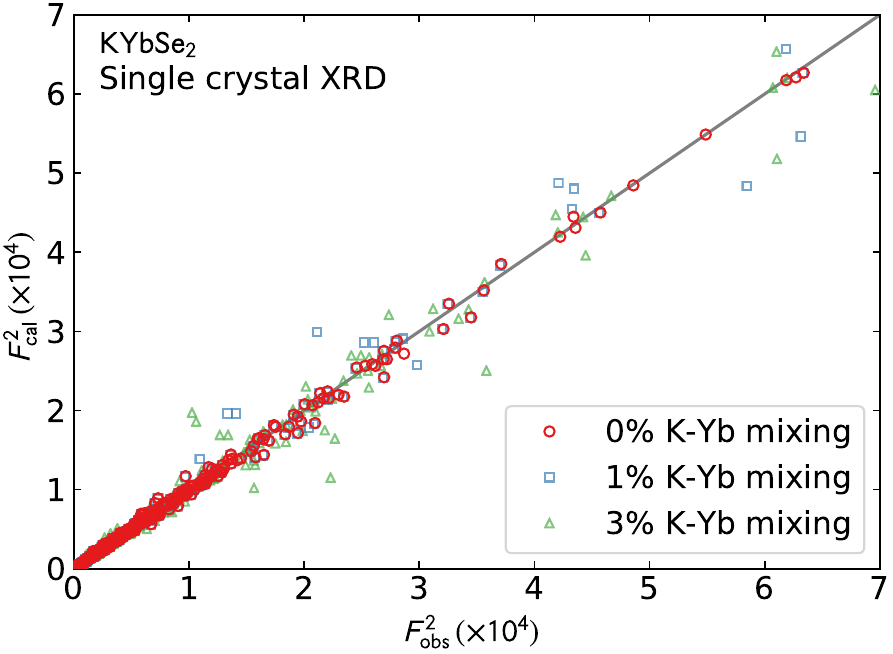}
		\caption{Single crystal KYbSe$_2$ X-ray diffraction, showing the observed peak intensities versus the peak intensities of a calculated model with no site mixing or disorder (red circles), 1\% Yb-K site mixing (blue squares), and 3\% Yb-K site mixing (green triangles). The 0\% site mixing visibly fits the data the best, indicating very high crystal quality.}
		\label{flo:XRD}
	\end{figure}
}

\section{Entanglement Witnesses}

\subsection{One tangle}

{
	One-tangle is calculated from the nuclear Schottky anomaly as explained in the main text. Dividing the total static moment by the in-plane $g$-tensor (as appropriate for $120^{\circ}$ order) gives a measure of the static spin: $\langle S \rangle = \langle \mu \rangle/g_{xx} = 0.193\pm0.013$, giving a one-tangle $\tau_1 = 1- 4\langle S \rangle^2 = 0.85 \pm 0.02$. 
	
	As noted in the main text, the static moment from nuclear Schottky fits indicates a moment 39(2)~\% of the maximal static moment. This is close to the fully static moment for the 2D pure $J_1$ model predicted by various theoretical techniques: 41\% from quantum Monte Carlo \cite{Capriotti_1999}, 41\% from DMRG \cite{PhysRevLett.99.127004}, 38\% from series expansion \cite{Zheng_2006}, and 44\% from Schwinger Boson theory  \cite{Ghioldi_2018}. 
	Experimentally, inter-plane and anisotropic exchange (such as are unavoidable in real materials) would tend to increase the static moment relative to the pure 2D isotropic case. The fact that KYbSe$_2$ orders at finite temperature is evidence of these effects. However, the addition of $J_2$ would tend to decrease the static moment relative to the 2D isotropic case. The fact that KYbSe$_2$, despite its 3D and weak anisotropic interactions, has a static moment slightly below the theoretical value for the pure 2D isotropic case is evidence that the $J_2$ is destabilizing the magnetic order. This concurs with our conclusion that KYbSe$_2$ is a proximate QSL.
}

\subsection{Two tangle}

We calculate Two-tangle from the real-space spin correlations, which we obtain from the Fourier transform of the energy-integrated $(hk0)$ plane scattering. For an isotropic $S=1/2$ system (which KYbSe$_2$ is to a good approximation---see Onsager reaction field fits), the two-tangle is defined as
\begin{align}
\tau_2  &=  8 \sum_{r \neq 0}\,\Bigg( \mathrm{max}\left\{\, 0,\, 2|g_{r}^{zz}| -\left|\frac{1}{4}+g_{r}^{zz}\right|\,\right\} \Bigg)^2,
\label{eq:twotangle}
\end{align}
where $g_{r}^{zz} = \langle S_i^z S_{i+r}^z\rangle$~\cite{PhysRevA.69.022304}. As eq. \ref{eq:twotangle} shows, $g_{r}^{zz}$ must exceed the classical threshold of 1/4 for two-tangle to be nonzero. As is shown in main text Fig. 4(b), none of the first four neighbor distances exceed this threshold, and thus two-tangle is zero for all temperatures in KYbSe$_2$.

\begin{figure}
	\centering\includegraphics[width=0.48\textwidth]{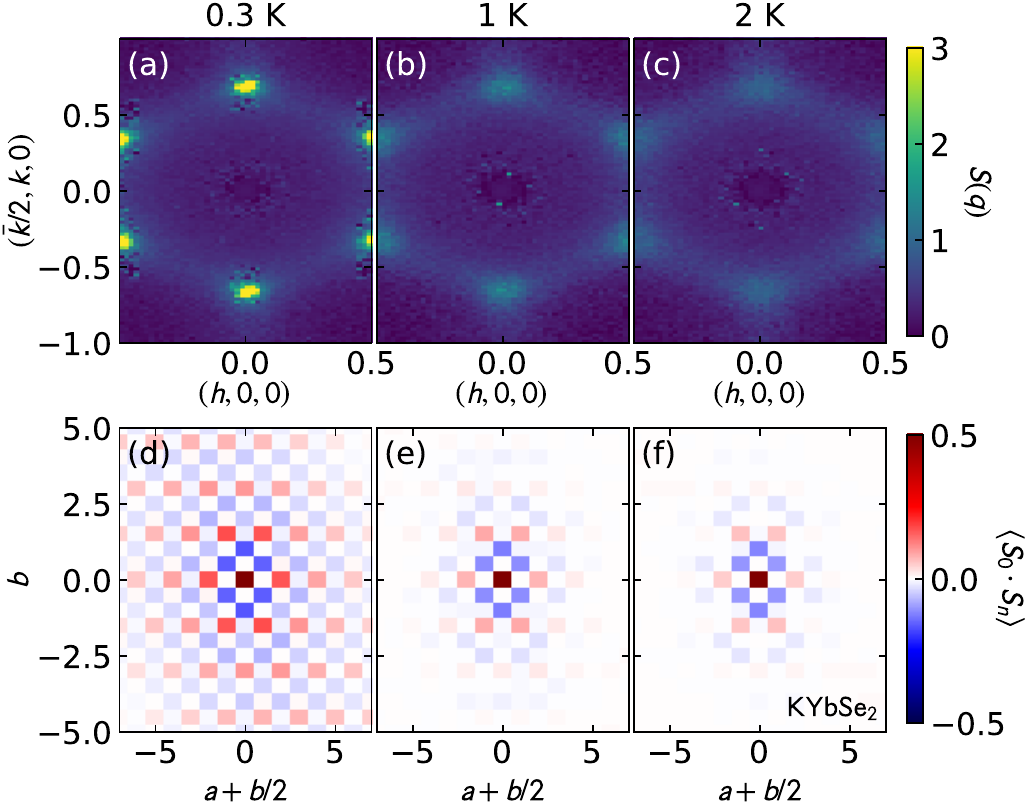}
	\caption{Energy-integrated KYbSe$_2$ scattering (a)-(c) Fourier transformed to obtain the static spin-spin correlation in real space (d)-(f). Red indicates ferromagnetic correlation, blue indicates antiferromagnetic correlation. Quasi-long-range order is visible at 0.3~K, but higher temperatures show a much shorter correlation length.}
	\label{flo:KYS_FourierTransform}
\end{figure}

We obtained the real space correlations in main text Fig. 4 for the two-tangle by taking Fourier transform of the energy-integrated data in the $(hk0)$ plane. To do this, we cut out a section of reciprocal space from $-0.5<h<0.5$ and $-1<k<1$. Data were corrected for the Yb$^{3+}$ form factor and background subtracted as described above. Empty data near $Q=0$ was filled in with data from the next Brillouin zones, and then the data were integrated over all energies, yielding the 2D slices shown in Fig. \ref{flo:KYS_FourierTransform}. 

We then took the 2D Fourier transform of the data to obtain the real-space spin-spin correlation function, which is shown in the bottom row of Fig. \ref{flo:KYS_FourierTransform}. Note that the correlations noticeably decrease as temperature increases, although they retain an overall antiferromagnetic correlation to the nearest neighbors. The average neighbor correlations for the two-tangle calculation in main text Fig. 4 were extracted from these plots.

The two-tangle is dependent upon the background subtraction scheme, and the uncertainty from this is difficult to estimate. We re-ran the analysis with no inelastic background subtraction whatsoever, and found a 0.3~K nearest neighbor two-tangle of -0.082 instead of the background subtracted value of -0.159. As a ``worst-case'' estimate of uncertainty, it indicates a 48\% error bar---which still leaves the nearest neighbor value below the threshold for witnessed bipartite entanglement. (This is most certainly an overestimate, as it leaves a fair amount of background and artifacts in the data near $Q=0$.) Thus, the two-tangle is most certainly zero for KYbSe$_2$. 

\subsection{Quantum Fisher Information}

Quantum Fisher Information can be calculated from the neutron spectrum by an integral over energy. When normalized by spin length, QFI is defined as 
\begin{equation}
{\rm nQFI} = \frac{\hbar}{3 \pi S^2} \int_0^{\infty} \mathrm{d} \left( \hbar\omega\right) \tanh \left( \frac{\hbar \omega}{2 k_B T} \right) \chi^{\prime\prime} (\hbar \omega, T), \label{eq:QFI}
\end{equation}
\cite{Hauke2016,scheie2021witnessing} 
where $\chi^{\prime\prime}({\bf q},\omega)$ is obtained via the fluctuation dissipation theorem 
$
\chi^{\prime\prime}(\omega) = \pi (1 -  e^{- \hbar \omega  \beta})  S(\omega)
$ 
\cite{PhysRevB.107.059902}. 
In our case, because the correlations are 120$^{\circ}$, we calculate the nQFI at ${\bf q}=(1/3,1/3)$ as shown in main text Fig. 4(b).
The magnetic excitations at ${\bf q}=(1/3,1/3)$ are gapless to within 40~$\rm \mu$eV and the QFI integral begins to diverge as temperature decreases.

	If we calculate nQFI from Schwinger boson theory and DMRG, we get $\rm nQFI=5.049$ for Schwinger Bosons and $\rm nQFI=0.107$ for DMRG (both assuming $T=0.3$~K).
	The tiny DMRG value is because of the large broadening present from finite size effects. We expect the Schwinger Boson result to more accurately reflect the theoretical QFI, and it is in good agreement with our experimental value of  3.4(2). (We expect the experimental value is smaller because of (i) experimental broadening, (ii) small 3D and anisotropic exchanges in KYbSe$_2$, and (iii) the fact that the experimental data is taken at finite temperature while the theoretical calculation is for $T=0$.) These experimental considerations notwithstanding, the 0.3~K nQFI
	far exceeds the unentangled threshold, even if the 0.3~K nQFI uncertainty were much larger. This allows us to conclusively rule out a trivial unentangled phase.

We also note that the nature of the nQFI integral guarantees that nQFI will increase as temperature decreases. This is for two reasons: (1) the $\tanh$ factor allows more low-energy intensity into the integral as $T \rightarrow 0$, leading to an increase in QFI as $T$ decreases because the most intense features are at the lowest energies. (2) scattering features generally sharpen as temperature decreases. This is certainly the trend in KYbSe$_2$ as we compare the 1~K, 2~K, and 0.3~K scattering. As the scattering features sharpen, the energy integrated intensity at $Q=K$ will increase as well. Therefore the $T \rightarrow 0 $ nQFI will certainly increase even more.

\section{Fitting the Roton Mode}

\begin{figure}
	\centering\includegraphics[width=0.47\textwidth]{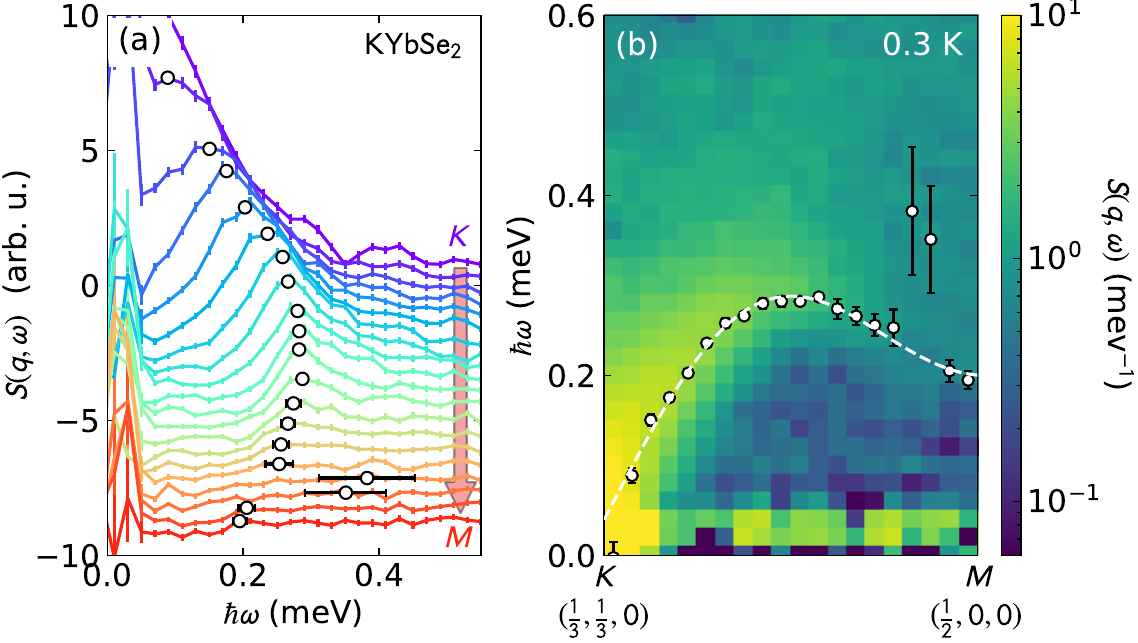}
	\caption{Fits to the KYbSe$_2$ roton mode. (a) shows constant $Q$ cuts between $K$ and $M$, and the fitted mode maximum. Each constant $Q$ slice is offset on the $y$ axis, and the colors show the variation from $K$ and $M$. (b) shows the mode maximum overplotted on the colormap data, along with a fitted sinusoidal dispersion function.}
	\label{flo:RotonMode}
\end{figure}

To quantify the extent and the gap of the roton-like mode, we fitted the intensity vs energy of many constant-$Q$ cuts as shown in Fig. \ref{flo:RotonMode}. We used an asymmetric Gaussian to model the mode, and with the exception of two data points near $M$, it picks out the peak maximum very well.
We then fitted these data points to a sinusoidal function $A \sin(Q) + B \sin(3Q) + C$ to estimate the mode maximum and minimum. These fits show a mode maximum of 0.288(12) meV, a roton minimum 0.200(13) meV, and a fitted gap of 0.059(7) meV. The fitted gap may be an artifact of the mode's deviation from the idealized sin function rather than an actual gap---the higher resolution scan in Fig.~8 of the main text do not reveal a clear gap above 40~$\rm \mu$eV.

\begin{figure}
	\centering\includegraphics[width=0.47\textwidth]{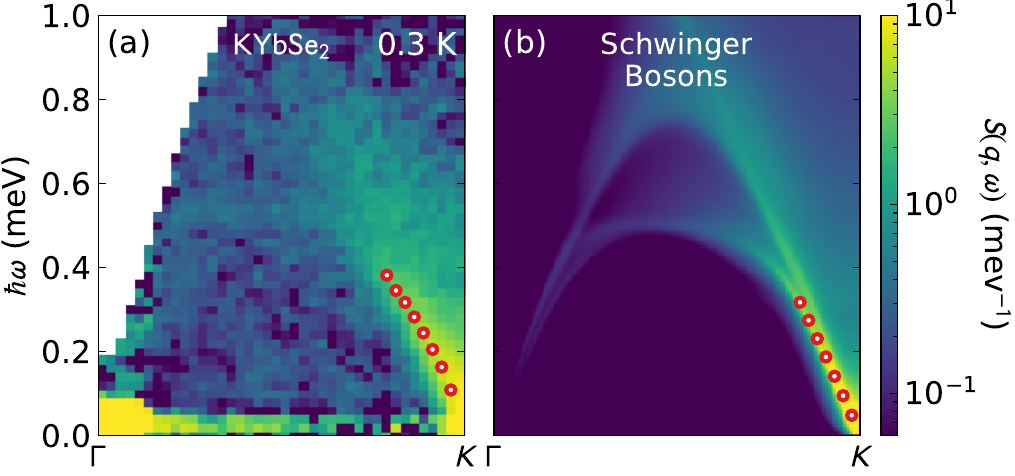}
	\caption{Fits to the mode maximum between $\Gamma$ and $K$ for both (a) KYbSe$_2$ and (b) Schwinger boson simulations. The red dots show the center of a fitted Gaussian, the slope of which was used to determine the energy scale for Schwinger boson simulations. The experimental slope appears to have a nonzero $\hbar \omega$ intercept, but is too small to be directly resolved with this experiment.}
	\label{flo:K-gammaMode}
\end{figure}

We also fitted the low-energy intense mode emanating from $K$ toward $\Gamma$ in order to match energy scales between theory and experiment, shown in Fig. \ref{flo:K-gammaMode}. We fit constant $Q$ cuts to Gaussian curves to define the center of the mode in both KYbSe$_2$ scattering data and Schwinger boson simulations. We then fit these fitted points to a sinusoidal curve between $\Gamma$ and $K$, and scaled the slope of the sin curves at $K$ so that theory matched experiment. This led to a fitted energy scale $J_1 = 0.56(3)$~meV.
The Schwinger boson simulations show the fitted maxima extrapolating toward $\hbar \omega = 0$ at $K$, but the KYbSe$_2$ mode maxima appear to have a nonzero intercept. Allowing for a gap in the fitted sin function, we estimate a KYbSe$_2$ fitted gap of 0.030(5)~meV---too small to be directly resolved using these data.

We can cross-check this fitted $J_1$ value by comparing to the saturation magnetization. Although saturation magnetization has not been measured for KYbSe$_2$, it has been measured for the sister compound NaYbSe$_2$, which has an $ab$-plane saturation magnetization of $\sim 12$~T~\cite{Ranjith2019_2}. Assuming the KYbSe$_2$ value to be close to 12~T, this gives an exchange energy scale of $J_1=5.36$~K $=0.462$~meV.

\section{Crystal electric field fits}

Here we describe the procedure used, and the results from the crystal electric field (CEF) fits to the KYbSe$_2$ ARCS data.

Because of the three-fold Yb$^{3+}$ rotational symmetry in KYbSe$_2$, there are six nonzero crystal field parameters in the Stevens operator formalism for the Yb$^{3+}$ ion: $B_2^0$, $B_4^0$, $B_4^3$, $B_6^0$, $B_6^3$, $B_6^6$~\cite{Hutchings1964}. The effective $J=7/2$ of Yb$^{3+}$ allows for four energetically-distinct Kramers doublet states, which means three crystal field excitation peaks should be visible in the neutron spectrum.

\begin{figure*}
	\centering\includegraphics[width=\textwidth]{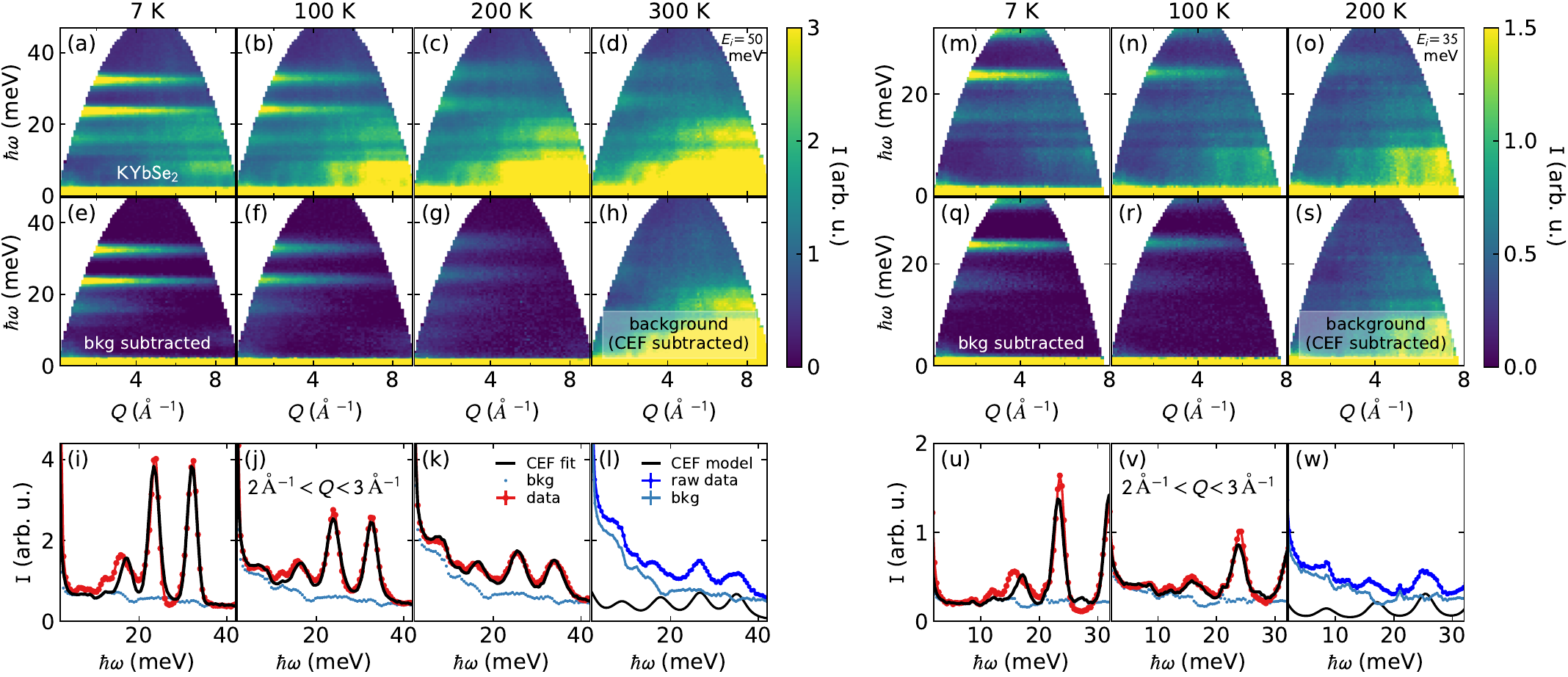}
	\caption{KYbSe$_2$ crystal field fit. The top row shows the raw data measured with $E_i = 50$ meV (left section) and $E_i = 35$ meV neutrons (right section). The middle row shows the background subtracted data, with the model-subtracted backgrounds shown in panels (h) and (s). The bottom row shows the fitted data between 2 \AA$^{-1}$ and  3 \AA$^{-1}$. Red data shows the raw data, light blue data shows the rescaled high-temperature background. The black line shows the CEF model plus the fitted background. The backgrounds are shown in panels (l) and (w).}
	\label{flo:KYS_CEF_bkg}
\end{figure*}

\begin{figure}
	\centering\includegraphics[width=0.45\textwidth]{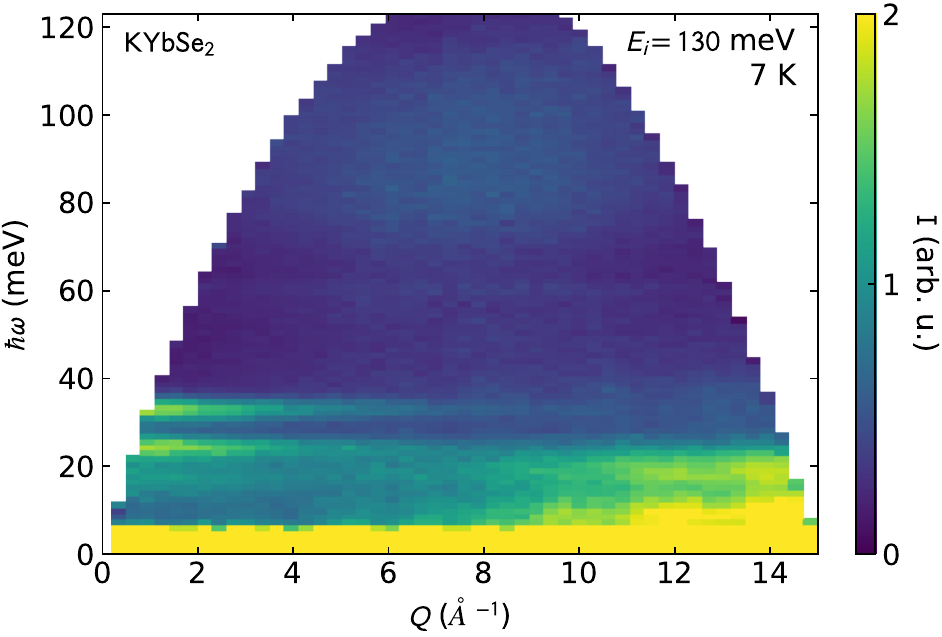}
	\caption{Crystal field spectrum of KYbSe$_2$ measured with $E_i=130$ meV neutrons. No crystal field excitations are visible above 40 meV, confirming the model derived in this study.}
	\label{flo:KYS_CEF_130}
\end{figure}

The crystal field excitations in Fig. \ref{flo:KYS_CEF_bkg} can be distinguished from the phonon background by the dependence upon $Q$: phonon intensity grows with $Q$ while magnetic intensity decreases with $Q$ according to the magnetic form factor. Three of the latter excitations are visible: one near 32 meV, one near 23 meV, and one near 17 meV.
Unfortunately, as shown in Fig. \ref{flo:KYS_CEF_bkg}, the 17 meV mode sits atop an intense flat phonon band which extends to low $Q$ 
{
	(it is clearly a phonon because its intensity grows with both temperature and $|Q|$), 
}
which potentially indicates coupling between phonons and the CEF excitation. (Alternatively, flat-band phonons can have $Q=0$ intensity from multiple scattering~\cite{aczel2012quantum}.) To verify that the low-energy mode is indeed the third CEF excited level, we measured the CEF spectrum up to $130$ meV [Fig.~\ref{flo:KYS_CEF_130}], and found no additional visible CEF levels. Furthermore, the observed energies are close to (i) point charge calculations which predict excited modes at 6.4 meV, 18.4 meV, and 33.0 meV, and (ii) measured crystal field excitations of sister compound NaYbSe$_2$ of 15.8 meV, 24.3 meV, and 30.5 meV~\cite{Zhang_2021_NYS}. Therefore, we are confident that the 17 meV, 23 meV, and 32 meV peaks are the three excited Yb$^{3+}$ CEF levels.

\subsubsection{Fitting procedure}

To fit the six crystal field parameters to the data, we started with a point-charge model calculation of the crystal field levels, which predicts energies at 6.4 meV, 18.4 meV, and 33.0 meV. We then used \textit{PyCrystalField} \cite{PyCrystalField} to fit the model to the neutron scattering data between 1~\AA$^{-1}$ and 2~\AA$^{-1}$ using the point charge model as starting values.

In order to isolate the crystal field excitations from the phonon background, we employed a self-consistent background subtraction scheme. We used the highest temperature data (300~K for $E_i=50$~meV and 200~K for $E_i=35$~meV) as background, but subtracted off the simulated CEF intensities and then rescaled the subtracted data to match the lower-temperature phonons. This way, the background improves as the CEF model improves, such that the best fit CEF model subtracts off the visible CEF excitations at high temperatures. Because the frequencies of the phonon spectrum are not precisely known, we created a phenomenological energy-dependent scale function to apply the background to lower temperatures. It was a step function of the form 
$$
\frac{a}{\exp[(\hbar \omega - \mu)/ k_B T] + 1} + b
$$
where $a$, $b$, and $\mu$ were fitted to the ratio of high-$T$ to low-$T$ scattering data at energy transfers where no crystal electric excitations are present. As shown in Fig. \ref{flo:KYS_CEF_bkg}, it produces a reasonable background for the fits.

\begin{figure}
	\centering\includegraphics[width=0.45\textwidth]{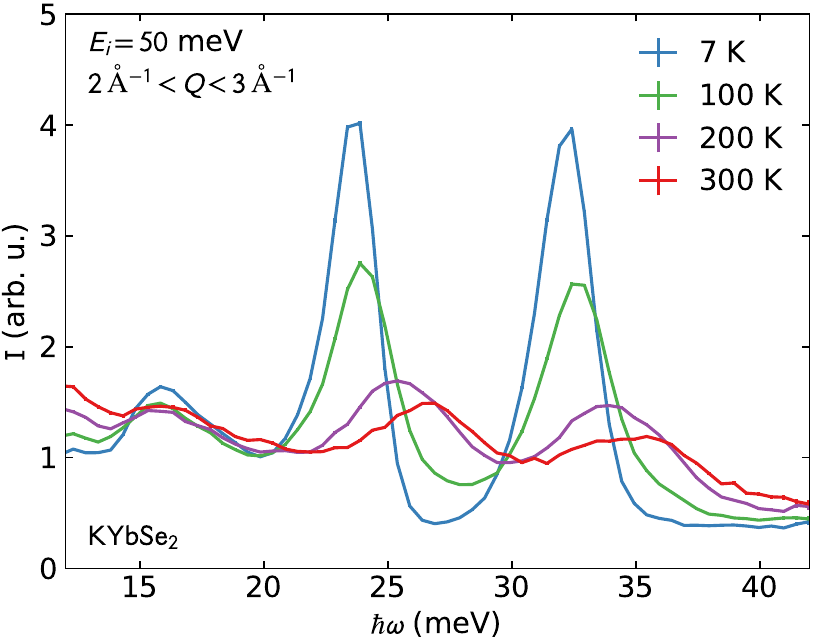}
	\caption{KYbSe$_2$ crystal field excitations as a function of temperature. The top two peaks noticeably shift to higher energies as temperature increases, while the bottom level stays constant. This effect was accounted for in the model fit.}
	\label{flo:KYS_CEF_EnergyShift}
\end{figure}

As in NaYbSe$_2$~\cite{Zhang_2021_NYS}, the crystal field levels broaden in energy and shift to higher energies as temperature increases, as shown in Fig. \ref{flo:KYS_CEF_EnergyShift}. The broadening indicates a shorter excitation lifetime, and is typical for crystal field levels at high temperatures. The shift in energy indicates CEF-phonon coupling, which is not surprising given that the lowest energy  CEF mode is at nearly the same energy as an intense phonon band.
To account for this in our fits, we applied an ad-hoc shift to the higher temperature energy eigenvalues so that they match the data. In theory, these shifts occur because of slight shifts in the CEF Hamiltonian and require a separate CEF fit---but in order to constrain the low-temperature Hamiltonian it was necessary to include the higher temperature data. Thus we assume that the slight shift in energy indicates a negligible change in the mode intensities, and the resulting fit matches the data very well.

We simulated the crystal field excitations with a Voigt profile, with a temperature-dependent Lorentzian width to account for finite lifetime and a Gaussian width to account for instrumental resolution. The Lorentzian widths were fitted to the two highest peaks for each temperature prior to the Hamiltonian fit and were kept fixed throughout the fit. The resolution function was treated as a fitted parameter, and was allowed to vary linearly as a function of energy transfer but not temperature. The fitted resolution was allowed to vary between $E_i = 50$~meV and $E_i = 35$~meV. Also, an overall scale factor was fitted to the data, one for each incident energy. We simultaneously fit the 7~K, 100~K, and 200~K $E_i = 50$~meV data and the 7~K and 100~K $E_i = 35$~meV data. $\chi^2$ minimization was performed with Powell's method~\cite{PowellsMethod} as implemented by Scipy~\cite{virtanen2020scipy}.
The best fit crystal field parameter values are given in Table \ref{tab:CEFparameters}, and the resulting crystal field Hamiltonian eigenstates are listed in Table \ref{tab:Eigenvectors}.
The best fit calculated $g$-tensor is given in the main text.

\begin{table}[]
	\caption{Best fit crystal field parameters for KYbSe$_2$. The middle column gives the nearest neighbor point charge model for KYbSe$_2$ and the  right column shows the best fit values.}
	\centering
	\begin{tabular}{c | c c}
		Crystal field parameter & Point charge model & Best fit\\
		\hline
		$B^{0}_{0}$  &  -0.556  &  -0.16(2)   \\
		$B^{0}_{0}$  &  0.0088  &  0.004(2)   \\
		$B^{3}_{3}$  &  -0.281  &  -0.15(2)   \\
		$B^{0}_{0}$  &  0.00005  &  0.00038(5)   \\
		$B^{3}_{3}$  &  0.0002  &  0.0143(5)   \\
		$B^{6}_{6}$  &  0.00043  &  0.0103(5) 
	\end{tabular}
	\label{tab:CEFparameters}
\end{table}

We calculated the uncertainty for these parameters by using a Monte Carlo stochastic search method to map out the $\chi^2$ contour around the best fit model~\cite{scheie2021quantifying}. Using a series of Markov chains, we generated several thousand solutions within $\Delta \chi^2 = 1$ of the best fit minimum $\chi^2_{red} =65.86$. This search was aided by principal component analysis of the valid solutions using Scikit~\cite{scikit}, such that the random guesses were more along principal component axes. The CEF parameter, $g$-tensor, and eigenvector uncertainties were calculated from the range of valid values in this set.

\begin{table*}
	\caption{Eigenvectors and eigenvalues for the best fit KYbSe$_2$ CEF Hamiltonian. Numbers in parenthesis are one standard deviation uncertainty.}
	\begin{tabular}{c|cccccccc}
		E (meV) &$| -\frac{7}{2}\rangle$ & $| -\frac{5}{2}\rangle$ & $| -\frac{3}{2}\rangle$ & $| -\frac{1}{2}\rangle$ & $| \frac{1}{2}\rangle$ & $| \frac{3}{2}\rangle$ & $| \frac{5}{2}\rangle$ & $| \frac{7}{2}\rangle$ \tabularnewline
		\hline 
		0.0 & 0.0 & 0.78(3) & 0.0 & 0.0 & -0.44(4) & 0.0 & 0.0 & -0.44(3) \tabularnewline
		0.0 & -0.44(3) & 0.0 & 0.0 & 0.44(4) & 0.0 & 0.0 & 0.78(3) & 0.0 \tabularnewline
		17.1(3) & 0.0 & -0.09(3) & 0.0 & 0.0 & 0.61(3) & 0.0 & 0.0 & -0.79(2) \tabularnewline
		17.1(3) & -0.79(2) & 0.0 & 0.0 & -0.61(3) & 0.0 & 0.0 & -0.09(3) & 0.0 \tabularnewline
		23.24(5) & 0.0 & 0.0 & 1.0 & 0.0 & 0.0 & 0.0 & 0.0 & 0.0 \tabularnewline
		23.24(5) & 0.0 & 0.0 & 0.0 & 0.0 & 0.0 & 1.0 & 0.0 & 0.0 \tabularnewline
		31.93(5) & 0.43(3) & 0.0 & 0.0 & -0.66(4) & 0.0 & 0.0 & 0.62(4) & 0.0 \tabularnewline
		31.93(5) & 0.0 & 0.62(4) & 0.0 & 0.0 & 0.66(4) & 0.0 & 0.0 & 0.43(3) 
	\end{tabular}
	\label{tab:Eigenvectors}
\end{table*}

\begin{figure}
	\centering\includegraphics[width=0.45\textwidth]{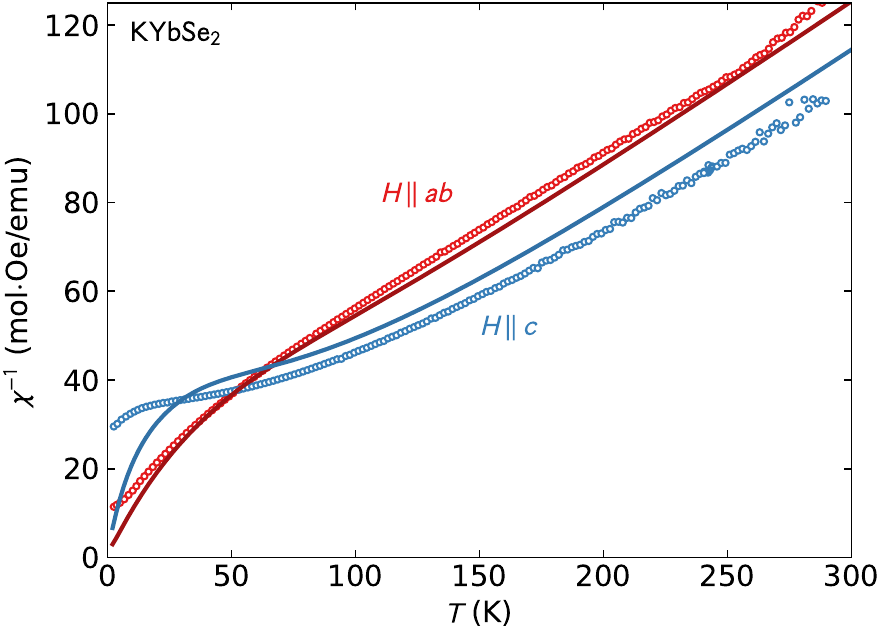}
	\caption{KYbSe$_2$ susceptibility compared to best fit CEF calculated single crystal susceptibility. Qualitatively, the simulation matches experiment, though the experiments show a higher susceptibility in the $c$ direction.}
	\label{flo:KYS_Susceptibility}
\end{figure}

It is often the case that crystal field fits to neutron data are underconstrained, and wildly different Hamiltonians can fit the data equally well~\cite{scheie2021quantifying,Scheie_2020}. The same is true here: two different models emerged from the fits, one with easy-axis magnetism and one with easy-plane. To select the correct Hamiltonian, we compared the calculated single-ion susceptibility to the measured susceptibility in Fig. \ref{flo:KYS_Susceptibility}. Measured susceptibility clearly shows an easy-plane magnetism at low temperature shifting to easy-axis magnetism at high temperature. The easy-plane model matches this behavior very well, and thus we select it as the correct model. However, this highlights the need to cross-check any fitted Hamiltonian with a different measure of magnetic anisotropy.

Although the calculated susceptibility qualitatively matches the measured susceptibility, the correspondence is not perfect (especially in the $c$ direction). Although this discrepancy is within error bars of the fitted Hamiltonian, two additional complications may also prevent perfect agreement: (i) magnetic exchange which shifts measured susceptibility values, and (ii) a shifting CEF Hamiltonian as a function of temperature. Because of these effects, we did not use susceptibility data to constrain the fit itself.

One final cross-check of the crystal field model can be made by comparing the calculated saturation magnetization to the measured KYbSe$_2$ 1/3 magnetization plateau. According to ref.~\cite{xing2021_KYS}, the 0.42~K 1/3 magnetization plateau occurs at $\mu_0 H = 4.2$~T,  $M = 0.569 \> \mu_B$. However, this value is offset by Van Vleck susceptibility which at low fields adds a linear offset to the ground state CEF magnetization. According to the fitted KYbSe$_2$ CEF model, the Van Vleck susceptibility is 0.0176~$\mu_B$/T at 0.42 K---which means we must subtract (0.0176~$\mu_B$/T)(4.2~T) = 0.074~$\mu_B$ from the measured plateau magnetization for a true 1/3 magnetization of 0.495~$\mu_B$. This is one third of the CEF predicted $ab$-plane saturation magnetization 1.48(9)~$\mu_B$. Thus we have high confidence that our fitted CEF model and associated $g$-tensor is correct.

\section{Onsager Reaction Field fits}

Here we outline the Onsager reaction field (ORF) approach for completeness. We consider the spin Hamiltonian
\begin{align}
\mathcal{H} & =\sum_{\left\langle i,j\right\rangle }\Bigl\{ J_{X}\left(S_{i}^{x}S_{j}^{x}+S_{i}^{y}S_{j}^{y}\right)+J_{Z}S_{i}^{z}S_{j}^{z}\nonumber \\
& +J_{A}\left[(S_{i}^{x}S_{j}^{x}-S_{i}^{y}S_{j}^{y})\cos\phi_{ij}-(S_{i}^{x}S_{j}^{y}+S_{i}^{y}S_{j}^{x})\sin\phi_{ij}\right],\nonumber \\
\label{eq:hamiltonian}
\end{align}
in which $\alpha \in \{x,y,z\}$ denote spin components with respect to Cartesian axes $\mathbf{x},\mathbf{y},\mathbf{z}$, and $\phi_{ij}\in\left\{ \frac{2\pi}{3},-\frac{2\pi}{3},0\right\} $ as specified in Ref.~\cite{Paddison_2020}.  
We use the Onsager reaction-field (ORF) approach \cite{Brout_1967,Hohlwein_2003,Wysin_2000} to calculate magnetic diffuse scattering patterns.
The Fourier transform of the interactions is given by
\begin{equation}
J_{\alpha\beta}(\mathbf{Q})\equiv-\sum_{\mathbf{R}}J_{\alpha\beta}(\mathbf{R})e^{-\mathrm{i}\mathbf{Q}\cdot\mathbf{R}},
\end{equation}
where $J_{\alpha\beta}(\mathbf{R})$ is the coefficient of $S_{i}^{\alpha}S_{j}^{\beta}$
in Eq.\,(\ref{eq:hamiltonian}) for sites $i$ and $j$ separated
by a lattice vector $\mathbf{R}$. 

The magnetic diffuse scattering
intensity is given, in the reaction-field approximation, by
\begin{equation}
I_{\mathrm{ORF}}(\mathbf{Q})\propto {[f(Q)]^{2}}\sum_{\mu=1}^{3}\frac{|\mathbf{s}_{\mu}(\mathbf{Q})|^{2}}{1-\chi_{0}(\lambda_{\mu}(\mathbf{Q})-\lambda)},\label{eq:intensity_onsager}
\end{equation}
where $\chi_{0}=1/3T$ is the Curie susceptibility and $\lambda_{\mu}$ denotes the eigenvalues of the interaction matrix, where $\mu$ labels its $3$ eigenmodes. The structure factor 
\begin{equation}
\mathbf{s}_{\mu}(\mathbf{Q})=\sum_{\alpha}(\hat{\mathbf{n}}_{\alpha}-\mathbf{Q}\thinspace\hat{\mathbf{n}}_{\alpha}\cdot\mathbf{Q}/Q^{2})g_{\alpha}U_{\mu}^{\alpha}, 
\end{equation}
where $\hat{\mathbf{n}}_{\alpha}\in\{\mathbf{x},\mathbf{y},\mathbf{z}\}$, $g_\alpha$ denotes components of the diagonal $g$-tensor, and $U_{\mu}^{\alpha}$ denotes the eigenvector components of the interaction matrix.
At each temperature, we obtain the reaction field $\lambda$ self-consistently by enforcing that $\sum_{\mu,\mathbf{q}}[1-\chi_{0}(\lambda_{\mu}(\mathbf{q)}-\lambda)]^{-1}=3N_{\mathbf{q}}$
for a grid of $N_{\mathbf{q}}=40^{3}$ wavevectors in the Brillouin
zone. 
The best fit values are given in the methods section of the main text.

We also performed the fit including the off-diagonal $J_B$ component.~\cite{Paddison_2020}. This quantity is difficult to determine because it depends upon distinguishing $K$ from $K'$, and there is some degree of twinning in KYbSe$_2$ which means we can only fit the magnitude of $J_B$. Nevertheless, for completeness we performed the ORF fit assuming a twinning model and found 
\begin{align}
J_X = 2.33(10) \> {\rm K} \quad & \quad J_Z = 2.28(10) \> {\rm K} \nonumber \\
J_A = -0.018(8) \> {\rm K} \quad & \quad J_2 = 0.11(2) \> {\rm K} \\
|J_B| = 0.00(5) \> {\rm K}.
\end{align}
The error bar indicates $|J_B|$ could be larger than $|J_A|$, but this is still much smaller than $J_X$ and $J_Y$, indicating that the Heisenberg model is still appropriate for KYbSe$_2$.

\section{Schwinger Boson calculations}

\begin{figure}
	\centering\includegraphics[width=0.48\textwidth]{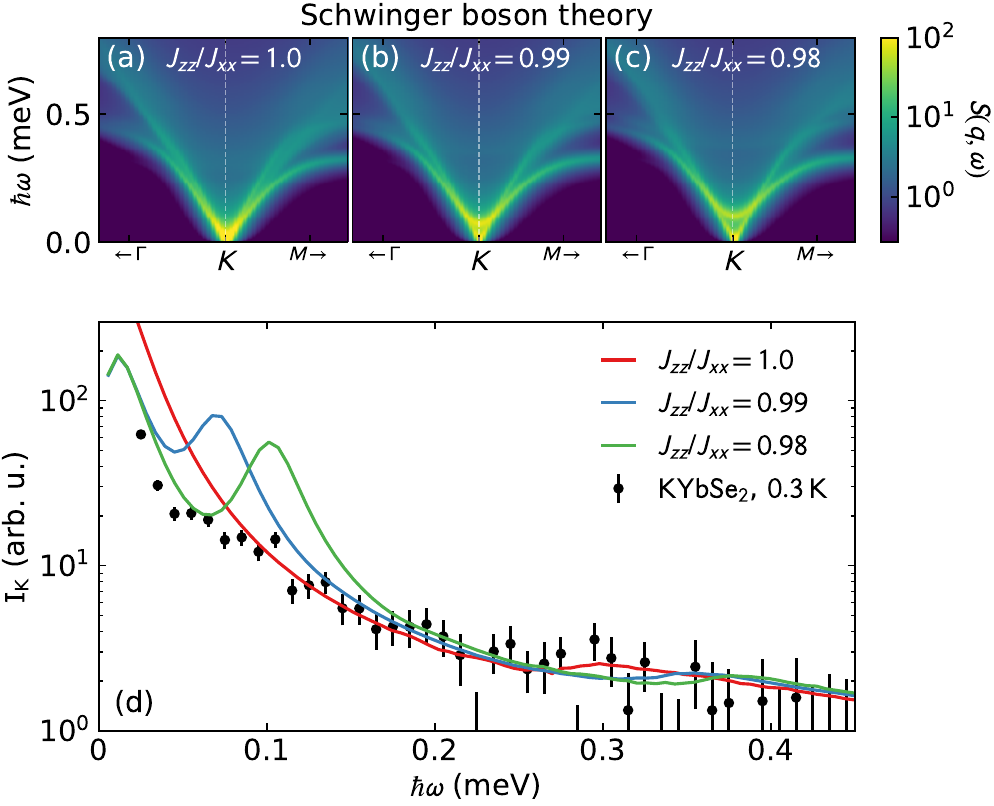}
	\caption{Effects of exchange anisotropy on the magnetic structure factor. Panels (a)-(c) show the low-energy structure factor predicted by Schwinger boson theory for different values of exchange anisotropy, where $J_{X} = J_{Y}$. As anisotropy increases, a well-defined mode becomes gapped. Panel (d) shows the energy-dependent scattering at $K$ compared to KYbSe$_2$ scattering at 0.3~K and $E_i = 1$~meV (0.02~meV FWHM). No finite-energy modes are seen on the scale that is predicted by Schwinger boson theory, suggesting a highly isotropic nearest neighbor exchange.}
	\label{flo:SBAnisotropy}
\end{figure}

Figure \ref{flo:SBAnisotropy} shows additional Schwinger boson calculated spectra for various values of nearest neighbor exchange anisotropy $J_{Z}/J_{X}$, where $J_{X} = J_{Y}$. As anisotropy increases, a low-energy mode at $K$ becomes gapped. However, high-resolution experimental scattering shows no such mode, even at the highest resolution setting (0.02~meV FWHM). This is consistent with the Onsager Reaction Field fits, which show $J_{zz}=J_{xx}$ to within uncertainty.

\begin{figure}
	\centering\includegraphics[width=0.47\textwidth]{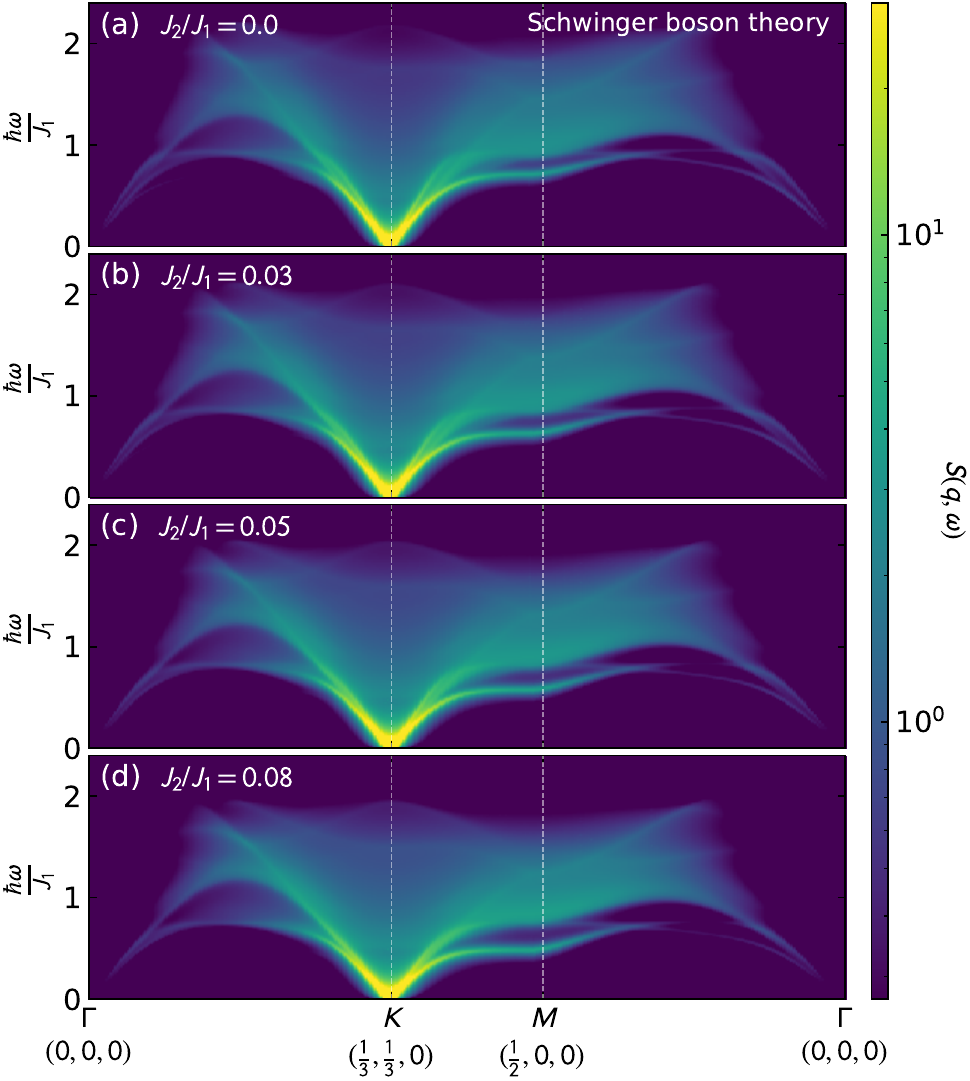}
	\caption{Schwinger boson calculations for the Heisenberg triangular lattice antiferromagnet for varying values of second nearest neighbor exchange.}
	\label{flo:SchwingerBosonsSI}
\end{figure}

Figure \ref{flo:SchwingerBosonsSI} shows the calculated Schwinger Boson spectra for different values of $J_2/J_1$. As $J_2/J_1$ increases, the spectra bandwidth slightly decreases, while the gap at $M$ grows smaller. 

\newpage

\end{document}